\documentclass[aps,pre,preprint,numerical,a4paper]{revtex4-1}

\usepackage[margin=1.1in]{geometry}
\usepackage{color} 
\usepackage{mathrsfs}
\usepackage{graphicx}
\usepackage{amsmath}
\usepackage{bbm}
\usepackage{amsfonts}
\usepackage{amssymb}
\usepackage{amsbsy}
\usepackage[T1]{fontenc}
\usepackage{theorem}
\usepackage{stmaryrd}
\usepackage{tipa}
\usepackage{textcomp}
\usepackage{mathcmd}
\usepackage{subfigure}
\usepackage{caption}
\usepackage{epsfig}
\usepackage{lmodern}
\usepackage{framed}
\usepackage{subeqnarray}
\usepackage{alphalph}
\usepackage[refpage,cfg,prefix]{nomencl}
\usepackage{multirow}
\usepackage{xifthen}
\usepackage{enumerate}
\usepackage{color}
\usepackage{wasysym}
\usepackage{bibentry}
\usepackage{todonotes}
\usepackage{verbatim}
\usepackage{booktabs}
\usepackage[hyperindex,breaklinks]{hyperref}
\usepackage{array}
\usepackage{pifont}
\hypersetup{
colorlinks=true,
urlcolor=blue,
linkcolor=black,
citecolor=black,
}
\usepackage{breakurl}
\usepackage{url}
\usepackage{setspace}
\usepackage{xr}
\usepackage{dsfont}
\usepackage{makecell}
\usepackage{tcolorbox}
\usepackage{fancyhdr}
\usepackage{subcaption}
\usepackage{float}
\usepackage{placeins}
\makenomenclature

\newcommand{\partder}[2]{\frac{\partial #1}{\partial #2}}
\newcommand{\ordder}[2]{\frac{\text{d} #1}{\text{d} #2}}

\newcommand{\secpartder}[2]{\frac{\partial^2 #1}{\partial #2 ^2}}

\renewcommand{\phi}{\varphi}

\graphicspath{{./Images/}}

\begin{document}
\raggedbottom

\title[Theoretical and experimental implications of the multi-Stage model for cell proliferation]{Multi-stage volume exclusion models for cell proliferation}
\author{John Carlo Dimaculangan, Cameron A. Smith, Christian A. Yates}
\affiliation{Department of Mathematical Sciences, University of Bath, BA2 7AY Bath, United Kingdom}
\date{\today}

\begin{abstract}
Cell proliferation and cell movement are fundamentally stochastic processes which lead to variability in the growth and spatial structure of cell populations in many biological settings, such as cell invasion, wound healing, and tumour growth. We develop stochastic, on-lattice agent-based models (ABMs) which incorporate volume exclusion, random movement, and multi-stage representations of the cell cycle. The multi-stage framework enables a more realistic representation of true cell cycle time distributions. We also introduce a novel form of myopic behaviour, in which cells sense their local environment when attempting to proliferate. For each ABM, we derive a corresponding continuum partial differential equation (PDE) description under the mean-field approximation. Using numerical simulations, we investigate how different proliferation mechanisms influence population-level dynamics in both the discrete and continuum models. In particular, we consider biologically relevant contexts of growth-to-confluence assays (using uniform initial conditions) and travelling wave behaviour associated with cell invasion. We examine how the PDE solutions compare with the behaviour of the corresponding ABMs averaged over many realisations. 
\end{abstract}

\maketitle

\section{Introduction}
Cell proliferation is a fundamental biological process which underpins tissue development, immune responses \cite{akberts2002mbc}, wound healing  \cite{hanahan2011hallmarks}, and cancer growth \cite{Gurtner2008wrr}. Understanding the cell cycle, the framework which regulates when and how cell proliferation occurs, is crucial to developing biologically realistic models of such processes. The cell cycle consists of four distinct phases: $\text{G}_1$ phase (Gap 1) where cells grow and prepare to replicate their DNA, S phase (Synthesis) where cells replicate their DNA,  $\text{G}_2$ phase (Gap 2) where cells prepare for mitosis and continue to grow, and M phase (Mitosis) where a cell divides \cite{schafer1998ccr}. Cells are able to temporarily arrest at key checkpoints prior to progressing into the S or  $\text{G}_2$ phases if they are sufficiently damaged or conditions for proliferation are unfavourable \citep{pietenpol2002ccc, johnson1999ccc}. In particular, it has been shown that crowded environments can cause a cell to become quiescent \citep{meerson2004pmd, leontieva2014cih}. Therefore, accurate mathematical models of such cell proliferation should incorporate density dependent behaviour within the cell cycle.

In this paper, we consider two complementary approaches to modelling the cell cycle. The first approach uses on-lattice ABMs which are able to capture individual cell behaviour and stochasticity, which is inherent to cell cycle and to cell movement. ABMs can be used to understand the spatial heterogeneity which can arise within a growing population of cells. Furthermore, it is relatively simple and intuitive to implement specific cellular behaviour or cell-cell interactions within these frameworks \cite{glen2019abm}. However, ABMs have the disadvantage of a high computational cost, and limited analytical tractability \cite{an2009abm}. 

The ability to accurately capture cell-cycle time distributions is essential for constructing biologically realistic models of cell proliferation. Previous work  often makes the simplifying assumption that cell cycle times follow an exponential distribution \citep{baar2016smi, mort2016rdm, turner2009cbc}. This assumption is convenient as it allows for simulations to be carried out using Gillespie's algorithm \cite{gillespie1977ess}. However, it can be shown that cell cycle times are not well represented by the exponential distribution \cite{golubev2016applications}.  In contrast, the multi-stage model for the cell cycle partitions the cell cycle into multiple stages (distinct from the biological phases above) with exponential waiting times, which cells must progress through prior to dividing. This allows the multi-stage model to yield cell cycle time distributions which accurately reflect experimentally observed cell cycle times \cite{yates2017msr}. Importantly, because these stages are modelled independently of each other, the multi-stage model can be implemented using Gillespie's algorithm. 

ABMs incorporating both cell proliferation and cell movement have also been widely studied. Some models assume that cell cycle times follow an exponential distribution \citep{baker2010cmf,simpson2009mss,simpson2010cip}, whereas other models adopt the multi-stage representation of the cell cycle \citep{simpson2018smc, gavagnin2019isc}. Under standard multi-stage models, cells choose the site of placement of the daughter cell at random from its neighbours, with the event aborted if the site is occupied. In contrast, myopic behaviour grants the ability for cells to sense their local environment. Previously, myopic behaviour has been studied in the context of cell movement by Landman and Fernando \cite{landman2011mrw}.  In this paper, we will continue to investigate spatially extended multi-stage ABMs and compare their behaviour to spatial ABMs which use the simpler assumption of exponentially distributed cell cycle times. We will also adapt the approach taken by Landman and Fernando and study myopic behaviour in the context of cell proliferation. We will determine how incorporating myopic behaviour into ABMs changes their behaviour, and study the effects of myopic behaviour together with the multi-stage model.

The second approach we take to modelling density dependent cell proliferation and cell movement is to use continuum (non-spatial) ordinary and (spatial) partial differential equations (ODEs/PDEs) which can capture the macroscale dynamics of a system.  There is well established theory (such as steady-state and bifurcation analysis) for many continuum models, and it is computationally cheaper to numerically solve a system of ODEs/PDEs in comparison to simulating an equivalent population of cells using ABMs. However, continuum models are unable to capture any stochastic behaviour within a system, and can neglect the local structure \cite{baker2010cmf}.

Non-spatial models of cell proliferation are typically formulated using coupled systems of ODEs \citep{vittadello2019mmi, gavagnin2021sog}. These frameworks incorporate realistic cell-cycle time distributions by adopting an underlying multi-stage structure, in which each ODE represents the density of cells in each different stage of the cell cycle. Delay differential equations (DDEs) have also been proposed \citep{vittadello2021nmm}, in which a distributed delay kernel is used to impose specific cell-cycle time distributions.  DDEs are also able to capture realistic cell cycle time distributions, but at the cost of a more complicated model. To account for spatial effects such as migration and crowding, these approaches must be extended. Classical spatial models often take the form of reaction–diffusion equations, such as FKPP-type equations with logistic growth \citep{murray2007mbi}. However, in their simplest form, these models implicitly assume exponential waiting times between cell division attempts. More accurate representations of cell-cycle dynamics can be obtained by extending these formulations to systems of coupled PDEs, where each PDE represents cells in a different cell cycle stage \citep{vittadello2018mmc, falco2025qcc}. While such models are capable of capturing crowding effects and structured cell-cycle progression, they are typically limited to a relatively small number of stages, as the analysis and parameter inference for these systems becomes more challenging with increasing model complexity.

In this paper, we derive the continuum PDE limits of our discrete ABMs under various assumptions. When these assumptions are valid, we demonstrate that the resulting continuum PDEs provide an accurate description of the average behaviour across many realisations of the corresponding ABM. We also investigate parameter regimes in which the mean-field assumption breaks down, and show that, in such cases, the growth dynamics predicted by the ABMs and their associated continuum PDE limits can differ significantly. This enables us to distinguish a priori between scenarios in which PDEs successfully capture the spatial structure of the underlying ABM and those in which they do not. Finally, we use the continuum PDE framework to recover known theoretical minimum wave-speeds for travelling wavefronts \cite{gavagnin2019isc}, and perform numerical simulations to assess how well these theoretical predictions agree with the behaviour observed in our ABMs.

The paper is organised as follows. In Section \ref{Section A}, we present the different implementations of ABMs and perform numerical simulations to investigate the differences in their behaviour. In Section \ref{Sec III: The effects of Cell motility}, we derive the continuum limits of each of our ABMs, and give conditions under which these continuum descriptions are valid. In Section \ref{Sec IV: Travelling Waves}, we study cell invasion and compare numerical experiments of our models to known theoretical results on the speed of invasion. We conclude in Section V with a discussion of our results and potential areas for future work.

\section{Agent based models for cell proliferation incorporating volume exclusion} \label{Section A}
In this section, we will present several on-lattice ABMs incorporating density dependence through volume exclusion for cell proliferation. We will consider two different proliferation models (exponential and multi-stage) together with two mechanisms for handling failed proliferation events for the multi-stage model. We then extend these frameworks through the introduction of myopic behaviour, which is a form of local sensing that has been previously studied in the context of movement \cite{landman2011mrw}. For each of the cell cycle models, we give the results of our numerical simulations and discuss the effects each model has on the growth and spatial structure of cell populations.

\subsection{The multi-stage model and volume exclusion}
We begin by introducing biologically motivated agent-based models for cell proliferation. We first introduce and give known results of proliferation models in a non-spatial context \cite{yates2017msr}, then extend these to spatial models incorporating volume exclusion \citep{simpson2009dpg, simpson2010cip}. Through numerical simulations, we will demonstrate that the non-spatial theory does not explain the influence of density dependence on cell proliferation dynamics. 

Throughout this section, we assume that all cells have an average cell cycle time of $T$. One method for simulating the cell cycle is to assume that cell cycle times follow an exponential distribution with rate $1/T$. We represent this process as the single reaction,
\begin{equation}
    X \overset{1/T}{\rightarrow} 2X.
\label{Reaction: exponential}
\end{equation}
Under this assumption, we can use Gillespie's stochastic simulation algorithm (SSA) \cite{gillespie1977ess} to simulate cell proliferation. We will call this method of simulating the cell cycle the \textit{Exponential model}. 

The Exponential model is simple to implement and is  analytically tractable. However, it has been shown that exponential distributions provide poor fits to experimentally observed cell cycle times \cite{yates2017msr}. Consequently, the underlying assumption required to utilise Gillespie's algorithm is invalid. Instead, we introduce the \textit{multi-stage model} which retains the ability to be simulated via Gillespie's algorithm, whilst also sampling from more realistic cell cycle time distributions. For the multi-stage model, we decompose the cell cycle into multiple stages, each with exponential waiting times, through which a cell must progress before it may divide. Upon division, the two newly produced daughter cells return to the initial stage of the cell cycle. We can represent a $K$-stage model of the cell cycle with the following series of reactions,
\begin{equation}
    X_1 \overset{\lambda_1}{\rightarrow} X_2 \overset{\lambda_2}{\rightarrow} ... \overset{\lambda_{K-1}}{\rightarrow} X_K \overset{\lambda_K}{\rightarrow}2X_1,
\label{Reaction: exponential}
\end{equation}
where the transition rates satisfy $\sum_{s=1}^K 1/\lambda_s = T$ (so that the average time to cell division under this model is also $T$).
In contrast to the exponential model, cell cycle times under the multi-stage model follow a hypoexponential distribution, the sum of a number of independent exponential random variables. Throughout this paper, we will focus on the case for which all transition rates are equal $\lambda_s = \lambda$. In this special case, we have explicit rate $\lambda = K/T$ for a $K$-stage model. Under equal progression rates, cell cycle times are Erlang distributed (the special case of a gamma distribution with integer shape parameter) with rate parameter $1/T$ and shape parameter $K$. This has been shown to provide a good match to real cell cycle times \cite{yates2017msr}. Note that when $K=1$, the multi-stage model reduces to the Exponential model.

These two models for cell proliferation have implications for cell population dynamics. We can use the probability master equation (describing the probability of observing the number of cells in each stage of the cell cycle) to derive ODEs which describe the evolution of the mean number of cells over time under a $K-$stage model. Given equal transition rates, the ODEs are analytically tractable and the mean number of cells can be shown to grow more slowly as the number of stages $K$ increases \citep{yates2017msr, kendall1948orv}. This suggests that given an observation of cell counts after a known period of time, we would predict a higher average cell cycle time under the exponential model than the multi-stage model (for $K>1$), demonstrating the importance of using correct cell cycle model when dealing with experimental data.

We now formulate an on-lattice, volume exclusion model for cell proliferation which incorporates realistic cell cycle times through the multi-stage representation of the cell cycle. This model is similar to previous implementations in \citep{simpson2010cip} which uses exponentially distributed cell cycle times. Cells move and proliferate on a lattice and we denote these sites by $(i,j)$. The lattice sites are square with length $\Delta > 0$, and there are $L_x$ and  $L_y$ sites along the horizontal and vertical axes respectively. Each lattice site can be occupied by at most one cell. Cells attempt to move according to a motility rate $r_m$. When a cell is chosen to move, it selects a neighbouring site (in the von-Neumann sense) uniformly at random to attempt to move into. If the site is empty, the cell moves into the chosen site, otherwise the movement event is aborted and the simulation continues. Given an average cell cycle time $T$, cells have proliferation rate $r_p = 1/T$. Using this, we can write the rate of progression of a cell through each stage of a $K$-stage model as $\lambda = Kr_p$. When a stage-$K$ cell is chosen to undergo a progression event, it chooses a neighbouring site uniformly at random to attempt to proliferate into. If the site is empty, the mother cell places a stage-$1$ daughter cell in the chosen site and itself returns to the beginning of the cell cycle (stage-$1$). Otherwise, the proliferation attempt fails and we handle this in two distinct ways. In the \textit{Reset model} the cell returns to stage-$1$, the beginning of the cell cycle. In the \textit{Remain model} the cell remains in the final stage of the cell cycle. When $K=1$, the reset and remain model are equivalent as the first and last stage are the same. In simulations, we employ Gillespie's algorithm to generate the time to the next progression or movement event. We select an event to execute with probability proportional to the event rates $r_m$ and $\lambda$, and then choose a cell on which to execute this event uniformly from the cell population. It is important to note that if a movement or proliferation attempt fails we continue the simulation without choosing a replacement event.

\begin{figure}[H]
\centering
\includegraphics[width=0.5\linewidth]{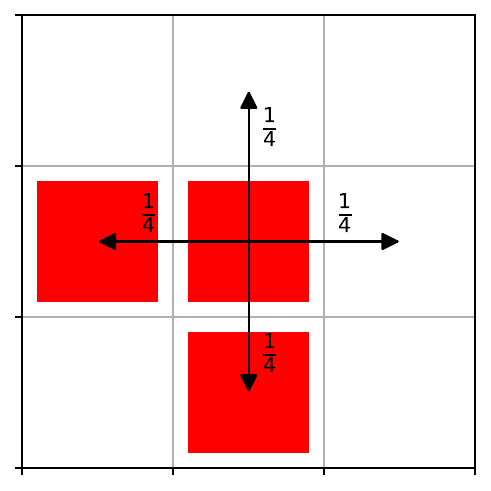}
\caption{\raggedright Schematic of how a cell chooses a neighbouring site during proliferation and movement events in our ABMs. Fractions denote the probability of moving in that direction.}
\label{schematics diffusion/prolif}
\end{figure}

\begin{figure}[H] 
\centering
\subfigure[][]{
	\includegraphics[width=0.29\textwidth]{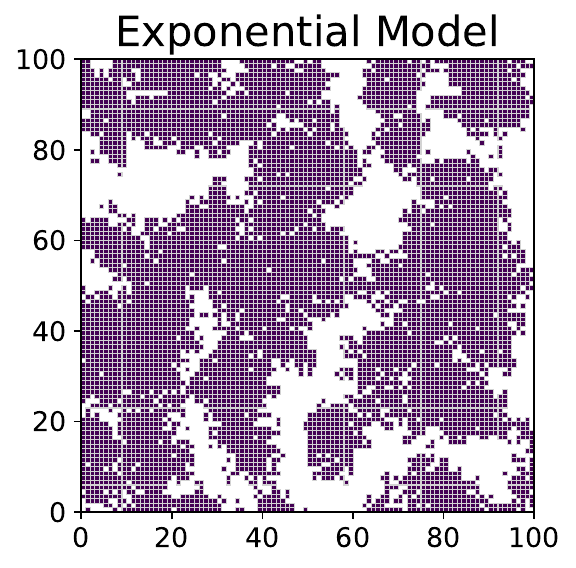}
    \label{exp:ABM}
}
\subfigure[][]{
	\includegraphics[width=0.29\textwidth]{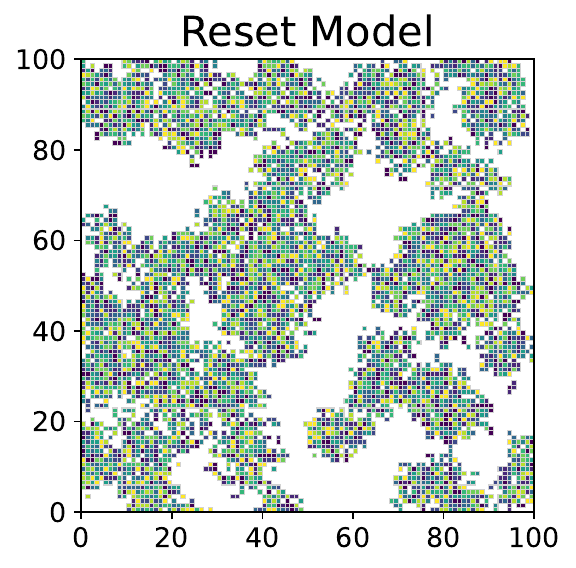}
    \label{reset:ABM}
}
\subfigure[][]{
	\includegraphics[width=0.35\textwidth]{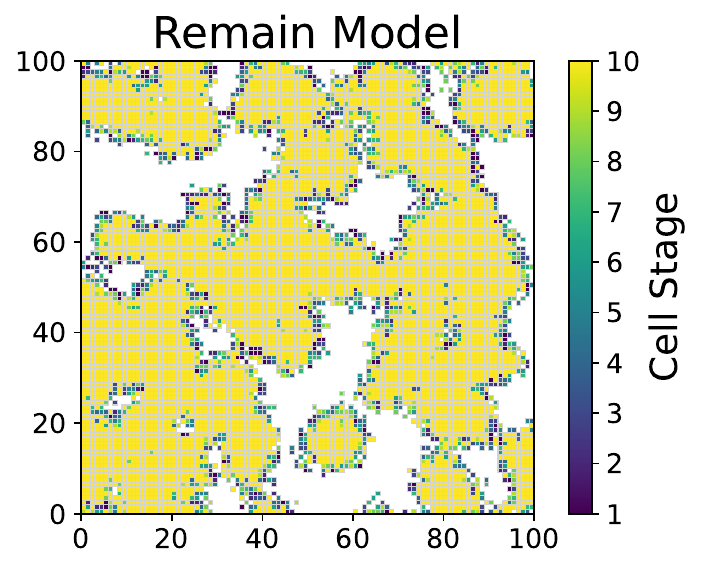}
    \label{remain:ABM}
}
\caption{Single realisations of various ABMs simulated over $t=10$ time units with motility and proliferation rate $r_m, r_p =1$ respectively. We simulate on a $100\times100$ lattice with lattice spacing $\Delta=1$ and periodic boundary conditions. For each simulation we initialise with an identically seeded uniformly populated lattice with $1\%$ density. Each cell begins in stage $1$, the beginning of the cell cycle. We display the final state of each simulation at time $t=10$ where occupied sites are coloured according to the current stage of the cell. (a) Exponential model. (b) Reset model with $K=10$ stages. (c) Remain model with $K=10$ stages.}
\label{Regular single realisations}
\end{figure}

We demonstrate in Figures~\ref{exp:ABM} and (b) below that under the Reset model all cells continue progressing through the cell cycle without arresting in high density environments. This is unrealistic behaviour as the implication of the Reset model is that cells which fail to proliferate would pass through all the cell cycle phases again and, for example, in real biological cells, duplicate their DNA again. Furthermore, experiments have shown that cells arrest in high density environments \citep{streichan2014scc, donker2022mgc}. However, under the Reset model, cells fail to arrest in crowded environments. 

In contrast, cells proliferating under the Remain model are able to arrest in high density environments. We observe this behaviour in the numerical simulations we carried out for cells proliferating under the multi-stage remain model. In Figure~\ref{remain:ABM}, most cells which are in the interior of a cluster have all arrested their cell cycle in stage $10$, the final stage. Note that arresting at the end of the cell cycle is not entirely biologically realistic behaviour either, as experimental evidence suggests that cells arrest prior to $S$ phase in high density environments. Although the Remain model provides a more realistic method of handling density dependence than the Reset and Exponential model, it is not a perfect representation of real cellular behaviour either.

\begin{figure}[H]
\centering
\subfigure[][]{
    \includegraphics[width=0.45\linewidth]{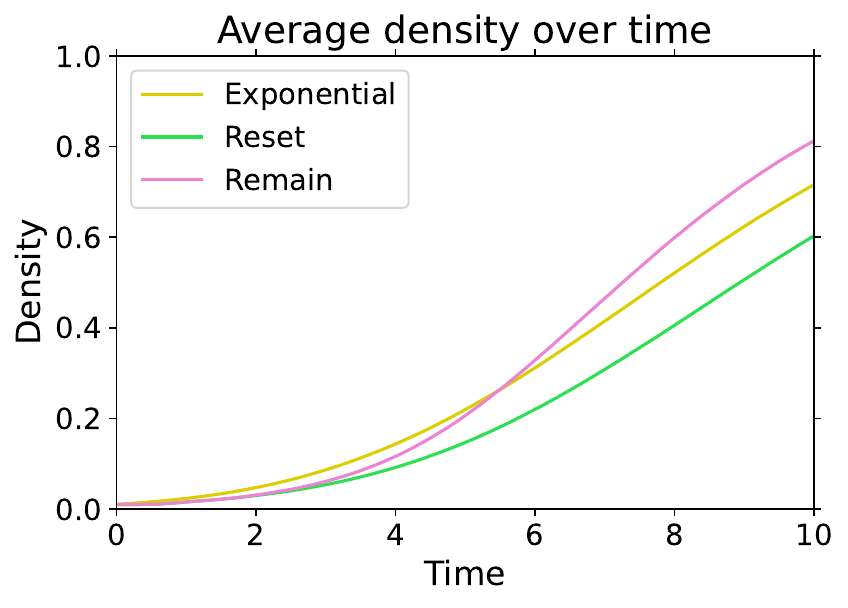}
    \label{non-myopic average density evolutions}
}
\subfigure[][]{
    \includegraphics[width=0.45\linewidth]{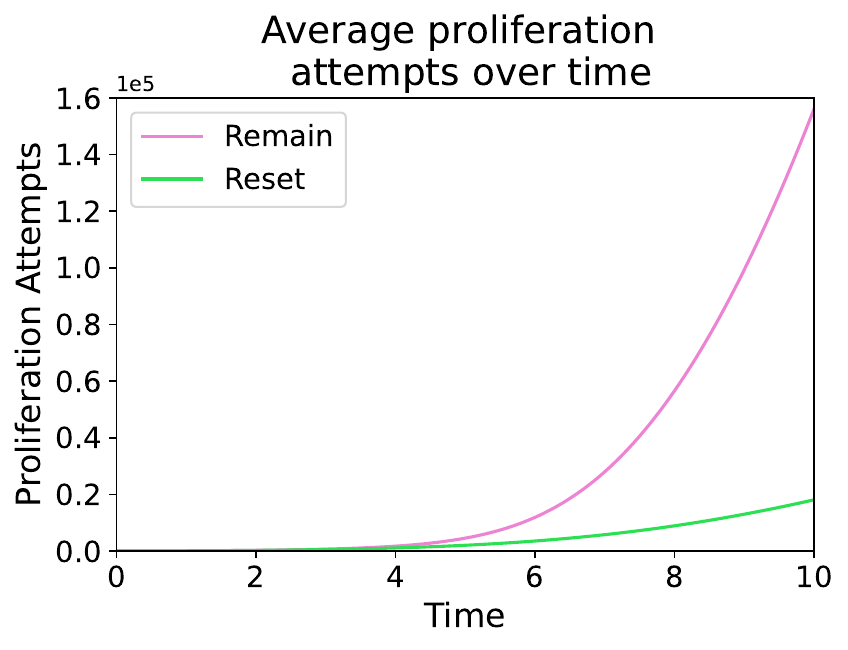}
    \label{num_atempts}
}
    \caption{(a) Density evolutions over $t=10$ time units for the exponential and multi-stage models averaged over $M=100$ realisations, with identical parameters to the simulations above. We initialise each realisation with different uniformly populated lattices with $1\%$ density as in Figure~\ref{Regular single realisations}. (b) Cumulative average number of proliferation attempts over time for the simulations in Figure \ref{non-myopic average density evolutions}.}
\end{figure}

The average density evolutions in Figure~\ref{non-myopic average density evolutions} for the spatial multi-stage and Exponential models exhibit markedly different rates of growth within their cell populations. We observe that the mean number of cells grows more quickly at intermediate densities in the multi-stage Remain model than in the Exponential model. This is because cells in the Remain model which fail to proliferate do not have to progress through the entire cell cycle again prior to another proliferation attempt \cite{yates2017msr}. We confirm that more proliferation attempts occur in the Remain model in  Figure~\ref{num_atempts}. When a cell fails to proliferate but still has empty neighbouring sites, the Remain model confers a proliferative advantage compared to the Exponential model and, by extension, the Reset model. This advantage grows as we increase the number of stages $K$ as a larger number of stages leads to a shorter amount of time between proliferation attempts in the Remain model (on average $T/K$ time units compared to $T$ for Exponential/Reset).

We have shown that the Remain multi-stage model is able to capture realistic cellular behaviour where cells arrest in high density environments. In contrast, the Reset multi-stage model continues to progress through the cell cycle. We have also observed that the Remain multi-stage model leads to quicker growth of cell populations than the Exponential model. This proliferative advantage is driven by cells at intermediate densities attempting to divide more frequently. However, an artifact of more proliferation attempts in the Remain model  is more failed proliferation attempts. This is because, in our current model, cells are unaware of their local environment. This is biologically unrealistic as it has been observed that groups of neural crest cells are able to coordinate their migration \cite{kulesa2010cnc}. We will now study an extension of these ABMs where we allow cells to sense their local environment. This behaviour will allow cells to be more likely to successfully proliferate at the proliferation stage.
\subsection{Myopic behaviour} \label{Sec II B}
We now introduce myopic behaviour into our ABMs which allows us to model the ability for cells to sense their local environment. Previously, myopic behaviour has been studied in the context of movement \cite{landman2011mrw}. In this paper we implement myopic behaviour for proliferative events and study the consequences on cell populations. Our simulations show that myopic behaviour leads to quicker growth of cell populations under all our cell cycle models. We also observe, through numerical simulations, that cell populations proliferating under the Myopic Reset and Myopic Remain model grow at the same rate. This contrasts the behaviour of the Reset and Remain models without myopic behaviour, where we observed that cells grow more quickly under the Remain model than under the Reset model.

\begin{figure}[H]
    \centering
    \includegraphics[width=0.5\linewidth]{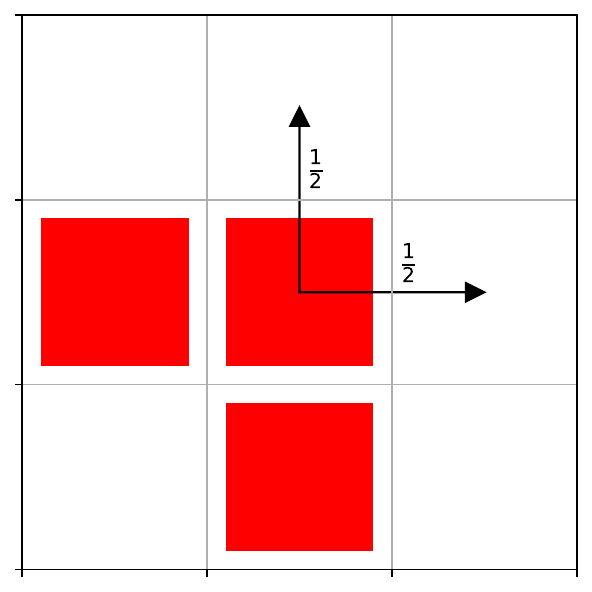}
    \caption{Example showing how a myopic cell chooses a neighbouring site during a proliferation event in our ABMs. Fractions denote the probability of choosing the site to proliferate into.}
    \label{Schematic for myopic proliferation}
\end{figure}

\begin{figure}[H] 
\centering
\subfigure[][]{
	\includegraphics[width=0.29\textwidth]{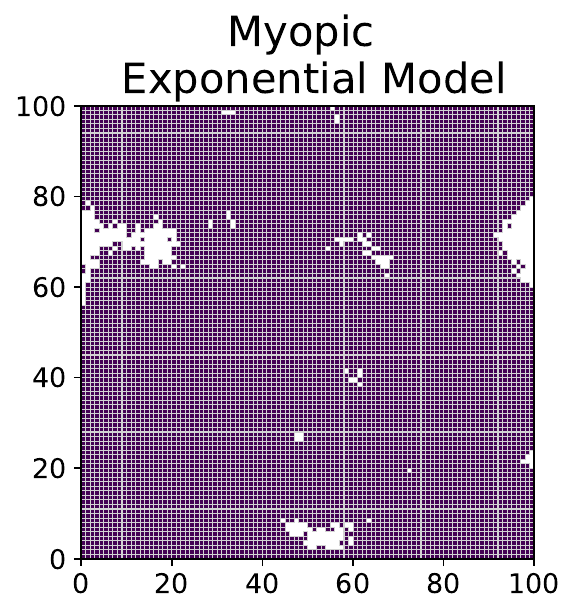}
}
\subfigure[][]{
	\includegraphics[width=0.29\textwidth]{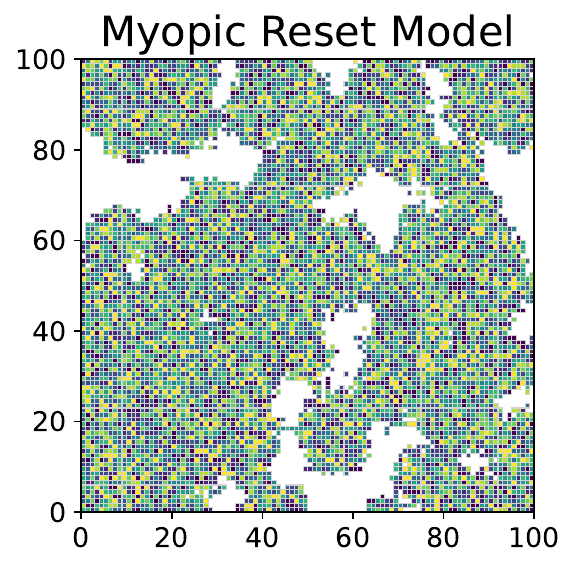}
}
\subfigure[][]{
	\includegraphics[width=0.35\textwidth]{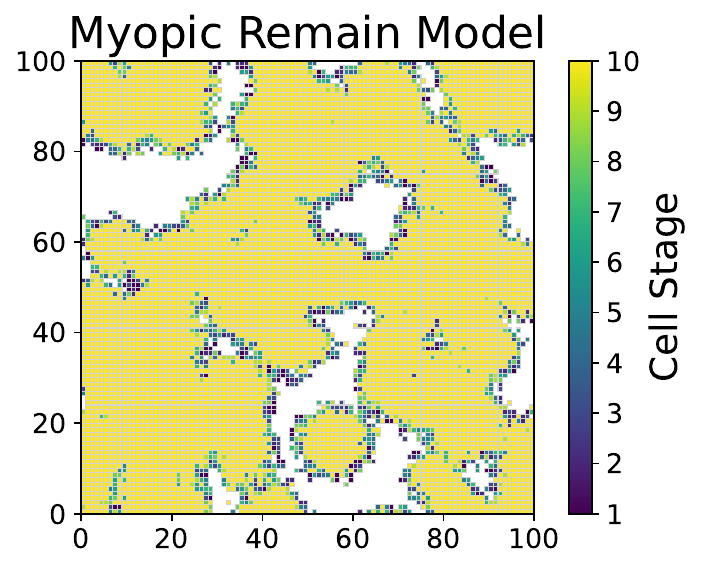}
}
\caption{Single realisations of various ABMs including myopic behaviour simulated over $t=10$ time units with identical parameters and identically seeded to Figure~\ref{Regular single realisations}. (a) Exponential model with myopic behaviour. (b) Reset model with myopic behaviour and $K=10$ stages. (c) Remain model with myopic behaviour and $K=10$ stages.}
\label{Myopic single realisations}
\end{figure}

In the previous simulations (exemplified in Figure~\ref{Regular single realisations}), cells are blind to the occupancy of neighbouring sites and choose between them randomly when attempting to proliferate. Now, when a final stage cell is chosen to proliferate, if there are unoccupied neighbouring sites, the cell will choose between these unoccupied sites uniformly at random (see Figure \ref{Schematic for myopic proliferation}). Otherwise, the proliferation attempt fails if there are no empty neighbours. The way in which we handle cells which have failed proliferation attempts in the three different cell cycle models stays the same. Consequently, we do not expect to see different cell arrest behaviour at very high local densities i.e. where many cells are completely surrounded by other cells. Indeed, Figure~\ref{Myopic single realisations} shows that cells that are completely surrounded arrest at the final stage in the Remain model but continue progressing through the cell cycle without arresting in high density environments in the Reset and Exponential models as was also seen under non-myopic proliferation (see Figure \ref{Regular single realisations}). This suggests that myopic behaviour neither improves nor worsens the realism of a cell cycle model at very high densities.

\begin{figure}[H]
    \centering
    \includegraphics[width=0.7\linewidth]{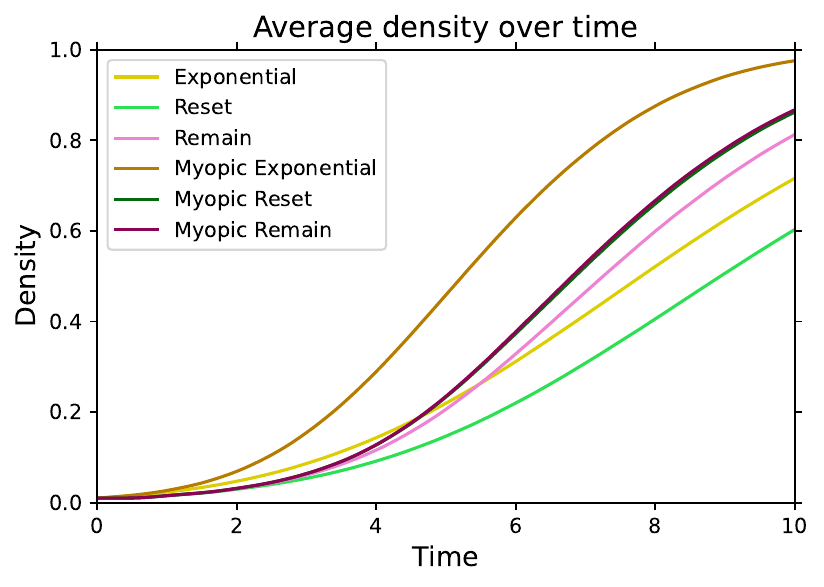}
    \caption{Average density evolutions over $t=10$ time units and $100$ realisations for every proliferation model with equal motility and proliferations rates $r_m=r_p=1$ as in Figure \ref{non-myopic average density evolutions}.  In each simulation, we initialise with a uniformly populated lattice with $1\%$ density. We compare the density evolutions of models including myopic behaviour to the evolutions without myopic behaviour. We distinguish between different proliferation models using different colours.}
    \label{myopic average density evolutions}
\end{figure}

In Figure~\ref{myopic average density evolutions}, we observe that every ABM with myopic behaviour leads to a higher population count than the corresponding ABMs without myopic behaviour. Indeed, at sufficiently large times, even the myopic proliferation strategy with the lowest density (Reset) outperforms the ``blind'' proliferation strategy with the highest density (Remain). This implies that myopic behaviour provides a larger proliferative advantage than remaining in the last stage of the cell cycle after a failed proliferation attempt. We can interpret this as myopic behaviour reducing the impact that high local densities have on cell proliferation.

Another observation we make is that, with myopic behaviour, cells proliferating under the exponential model now grow more quickly than cells proliferating under the multi-stage model. In the absence of failed proliferation events, cells which have just proliferated under the Exponential model are more likely to reattempt proliferation in short periods of time than under the multi-stage model. This is due to the larger variance in the exponential distribution compared to the Erlang distribution. Furthermore, myopic behaviour effectively erodes the difference in cell growth between the Reset and Remain models. The behaviour of the Remain model explains these observations. The Remain model's main advantage over the Reset model is that cells can attempt to proliferate again more quickly than those in the Reset model following a failed attempt. However, under the myopic model, cells which attempt but fail to proliferate have all four neighbours occupied. For intermediate cell densities (in which failed proliferation events become more prominent) and low motility relative to the proliferation rate, successful proliferation events are restricted by the motility of neighbouring cells and not by the frequency of proliferation attempts. A proliferating cell requires a neighbour to move away between proliferation attempts.

We have shown that myopic behaviour confers a noticeable advantage to proliferating cell populations. This is because cells are now able to sense their local environment, which is more biologically realistic than ABMs where cells are unaware of their surroundings. Myopic behaviour also maintains the ability for Remain model cells to arrest at high density environments. This suggests that the Myopic Remain model successfully captures three key characteristics of proliferating cells: realistic cell cycle durations, the ability to undergo cell cycle arrest, and sensitivity to the local environment.

In this section we have introduced spatial ABMs incorporating volume exclusion which simulate cell diffusion and progression through the cell cycle. Through decomposing the cell cycle into multiple stages, each with exponential waiting times, we are able to capture realistic cell cycle behaviour. We then introduced the Reset model, where cells return to the first stage of the cell cycle after a failed proliferation attempt, and the Remain model, where cells stay in the final stage of the cell cycle after a failed proliferation attempt. This captured the ability of cells to arrest at high density environments. Finally, we implemented a novel form of myopic behaviour at the proliferative stage, capturing the ability for cells to sense their local environment. In the next section, we will derive continuum PDEs from the ABMs we presented in this section. 

\section{Continuum PDE models } \label{Sec III: The effects of Cell motility}
In this section, we will derive corresponding continuum PDE limits for each ABM under the mean-field assumption. We seek a continuous time and space description for the average density of cells proliferating and diffusing under our various ABMs. Throughout this section, we assume that the average cell cycle time is fixed. Let the rate of proliferation be $r_p$. Recall that lattice spacing is given by $\Delta > 0$, $r_m$ is the motility rate, and that we have $K$ stages and rate of cell stage progression $\lambda = Kr_p$. Define $C_{i,j}^{(s)}(t)$ to be the density of stage $s$ cells in lattice site $(i,j)$ at time $t$ when averaged over many repeats over initially identically distributed simulations. We define $C_{i,j}(t) = \sum_{s=1}^K C_{i,j}^{(s)}(t)$ as the average occupancy of the site $(i,j)$ at time $t$. As each site contains at most one cell, it is immediately apparent that for any stage, $s$, the densities are bounded as follows $0 \leq C_{i,j}^{(s)} \leq C_{i,j} \leq 1$.

We now give the master equations for a $K$-stage Reset model following similar derivations in Simpson et al. \citep{simpson2009mss, simpson2010cip} where the authors consider cell proliferation and movement models which incorporate volume exclusion. The master equations are valid under the mean-field assumption, which assume that the lattice site occupancies are independent. Let $\tau$ be an infinitesimally small time step. By considering the probabilities that a cell can enter or exit a given site within the small time interval $[t,t+ \tau)$, we derive the following expression for the average density of stage $1$ cells at lattice site $(i,j)$ at time $t+\tau$. Note we omit the dependence of right hand side terms on $t$ for convenience.
\begin{align}
C_{i,j}^{(1)}(t+\tau)  &=  C_{i,j}^{(1)} \notag \\
&+ \frac{r_m \tau}{4}\left(1-C_{i,j}\right)\left(C_{i-1,j}^{(1)} + C_{i+1,j}^{(1)} + C_{i,j-1}^{(1)} + C_{i,j+1}^{(1)}\right) \notag\\
&- \frac{r_m \tau}{4} C_{i,j}^{(1)} \left(4 - C_{i-1,j} - C_{i+1,j} - C_{i,j-1} - C_{i,j+1}\right) \notag \\
&+ \lambda C^{(K)}_{i,j} \tau - \lambda C_{i,j}^{(1)} \tau 
\notag \\
&+ \frac{\lambda \tau}{4}(C_{i-1,j}^{(K)} + C_{i+1,j}^{(K)} + C_{i,j-1}^{(K)} + C_{i,j+1}^{(K)})(1-C_{i,j})  + O(\tau^2). \label{Reset stage 1 PDE derivation}
\end{align}
The second line on the right hand side of equation~(\ref{Reset stage 1 PDE derivation}) represents cells successfully moving into site $(i,j)$ whilst the third line represents cells successfully moving out of site $(i,j)$. The terms in the fourth line of this equation represent cells moving into stage $1$ from stage $K$, as a result of a successful proliferation attempt or as a result of a reset due to a failed proliferation attempt, and cells progressing out of stage $1$ to stage $2$. Note that there is no density dependent term capturing progression for stage $K$ cells as even after a failed proliferation event, a cell will return to stage $1$ under the Reset model. The final line of the equation represents a successful proliferation into site $(i,j)$ from a neighbouring site. 

Similarly for stage $s = 2,...,K$ cells, we have the following equation
\begin{align}
    C_{i,j}^{(s)}(t+\tau)  &=  C_{i,j}^{(s)}(t) \notag \\
    &+ \frac{r_m \tau}{4}\left(1-C_{i,j}(t)\right)\left( C_{i-1,j}^{(s)}(t) + C_{i+1,j}^{(s)}(t) + C_{i,j-1}^{(s)}(t)  + C_{i,j+1}^{(s)}\right) \notag\\
    &- \frac{r_m \tau}{4}C_{i,j}^{(s)}(t)\left(4 - C_{i-1,j}(t) - C_{i+1,j}(t) - C_{i,j-1}(t) - C_{i,j+1}(t)\right) \notag \\
    &+ \lambda C^{(s-1)}_{i,j}\tau- \lambda C_{i,j}^{(s)}\tau + O(\tau^2),
\label{Reset generic stage PDE derivation}
\end{align}
which is similar to equation~(\ref{Reset stage 1 PDE derivation}) but without the final line which resulted from proliferation. 

In order to determine the continuum limit, we first apply a Taylor expansion about site $(i,j)$ to the expressions for the occupancies of the neighbouring sites, $C_{i-1,j}, C_{i+1,j}, C_{i,j-1}, C_{i,j+1}$. Next, we subtract  $C_{i,j}^{(s)}(t)$ and divide through by $\tau$ in equations~(\ref{Reset stage 1 PDE derivation}) and (\ref{Reset generic stage PDE derivation}), and finally take the limits $\tau, \Delta \rightarrow 0$ whilst maintaining $r_m\Delta^2$ and $\lambda$ constant to obtain the following system of PDEs,
\begin{equation}
    \begin{cases}
        \partder{C^{(1)}}{t} = D[(1-C) \nabla^2C^{(1)} + C^{(1)}\nabla^2C] + \lambda C^{(K)}(2-C) - \lambda C^{(1)},  \\
        \partder{C^{(s)}}{t} = D[(1-C) \nabla^2C^{(s)} + C^{(s)}\nabla^2C] + \lambda C^{(s-1)} - \lambda C^{(s)}, \qquad \text{for }s=2,...,K, \\
    \end{cases}
    \label{Reset: PDEs (hold)}
\end{equation}
where the diffusion constant $D$ is given by the following limit,
\begin{equation}
        D = \underset{\tau, \Delta \rightarrow0}{\text{lim}} \frac{r_m\Delta^2}{4}.
    \label{Diffusion constant}
\end{equation}
In the PDEs~(\ref{Reset: PDEs (hold)}) above, the diffusion term is non-linear as the movement of any sub-population of cells is dependent on the density of total population of cells. If we sum over all the stages $s=1,..,K$, we obtain the following PDE for the density of cells under the Reset model
\begin{equation}
        \label{occupancy multi-stage PDE}
    \partder{C}{t} = D\nabla^2C + \lambda C^{(K)}(1-C).
\end{equation}
As opposed to the PDEs~(\ref{Reset: PDEs (hold)}), the diffusion term for the total cell density is linear in the above PDE~(\ref{occupancy multi-stage PDE}). Note that in the Exponential model (where $K=1$), we recover the Fisher-Kolmogorov-Petrovsky-Piskunov (FKPP) equation \cite{murray2007mbi}, a well studied PDE which is commonly used to model biological phenomenon such as wound healing \cite{maini2004travelling} and species invasion \cite{paul2022rfm}. 

We now give the master equations for the $K$-stage Remain model. For stage $1$ cells we have the following equation,
\begin{align}
    C_{i,j}^{(1)}(t+\tau) &= C_{i,j}^{(1)} \notag\\
    &+ \frac{r_m \tau}{4}(1-C_{i,j})(C_{i-1,j}^{(1)} + C_{i+1,j}^{(1)} + C_{i,j-1}^{(1)} + C_{i,j+1}^{(1)}) \notag \\
    &-
    \frac{r_m \tau}{4}C^{(1)}_{i,j} (4 - C_{i-1,j} - C_{i+1,j} - C_{i,j-1} - C_{i,j+1}) \notag \\
    &+ \frac{\lambda \tau}{4}C^{(K)}_{i,j}(4-C_{i-1,j}-C_{i+1,j}-C_{i,j-1}-C_{i,j+1}) -\lambda C^{(1)}_{i,j}\tau    \notag \\
    &+ \frac{\lambda \tau}{4}(1-C_{i,j})(C_{i-1,j}^{(K)} + C_{i+1,j}^{(K)} + C_{i,j-1}^{(K)} + C_{i,j+1}^{(K)}) + O(\tau^2). \label{Remain stage 1 PDE derivation}
\end{align}
This only differs from equation~(\ref{Reset stage 1 PDE derivation}) in the fourth line, where the term which represents stage $K$ cells progressing into stage $1$ is dependent on the density of neighbouring sites. This is because if a final stage cell fails to proliferate, then it remains in the final stage of the cycle under the Remain model. Similarly, the master equation for stage $K$ cells is altered due to the change in the way we handle failed proliferation events in the Remain model. This gives the following equation,
\begin{align}
    C_{i,j}^{(K)}(t+\tau) &= C_{i,j}^{(K)} \notag\\
    &+ \frac{r_m\tau}{4}(1-C_{i,j})(C_{i-1,j}^{(K)} + C_{i+1,j}^{(K)} + C_{i,j-1}^{(K)} + C_{i,j+1}^{(K)}) \notag \\
    &-
    \frac{r_m\tau}{4}C^{(K)}_{i,j} (4 - C_{i-1,j} - C_{i+1,j} - C_{i,j-1} - C_{i,j+1}) \notag \\
    &+ \lambda C^{(K-1)}_{i,j}\tau - \frac{\lambda \tau}{4}C^{(K)}_{i,j}(4-C_{i-1,j}-C_{i+1,j}-C_{i,j-1}-C_{i,j+1}) 
    \notag \\
    &+ O(\tau^2),   \label{Remain stage K PDE derivation}
\end{align}
where the term in the fourth line representing cell progression is density dependent as opposed to equation~(\ref{Reset generic stage PDE derivation}). For stage $s=2,..,K-1$ cells, the equation for $C_{i,j}^{(s)}(t+\tau)$ is precisely the same as in the Reset model, which is given by equation~(\ref{Reset generic stage PDE derivation}). From these equations~((\ref{Reset generic stage PDE derivation}), (\ref{Remain stage 1 PDE derivation}), (\ref{Remain stage K PDE derivation})) we can derive the continuum PDEs for the Remain model, using the same methodology and under the same assumptions as the Reset PDEs~(\ref{Reset: PDEs (hold)}):
\begin{equation}
    \begin{cases}
        \partder{C^{(1)}}{t} =  D[(1-C) \nabla^2C^{(1)} + C^{(1)}\nabla^2C] + 2\lambda C^{(K)}(1-C)  - \lambda C^{(1)},  \\
        \partder{C^{(s)}}{t} =  D[(1-C) \nabla^2C^{(s)} + C^{(s)}\nabla^2C] +  \lambda C^{(s-1)} - \lambda C^{(s)},  \qquad \text{for } s=2,..,K-1, \\
        \partder{C^{(K)}}{t} =  D[(1-C) \nabla^2C^{(K)}+ C^{(K)}\nabla^2C] +  \lambda C^{(K-1)} - \lambda C^{(K)}(1-C). \\
    \end{cases}
    \label{Remain: PDEs (hold)}
\end{equation}
We observe that the PDE describing the dynamics of stage $K$ cells has a density dependent progression term. This means that for high densities of cells, we expect that cells should arrest in stage $K$ which matches the behaviour observed in our ABM simulations of the Remain model.

If we sum over all stages, we obtain a PDE for the density of cells under the Remain model. 
\begin{equation}
    \label{occupancy multi-stage PDE 2}
    \partder{C}{t} = D\nabla^2C + \lambda C^{(K)}(1-C).
\end{equation}
Note that this PDE has the same functional form as the PDE~(\ref{occupancy multi-stage PDE}) describing the evolution of the total cell population under the Reset model. However, we do not expect the cell densities under the Reset and Remain model PDEs to evolve identically as PDEs~(\ref{occupancy multi-stage PDE}) and (\ref{occupancy multi-stage PDE 2}) are dependent on the population of stage-$K$ cells. As the PDEs~(\ref{Reset: PDEs (hold)}) and (\ref{Remain: PDEs (hold)}) which describe the dynamics of stage $K$ cells are different, this will result in different solutions for PDEs~(\ref{occupancy multi-stage PDE}) and (\ref{occupancy multi-stage PDE 2}). Under the Remain model, there is a density dependence in the proliferation term. Explicitly, proliferation depends on the density of stage-$K$ cells. As already mentioned, we expect more cells to accumulate in stage-$K$ in the Remain model than in the Reset model. Consequently, we expect to see cell counts growing more quickly under the Remain model PDEs, which matches our ABM simulations in Figures~\ref{myopic average density evolutions} and \ref{non-myopic average density evolutions}.

Now, we derive continuum PDE limits for ABMs which incorporate myopic behaviour. Let $P((i,j)\mid (i',j'))$ denote the probability that a myopic cell at site $(i',j')$ chooses to attempt to proliferate into site $(i,j)$. Using this notation, we have the following equation for stage $1$ cells under the Myopic Reset model,
\begin{align}
    C_{i,j}^{(1)}(t+\tau)  &=  C_{i,j}^{(1)} \notag \\
    &+ \frac{r_m\tau}{4}\left(1-C_{i,j}\right)\left(C_{i-1,j}^{(1)} + C_{i+1,j}^{(1)} + C_{i,j-1}^{(1)} + C_{i,j+1}^{(1)}\right) \notag\\
    &- \frac{r_m\tau}{4}C_{i,j}^{(1)}\left(4 - C_{i-1,j} - C_{i+1,j} - C_{i,j-1} - C_{i,j+1}\right) \notag \\
    &+ \lambda C^{(K)}_{i,j}\tau- \lambda C_{i,j}^{(1)}\tau
    \notag \\
    &+ \lambda \tau \left( C_{i-1,j}^{(K)}P((i,j)\mid(i-1,j)) +
    C_{i+1,j}^{(K)}P((i,j)\mid(i+1,j)) 
    \right)
    \notag \\
    &+ \lambda \tau \left( C_{i,j-1}^{(K)}P((i,j)\mid(i,j-1)) +
    C_{i,j+1}^{(K)}P((i,j)\mid(i,j+1))\right)
    \notag \\
    &+ O(\tau^2). \label{Myopic reset stage 1 PDE derivation}
\end{align}
The terms representing cell movement and stage progression in the above equation~(\ref{Myopic reset stage 1 PDE derivation}) are the same as in equation~(\ref{Reset stage 1 PDE derivation}) for stage 1 cells under the Reset model without myopic behaviour. However, the terms on the fourth and fifth lines of the right hand side, corresponding to cells proliferating into site $(i,j)$, are generalised. In our derivations for continuum PDEs corresponding to ABMs without myopic behaviour, $P((i,j)\mid(i',j')) = \frac{1}{4}(1-C_{i,j})$ for any neighbouring site $(i',j')$ of site $(i,j)$. Under Myopic models, this probability is now dependent on the occupancy of the neighbours of sites $(i',j')$ i.e. the neighbours of the neighbours of site $(i,j)$. Below, we give an equation for $P((i,j)\mid(i\pm1,j))$. 
\begin{align}
&P((i,j)\mid(i\pm 1,j)) 
= (1 - C_{i,j}) \times\sum_{l=0}^{3} \mathcal{N}_l(i\pm1, j),
\label{Myopic proliferation probability}
\end{align}
 where $ \mathcal{N}_l(i\pm1,j)$ represents the terms for the probability of proliferating into site $(i,j)$ from $(i\pm1, j)$ given that site $(i\pm 1,j )$ has $l$ occupied neighbouring sites. For $l=0,1,2,$ or $3$ occupied neighbours of site $(i\pm1, j)$ $\mathcal{N}_l =\mathcal{N}_l(i\pm1,j)$ takes the following form: 
 \begin{align}
     \mathcal{N}_0 &= \frac{1}{4} \left( 
    (1 - C_{i\pm2,j})(1 - C_{i\pm1,j-1})(1 - C_{i\pm1,j+1}) 
\right) \notag \\
    \mathcal{N}_1 &= \frac{1}{3} 
    C_{i\pm2,j}(1 - C_{i\pm1,j+1})(1 - C_{i\pm1,j-1}) \notag \\
    &+ \frac{1}{3}C_{i\pm1,j-1}(1 - C_{i\pm2,j})(1 - C_{i\pm1,j+1})  \notag \\
    &+ \frac{1}{3}C_{i\pm1,j+1}(1 - C_{i\pm2,j})(1 - C_{i\pm1,j-1}) \notag \\
    \mathcal{N}_2 &= \frac{1}{2} 
    C_{i\pm2,j}C_{i\pm1,j-1}(1 - C_{i\pm1,j+1}) \notag \\
    &+ \frac{1}{2}C_{i\pm2,j}C_{i\pm1,j+1}(1 - C_{i\pm1,j-1})  \notag \\
    &+ \frac{1}{2}C_{i\pm1,j+1}C_{i\pm1,j-1}(1 - C_{i\pm2,j}) \notag \\
    \mathcal{N}_3 &= C_{i\pm2,j}C_{i\pm1,j-1}C_{i\pm1,j+1}.
    \label{Neighbours: (i+1,j)}
 \end{align}
We can define $P((i,j)\mid(i,j\pm 1))$ analogously by using equation~(\ref{Myopic proliferation probability}). Substituting $P((i,j)\mid(i-1,j)), \ P((i,j)\mid(i+1,j)), \ P((i,j)\mid(i,j-1)), \text{ and }  P((i,j)\mid(i,j+1))$ into equation~(\ref{Myopic reset stage 1 PDE derivation}) using expressions similar to equation~(\ref{Myopic proliferation probability})), then Taylor expanding and taking $\Delta, \tau \rightarrow 0$ yields the following PDE for stage-$1$ cells under the Myopic Reset model (see Appendix \ref{Appendix: Derivation} for more details),
\begin{equation}
\partder{C^{(1)}}{t} = D[(1-C) \nabla^2C^{(1)} + C^{(1)}\nabla^2C] + \lambda C^{(K)}(2-C^4) - \lambda C^{(1)},
\end{equation}
where $D$ is given by the same limit as in equation~(\ref{Diffusion constant}). The PDEs describing the evolution of cells in stage $s = 2,..,K$  are the same as those in the Reset model without myopic behaviour~(\ref{Reset: PDEs (hold)}). This is because myopic behaviour only affects the probability of a successful proliferation event. Below, we give the full system of PDEs describing the Myopic Reset model.
\begin{equation}
    \begin{cases}
        \partder{C^{(1)}}{t} = D [(1-C)\nabla^2C^{(1)} + C^{(1)}\nabla^2C] + \lambda C^{(K)}(2-C^4)  - \lambda C^{(1)} , \\
        \partder{C^{(s)}}{t} = D [(1-C)\nabla^2C^{(s)} + C^{(s)}\nabla^2C] +  \lambda C^{(s-1)} - \lambda C^{(s)}, \qquad \text{for } s=2,..,K. \\
    \end{cases}
    \label{Myopic reset: PDEs (hold)}
\end{equation}
Summing over all cell stages gives a PDE for the total cell density,
\begin{equation}
    \label{occupancy myopic PDE}
    \partder{C}{t} = D\nabla^2C + \lambda C^{(K)}(1-C^4).
\end{equation}
The proliferation term is now $\lambda C(1-C^4)$ which differs from the PDE~(\ref{occupancy multi-stage PDE}) describing the total density of cells under the Reset model without myopic behaviour. As $0 \leq C \leq 1$, we have $\lambda C^{(K)}(1-C^4) \geq \lambda C^{(K)}(1-C)$ and we therefore expect a population of cells exhibiting myopic behaviour to proliferate more quickly than populations of cells without myopic behaviour. Indeed, we observed this behaviour in our ABMs for non-myopic and myopic behaviour under the Reset model in Figure \ref{myopic average density evolutions}.

We finish this section by deriving the continuum PDEs for the Myopic Remain model. For stage $1$ cells we have the following equation,
\begin{align}
    C_{i,j}^{(1)}(t+\tau)  &=  C_{i,j}^{(1)} \notag \\
    &+ \frac{r_m\tau}{4}\left(1-C_{i,j}\right)\left(C_{i,j+1}^{(1)} + C_{i,j-1}^{(1)} + C_{i+1,j}^{(1)} + C_{i-1,j}^{(1)}\right) \notag\\
    &- \frac{r_m\tau}{4}C_{i,j}^{(1)}\{4 - C_{i,j+1} - C_{i,j-1}(t) - C_{i+1,j} - C_{i-1,j}\} \notag \\
    &+ \lambda C^{(K)}_{i,j}(1-C_{i-1,j}C_{i+1,j}C_{i,j-1}C_{i,j+1})\tau- \lambda C_{i,j}^{(1)}\tau 
    \notag \\
    &+ \lambda \tau \left( C_{i-1,j}^{(K)}P((i,j)\mid(i-1,j)) +
    C_{i+1,j}^{(K)}P((i,j)\mid(i+1,j)) 
    \right)
    \notag \\
    &+ \lambda \tau \left(C_{i,j-1}^{(K)}P((i,j)\mid(i,j-1)) +
    C_{i,j+1}^{(K)}P((i,j)\mid(i,j+1))\right)
    \notag \\
    &+ O(\tau^2). \label{Myopic remain stage 1 PDE derivation}
\end{align}
This is the same as the master equation~(\ref{Myopic reset stage 1 PDE derivation}) for stage-$1$ cells under the Myopic Reset model except for the fourth line on the right hand side. There is an extra density dependence (corresponding to the probability that there is an empty neighbour of site $(i,j)$) on the term representing stage-$K$ cells returning to the beginning of the cell cycle. This is because cells proliferating under the Remain model only return to the beginning of the cell cycle after a successful proliferation.

Similarly to the derivation of the system of PDEs for the Remain model without myopic behaviour, the master equations and corresponding PDEs for the Myopic Remain model are the same as in the Myopic Reset model for cells in stage $s=2,..,K-1$ (see equation~(\ref{Reset generic stage PDE derivation})). The master equation for stage $K$ cells is as follows:
\begin{align}
    C_{i,j}^{(K)}(t+\tau) &= C_{i,j}^{(K)} \notag\\
    &+ \frac{r_m \tau}{4}(1-C_{i,j})(C_{i-1,j}^{(K)} + C_{i+1,j}^{(K)} + C_{i,j-1}^{(K)} + C_{i,j+1}^{(K)}) \notag \\
    &-
    \frac{r_m \tau}{4}C^{(K)}_{i,j} (4 - C_{i-1,j} - C_{i+1,j} - C_{i,j-1} - C_{i,j+1}) \notag \\
    &+ \lambda C^{(K-1)}_{i,j}dt - \lambda \tau C^{(K)}_{i,j}(1-C_{i-1,j}C_{i+1,j}C_{i,j-1}C_{i,j+1}) 
    \notag \\
    &+ O(\tau^2),   \label{Myopic remain stage K PDE derivation}
\end{align}
This is the same as equation~(\ref{Remain stage K PDE derivation}) for stage $K$ cells under the Remain model without myopic behaviour except from the term on the fourth line of the right hand side as the probability that a cell exhibiting myopic behaviour successfully proliferates is the probability that a cell has an empty neighbour. Using equations~(\ref{Reset generic stage PDE derivation}), (\ref{Myopic remain stage 1 PDE derivation}), and (\ref{Myopic remain stage K PDE derivation}), we can derive the following system of PDEs as we have done for our previous models,
\begin{equation}
    \begin{cases}
        \partder{C^{(1)}}{t} = D [(1-C)\nabla^2C^{(1)} + C^{(1)}\nabla^2C] + 2\lambda C^{(K)}(1-C^4)  - \lambda C^{(1)} , \\
        \partder{C^{(s)}}{t} = D [(1-C)\nabla^2C^{(s)} + C^{(s)}\nabla^2C] +  \lambda C^{(s-1)} - \lambda C^{(s)}, \qquad \text{if } s=2,..,K-1, \\
        \partder{C^{(K)}}{t} = D [(1-C)\nabla^2C^{(K)} + C^{(K)}\nabla^2C] +  \lambda C^{(K-1)} - \lambda C^{(K)}(1-C^4). \\
    \end{cases}
    \label{Myopic remain: PDEs (hold)}
\end{equation}
As is the case for the models without myopic behaviour (see equations~(\ref{occupancy multi-stage PDE}) and (\ref{occupancy multi-stage PDE 2})), the PDE describing the evolution of the total density of cells proliferating under the Myopic Remain model is the same as PDE~(\ref{occupancy myopic PDE}) which describes the density of cells proliferating under the Myopic Reset model:

\begin{equation}
    \label{occupancy myopic PDE 2}
    \partder{C}{t} = D\nabla^2C + \lambda C^{(K)}(1-C^4).
\end{equation}

For the $K$-stage model, we have derived four corresponding systems of PDEs~(\ref{Reset: PDEs (hold)}), (\ref{Remain: PDEs (hold)}), (\ref{Myopic reset: PDEs (hold)}), (\ref{Myopic remain: PDEs (hold)}) which capture the full combination of reset versus remain and myopic versus non-myopic behaviours. From these PDEs, we can immediately observe behaviour that we had observed previously from their respective ABMs. The Remain model PDEs ((\ref{Remain: PDEs (hold)}), (\ref{Myopic remain: PDEs (hold)})) contain a density dependent proliferation term in the equation for stage-$K$ cells. This means we expect cells to arrest in stage-$K$ which agrees with our ABM simulations. From the PDEs~(\ref{occupancy multi-stage PDE}) and (\ref{occupancy myopic PDE}) which describe the evolution of the total density of cells over time under the non-myopic and myopic models respectively, we see that cell numbers grow more quickly under the Remain model than the Reset model as we expect to see more cells in stage $K$ under the Remain model than in the Reset model. We further observe that myopic cells have a proliferative advantage over non-myopic cells due to the weaker density dependence in the proliferative terms $(\lambda (1-C)$ for PDEs~(\ref{occupancy multi-stage PDE}) compared to $(\lambda (1-C^4))$ PDEs~(\ref{occupancy myopic PDE})). 

Although these PDEs have characteristics which match their respective ABMs, it is important to note that we have derived these continuum PDE limits under various assumptions. In particular, we assume that as $\tau, \Delta \rightarrow 0$ we have that $r_m\Delta^2$ and $\lambda$ are held constant imply that 
\begin{equation}
    \underset{\tau,\Delta \rightarrow0}{\text{lim}} \frac{\lambda}{r_m} = 0.
    \label{Parameter agreemeent}
\end{equation}
This condition requires that the motility and progression rate parameters to satisfy $r_m \gg \lambda$ in order to achieve good agreement between the ABM simulations and the corresponding continuum PDE limit. 

In this section, we have derived continuum PDEs (under the mean-field approximation) which correspond to the ABMs we presented in Section \ref{Section A}. We observed how incorporating the multi-stage representation of the cell cycle changes the system of PDEs. Reset, remain and myopic behaviour within our ABMs lead to different density dependent terms within the derived systems of related PDEs. In the next section, we will use the continuum PDEs to investigate the behaviour of the various proliferation models we have presented under different parameter regimes. We also identify parameter values under which the continuum models yield close agreements with their corresponding ABMs.

\section{The effects of cell motility} \label{Sec III}
It is well known that mean-field models fail to accurately approximate the behaviour of discrete ABMs when the motility rate of a system is insufficiently high compared to the proliferation rate \citep{baker2010cmf, markham2014cmm}. We will demonstrate that this is the case by comparing the average ABM behaviour over many realisations and their continuum limits at high densities. We investigate the behaviour of numerical solutions of the derived PDEs under a uniform initial conditions. Using the pair correlation function (PCF), a measure for the pairwise correlation of a lattice, we will qualitatively predict the degree of agreement between the ensemble averages of ABMs and their corresponding continuum limits. We will compare the average ABM behaviour in our different proliferation models to the continuum limit, and demonstrate through numerical simulations that the PCF can be used to explain the observed discrepancies. 

Throughout this section, we consider a uniform initial condition with periodic boundary condition in a box $[0,L_x] \times [0,L_y]$ and compare the solutions of our PDEs to the behaviour of their corresponding ABMs from Section \ref{Section A}. Studying uniform initial conditions is relevant to modelling growth-to-confluence assays \citep{cai2007msm, parker2018ied}. Given a constant initial density of cells $C^{(s)}(x,y,0) = \kappa^{(s)}$ for some $\kappa^{(s)} \in (0,1)$ in each stage $s=1,..,K$ and periodic boundary conditions, the solutions to the systems of PDEs~(\ref{Reset: PDEs (hold)}), (\ref{Remain: PDEs (hold)}), (\ref{Myopic reset: PDEs (hold)}), (\ref{Myopic remain: PDEs (hold)}) are not spatially dependent. This means that solutions are of the form $C^{(s)}(x,y,t) = C^{(s)}(t) $ and we may reduce the problem of solving a system of PDEs to solving a system of ODEs. We can do this by dropping the spatial terms from the PDEs~(\ref{Reset: PDEs (hold)}), (\ref{Remain: PDEs (hold)}), (\ref{Myopic reset: PDEs (hold)}), (\ref{Myopic remain: PDEs (hold)}). Notably, under a uniform initial condition, the solutions are independent of the diffusion coefficient $D$. This is because the uniform initial condition combined with the translationally invariant mechanisms, preserves uniformity through all time. 

For general initial conditions, as we increase the motility rate $r_m$ in ABM simulations the mean-field assumption becomes more appropriate. Therefore, for a fixed progression rate $\lambda$ and a given uniform initial condition, we may interpret the solutions of the continuum PDEs as describing the limiting behaviour of the corresponding ABMs as we take the motility rate $r_m \rightarrow \infty$. 

In Figure~\ref{uniform PDE solutions} below, we compare the density evolutions for each PDE to their corresponding ABM using the simulations we performed in Figure \ref{myopic average density evolutions} where we set the motility rate to be $r_m=1$. Every continuum PDE predicts a quicker rate of growth than the corresponding ABM. The poor agreements between our PDE models and ABMs can be explained by the low motility rate relative to the progression rate in the ABM. Due to the relatively low motility rate, pairwise spatial correlations build up during ABM simulations, which means that the neighbouring sites of a cell are more likely to be occupied on a lattice of the same density in which cell positions are uncorrelated. This results in populations of cells growing more slowly in our ABMs compared to our PDEs. 

\begin{figure}[H]
    \centering
    \includegraphics[width=0.8\linewidth]{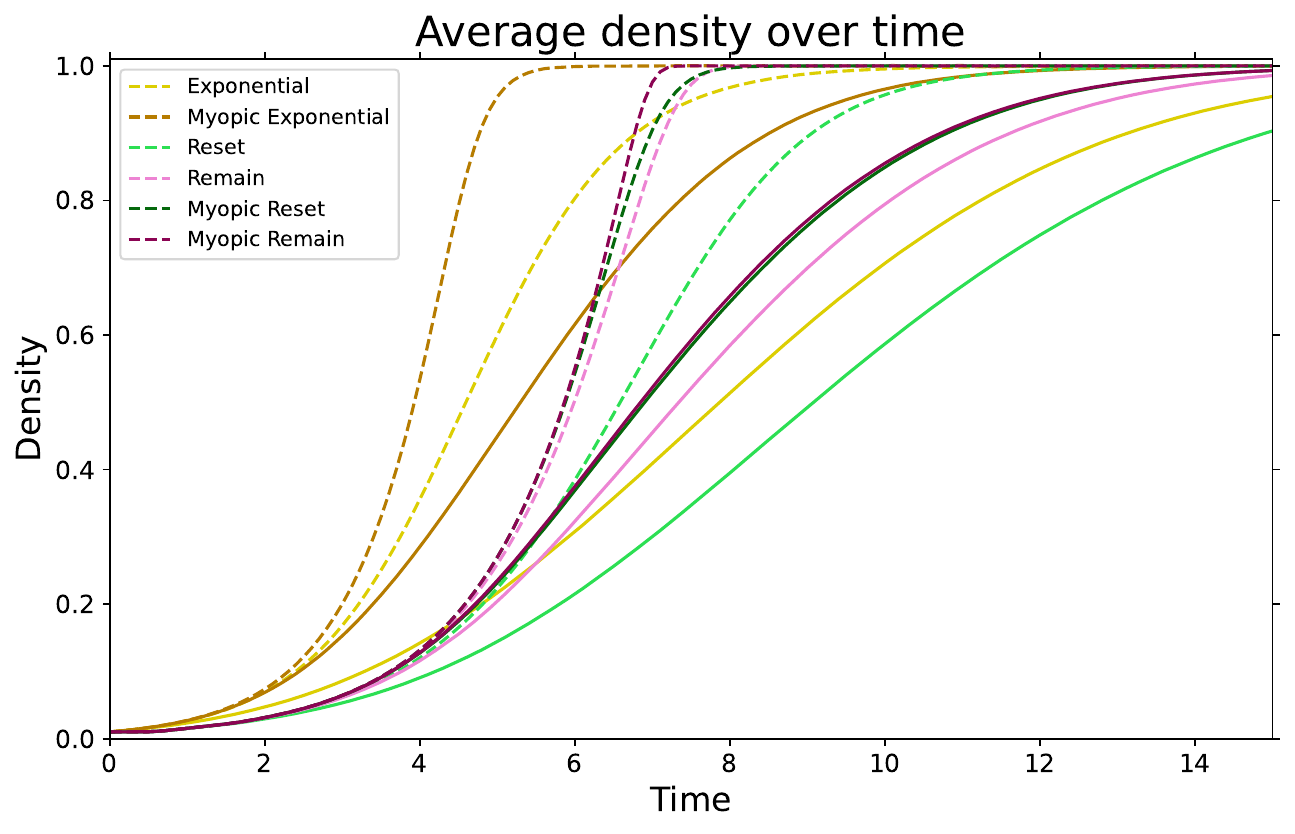}
    \caption{Comparing the numerical solutions of the PDEs~(\ref{occupancy multi-stage PDE}), (\ref{occupancy multi-stage PDE 2}), (\ref{occupancy myopic PDE}), (\ref{occupancy myopic PDE 2}) to the average ABM behaviour over $15$ time units. We initialise with the following uniform initial conditions $C^{(1)}(0) = C(0) = 0.01$ and $C^{(s)} = 0$ for $s=2,..,K$, and parameters $r_m=r_p=1$, so that the initial conditions and parameters for the PDEs and ABMs are identical to the ABM simulations in Figure~\ref{myopic average density evolutions}. We obtain the solutions to the corresponding ODE systems using the $\texttt{RK45}$ method from the $\texttt{solve\_ivp}$ function from the $\texttt{Scipy}$ library in $\texttt{Python}$. PDE solutions are denoted using dashed lines and we distinguish between different models by using different colours. For comparison to the ABMs, we plot the average density evolutions over time from Figure~\ref{myopic average density evolutions} using the same colour scheme to denote models as in the PDEs, but in solid lines.}
    \label{uniform PDE solutions}
\end{figure}

We further observe that the proliferative advantage over the non-myopic Exponential model conferred by myopic and remain behaviour in the multi-stage model becomes apparent only at higher densities (at around density $0.9$) in the continuum PDEs when compared to the corresponding ABM simulations (at around density $0.2$). Since the proliferative advantages of myopic and remain behaviour arises exclusively when a cell attempts to move from stage $K$ to stage $1$, the impact of these models depend strongly on local cell density. Given a fixed cell density, cells in well-mixed systems (corresponding to ABMs with high motility rates) are less likely to have occupied neighbouring sites than in ABMs initialised with low motility rates. This implies that proliferation is less restricted by local density in the continuum PDE framework than in the low motility ABM simulations. Consequently, the relative proliferative benefit of myopic and remain behaviour is reduced compared with systems that exhibit pairwise spatial correlations. In the absence of volume exclusion effects, we know cell populations grow more rapidly under the exponential model than in multi-stage models. We suggest that the Exponential model maintains a higher growth rate for a longer period of time in the continuum PDE compared to the corresponding ABM due to the weakened density dependence. 

Another difference between the PDE solutions and ABM simulations is that, at higher densities, we are able to distinguish between the Myopic Reset and Myopic Remain models more easily in the PDE evolutions than in the average ABM density evolutions. Recall that when cells proliferate under myopic behaviour, they only fail to proliferate if all adjacent sites are occupied. It is therefore unlikely that a cell fails to proliferate until reaching higher density environments when exhibiting myopic behaviour. Under the assumption that there are no spatial correlations, cell proliferation is not limited by the ability for neighbouring cells to move out of the way but by the global density of the environment. This means that the more frequent proliferation attempts under the Myopic Remain model will lead to quicker growth than in the Myopic Reset model but not until higher densities. 

Despite these differences, we note there are also similarities when comparing the qualitative behaviour of the PDE solutions to that of the average ABM behaviour over many realisations. For example, we can see that myopic behaviour leads to quicker population growth rates under all cell proliferation models in both the PDE models and average ABM behaviour. Furthermore, we can see that the Remain model does offer a proliferative advantage over the Reset model in the continuum PDEs as well as in the ABMs. We also see that at low densities, cells grow more quickly under the Exponential model than in the multi-stage models in the solutions of PDEs and average ABM behaviour.

We have demonstrated that under low motility rates, the continuum descriptions of our cell proliferation models provide poor agreements with their associated ABMs. This is due to pairwise spatial correlations which build up over time. We would like a method to quantify the degree of spatial correlation which arises from a given ABM. We will use this method to investigate how spatial correlations impact the degree of agreement between ABMs and their continuum PDE solutions under a uniform initial condition.

\subsection{Pair correlation function}
Pair correlation functions (PCFs) can be used to determine the degree of spatial correlation in a lattice filled with cells. We define a PCF following previous notation from work analysing different PCFs \cite{gavagnin2018pcf}. Given a $(X \times Y)$ lattice of cells, let $M$ be the occupancy matrix whose entries denote whether a lattice site is occupied or not. If a site $(x,y)$ is occupied, we set $M_{xy} = 1$, otherwise we set $M_{xy} = 0$. Given an occupancy matrix $M$, let $\psi^M$ be the set of all pairs of cells in the lattice which can be written as follows,
\begin{equation}
\psi^M = \left\{ (\mathbf{a}, \mathbf{b}) \in \mathbb{L} \times \mathbb{L} : \mathbf{a} \neq \mathbf{b}, \ M_\mathbf{a} = M_\mathbf{b} = 1 \right\},
\label{pairs of cells}
\end{equation}
where $\mathbb{L} = \{ 1,..,X\} \times \{ 1,..,Y\}$ denotes the set of possible sites of a lattice. In order to measure the degree of spatial correlations within a lattice, we need to determine a metric for our lattices. As our ABMs use the von-Neumann neighbourhood for both proliferation and movement, and they are periodic, the appropriate metric to use for distance is the metric induced by the $L^1$ norm i.e. for $(\mathbf{a}, \mathbf{b}) \in \mathbb{L} \times \mathbb{L} $
\begin{align}
    ||\mathbf{a} - \mathbf{b}||=|| (x_1,y_1) - (x_2,y_2)|| &=  \min \left\{|x_1-x_2|, \ L_x - |x_1-x_2|\right\}
    \notag \\
    & + \min \left\{|y_1-y_2|, \ L_y - |y_1-y_2|\right\}.
    \label{L1 norm}
\end{align}
Using definitions (\ref{pairs of cells}) and (\ref{L1 norm}) we can define the subsets of cell pairs separated by distance $d$ in matrix $M$ to be

\begin{equation}
    C^M(d) = \left\{ (\textbf{a}, \textbf{b}) \in \psi^M: || \textbf{a}- \textbf{b}|| = d\right\}.
\end{equation}
We define the PCF for a given matrix $M$ to be the ratio between the number of pairs of cells separated by distance $d$, and the expected number of pairs separated by distance $d$ within a uniformly populated lattice at the same density. Let $N$ be the number of occupied sites in the matrix $M$ (i.e. the number of non-zero entries), and $U$ be a random matrix with the same dimensions as $M$ with entries $U_{xy} = 1$ for $N$ sites chosen uniformly at random without replacement and $U_{xy}=0$ otherwise. We define the PCF for a matrix $M$ as follows
\begin{equation}
    f^M(d) = \frac{|C^M(d)|}{\mathbb{E}[C^U(d)]},
    \label{PCF Definition}
\end{equation}
\citet{gavagnin2018pcf} show that the expected number of sites separated by distance $d$ under metric (\ref{L1 norm}) is $\mathbb{E}[C^U(d)] = \frac{2dN(N-1)}{L_xL_y -1}$. This means we can rewrite the PCF (\ref{PCF Definition}) as
\begin{equation}
f^M(d) = \frac{(XY-1)|C^M(d)|}{2dN(N-1)}.
    \label{PCF simplified}
\end{equation}

Given lattices from multiple realisations of a cell proliferation ABM, we can calculate the PCF for each individual lattice and find the average value of the PCF, $\hat f(d)$ over all realisations. If $\hat f(d) = 1$ for all distances $d$, this means that the occupied sites of the lattice under the ABM are uniformly distributed which implies there are no spatial correlations. If $\hat f(d) > 1$ for small distances $d$, then there are more pairs of occupied sites which are close together than we would expect under a uniformly distributed lattice. This would imply the ABM gives rise to positive spatial correlations at short distances. Similarly, an average PCF $\hat f(d) < 1$ for small $d$ suggests the ABM causes negative spatial correlations at short distances. More generally, larger quantities $|\hat f(d) -1|$ suggests a stronger spatial correlation. 

\begin{figure}[H] 
\centering
\subfigure[][]{
	\includegraphics[width=0.45\textwidth]{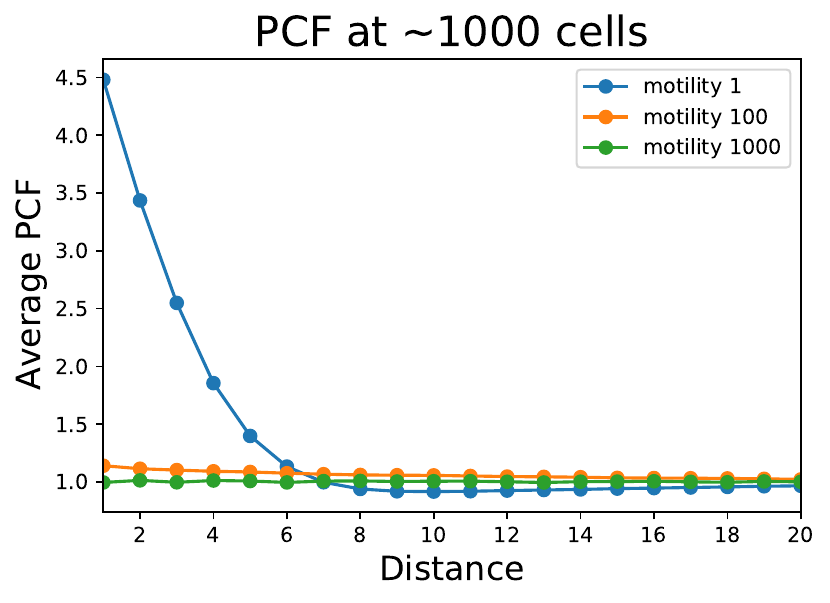}
}
\subfigure[][]{
	\includegraphics[width=0.45\textwidth]{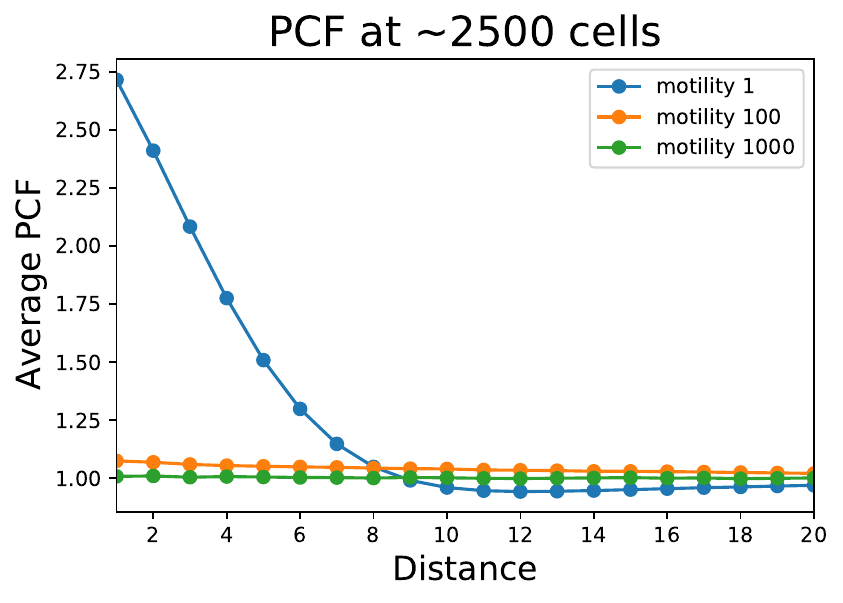}
}
\subfigure[][]{
	\includegraphics[width=0.45\textwidth]{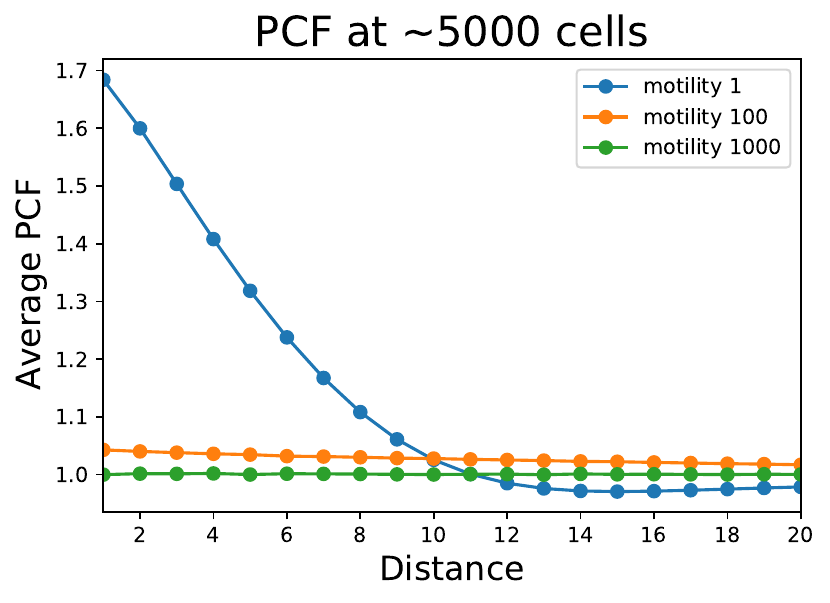}
}
\caption{Plotting the PCFs $\hat f(d)$ for $d=1,2,...,20$ averaged over $20$ realisations of the Myopic Remain ABM for motility rates $r_m=1,100,1000$ with a fixed proliferation rate $r_p = 1$ and number of stages $K=10$. Panels (a), (b), and (c) give the average PCF at a time where the average number of cells over the $20$ realisations (for the different motility rates) first exceeds $1000,2500,$ and $5000$ respectively.}
\label{PCF comparisons}
\end{figure}

\begin{figure}[H] 
\centering
\subfigure[][]{
	\includegraphics[width=0.3\textwidth]{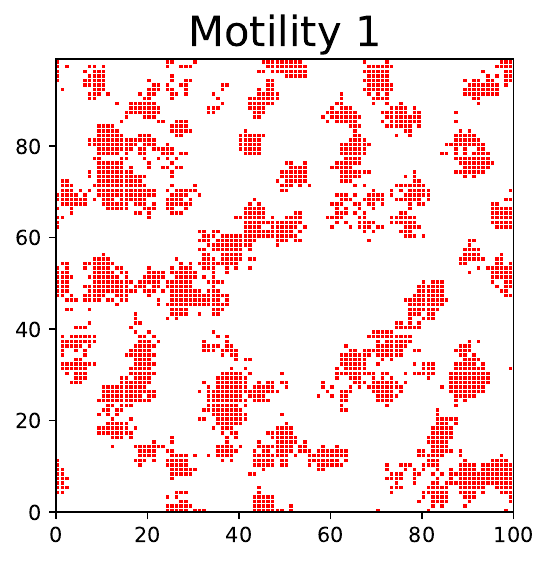}
}
\subfigure[][]{
	\includegraphics[width=0.3\textwidth]{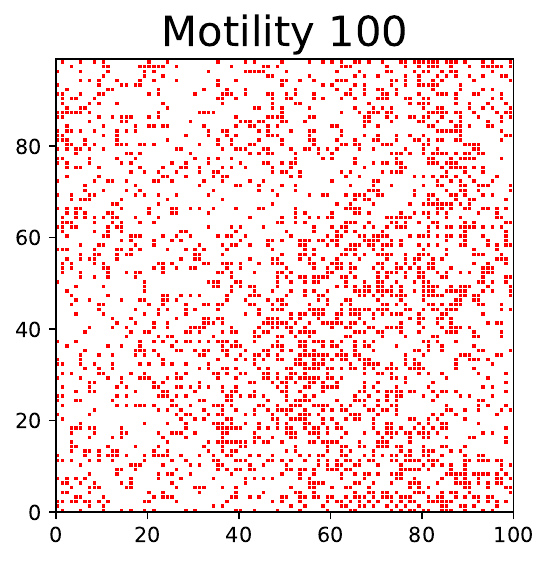}
}
\subfigure[][]{
	\includegraphics[width=0.3\textwidth]{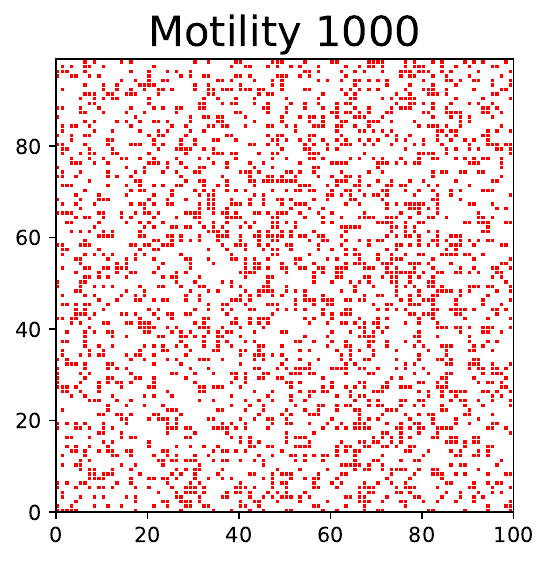}
}
\caption{Example occupancy matrices, $M$, from the simulations in Figure \ref{PCF comparisons} (b). Each lattice is taken at the first time step where the cell count first exceeds $2500$ cells. Panels (a), (b), and (c), correspond to motility rates $r_m = 1,100, $ and $1000$ respectively.}
\label{PCF Lattices}
\end{figure}

We see in the plots in Figure \ref{PCF comparisons} below that for motility rate $1$, the average PCFs for $d=1,.., 7$ are clearly above $1$ which suggests the ABM causes strong spatial correlations. Indeed, in Figure \ref{PCF Lattices} (a), we see that an example realisation of a Myopic Remain ABM with motility rate $1$ exhibits strong positive correlations at short distances when we reach around $2500$ cells. Increasing the motility rate to $r_m = 100$ lessens the degree of spatial correlation when compared to motility rate $r_m = 1$, however the PCF plots in Figure \ref{PCF comparisons} show that $\hat f(d) > 1$ for small distances $d$. This suggests there is still some spatial correlation at motility $r_m = 100$, which can be seen in Figure \ref{PCF Lattices} (b) where the spatial correlation is present but clearly much weaker than in Figure \ref{PCF Lattices} (a). When we increase the motility rate to $r_m = 1000$, then we can see in Figure \ref{PCF comparisons} that the average PCF satisfies $\hat f(d) \approx 1$ for all distances $d = 1,...,20$. This suggests that motility rate $r_m = 1000$ is sufficiently high to remove almost all spatial correlations within the lattice for this proliferation rate. We can observe that this holds in the example lattice in Figure \ref{PCF Lattices} (c), where there appears (by eye) to be no spatial correlation beyond what would be expected on a lattice of uniformly distributed cells. In Appendix \ref{Appendix: B PCF}, we demonstrate that the PCF behaves similarly for every other ABM.

Now that we have a method to measure the degree of spatial correlation of an ABM, we are able to predict how well the solution to a continuum PDE approximates the average behaviour over many realisations of its corresponding ABM. We will demonstrate that PCF allows us to successfully predict which motility rates give good agreement between the ABM and continuum PDE. 

\subsection{Comparing ABMs to continuum PDEs}
Throughout this section we will only consider the Myopic Remain model as this is the most complex model. We can perform analysis of every other model using similar methods. In order to determine a sufficiently high motility rate for good agreement between the continuum PDEs and corresponding ABMs for the different cell proliferation models which we have presented, we can calculate the average PCFs under these ABMs over many realisations for different motility rates until $\hat f(d) \approx 1$ for all distances $d$.

\begin{figure}[H]
    \centering
    \includegraphics[width=0.9\linewidth]{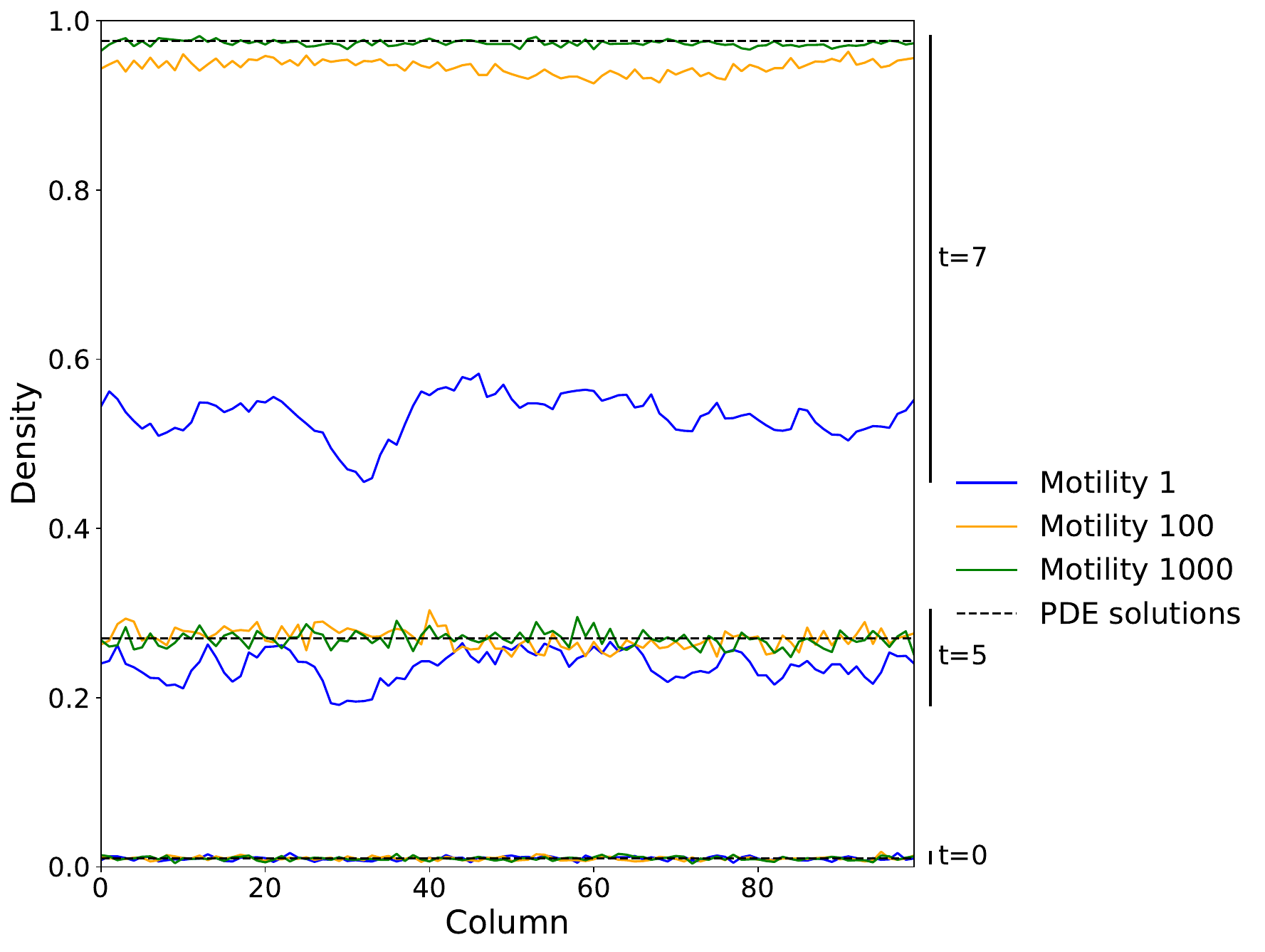}
    \caption{Comparing the solutions to PDE (\ref{occupancy myopic PDE 2}) to the average ABM column densities over $20$ realisations at times $t=0,5,$ and $7$. Parameters for the ABMs are the same as in Figure~\ref{PCF comparisons} and we initialise with a uniform density of $1\%$ as in the simulations for Figures \ref{myopic average density evolutions} and \ref{uniform PDE solutions}. We distinguish between motility rates $r_m=1,100,1000$ in ABMs by using different colours and denote the constant solution to PDE (\ref{occupancy myopic PDE 2}) over these columns with dashed black line at each time.}
    \label{Uniform initial condition: PDE vs ABM w/ different motility rates}
\end{figure}

In Figure \ref{Uniform initial condition: PDE vs ABM w/ different motility rates}, we plot the average column density in each ABM. Given occupancy matrices at time $t$ $M(t)$ of a realisation, we can calculate the density of column $i$, $\hat L_i(t)$, of a lattice at time $t$ as follows:    
\begin{equation}
    \hat L_i(t) = \frac{1}{Y}\sum_{j=0}^{Y} M_{ij}.
\end{equation}
We begin by noting that the average column density for ABMs is not constant due to the inherent stochasticity of our cell proliferation models. This is in contrast to the uniform profile over the whole domain when cell populations evolve under the system of PDEs. The variance in column densities is larger for smaller motility rates due to the the larger pairwise spatial correlations that build up at short distances for these smaller motility rates. When we average over a finite number of repeats, the column densities are not spread out as effectively for low motilities as they are for high motilities. Increasing the number of repeats we perform, we would see that the average column densities become increasingly uniform. 

We also observe that the agreement between the ABM and continuum PDE improves as we increase the motility rate. This is expected behaviour given the PCFs in Figure \ref{PCF comparisons}, which show $|\hat f(d) -1|$ decreases as we increase the motility rate for all distances $d$. At motility rate $r_m=1$, the agreement is poor at $t=5$ and very poor at $t=7$, which we expect given that the PCF $|\hat f(d)|$ is much larger than $1$ under motility rate $r_m=1$. Increasing the motility rate to $r_m=100$ in the ABM gives a better agreement to the continuum PDE for $t=5$, but a poor agreement at $t=7$ (although still an improvement on motility rate $r_m=1$). This is because the PCF satisfies $|\hat f(d) -1| > 0$ for $d\leq 20$, which means that there are postive spatial correlations in the ABM at these distances. Eventually, these build up enough to slow down the growth of the ABM when compared to the PDE. We see that motility rate $r_m=1000$ is sufficient to give good agreement between the ABM and the PDE at both times $t=5,7$. As the PCF $\hat f(d) \approx 1 $ at this motility rate, this suggests that the system is well-mixed, which explains the good agreement between the ABM and PDE.

In Figure \ref{uniform PDE solutions high_mot} below, we observe across all models, that the error between the discrete and continuum models decreases when we increase from motility rate $r_m= 1$ to motility rate $r_m=100$.

\begin{figure}[H]
    \centering
    \includegraphics[width=0.8\linewidth]{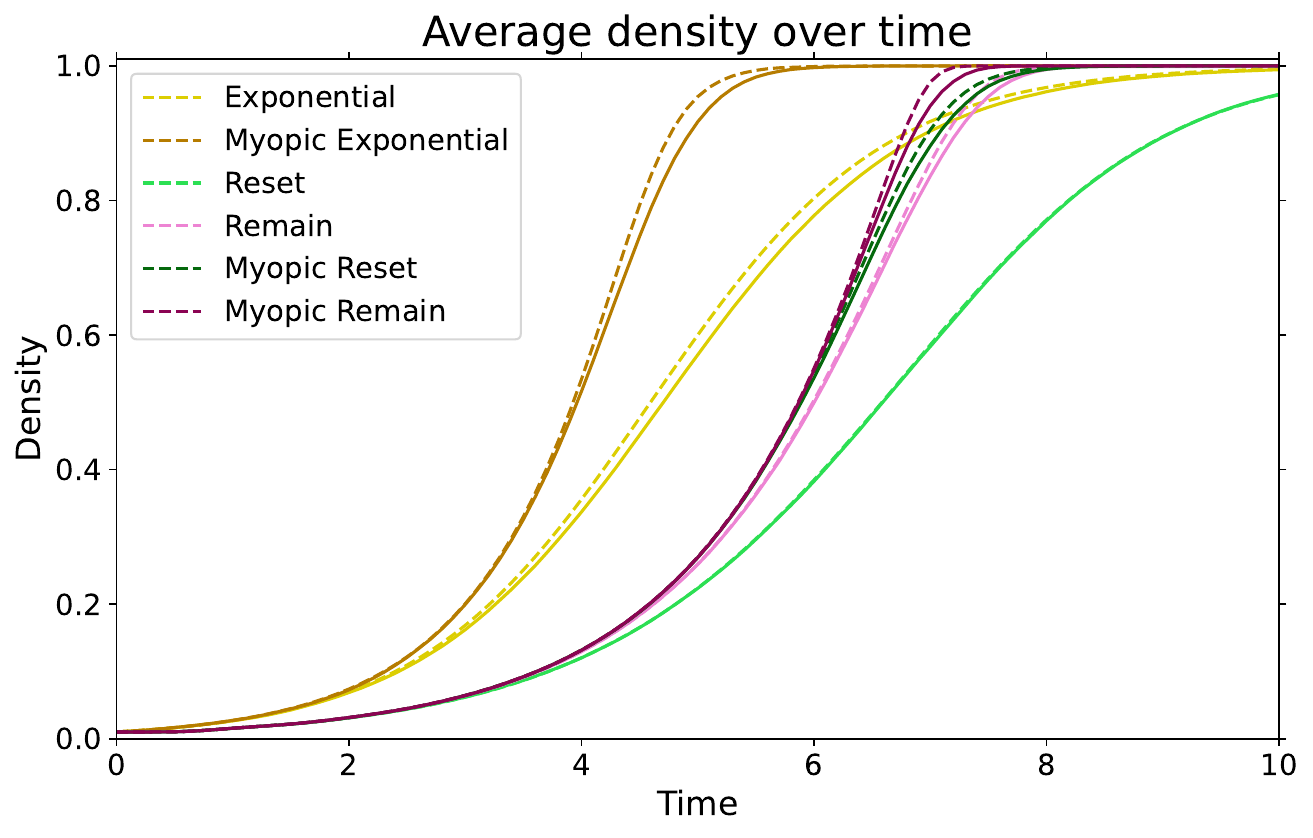}
    \caption{As in Figure \ref{uniform PDE solutions}, we compare the average behaviour of ABMs to the numerical solutions of the corresponding PDEs (\ref{occupancy multi-stage PDE}), (\ref{occupancy multi-stage PDE 2}), (\ref{occupancy myopic PDE}), (\ref{occupancy myopic PDE 2}). We initialise identically to the simulations in Figure \ref{uniform PDE solutions} and increase to motility rate $r_m = 100$ in our ABM simulations.  }
    \label{uniform PDE solutions high_mot}
\end{figure}

\begin{figure}[H] 
\centering
\subfigure[][]{
	\includegraphics[width=0.45\textwidth]{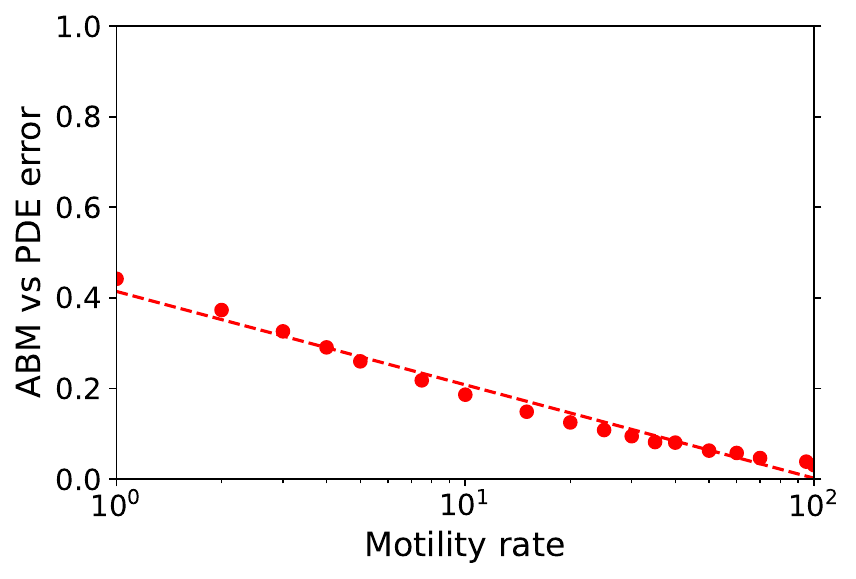}
}
\subfigure[][]{
	\includegraphics[width=0.45\textwidth]{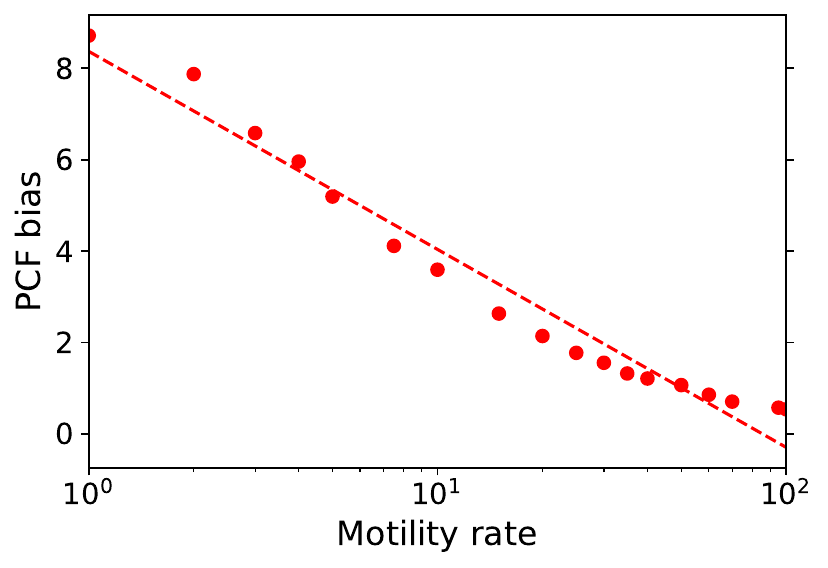}
}
\subfigure[][]{
	\includegraphics[width=0.45\textwidth]{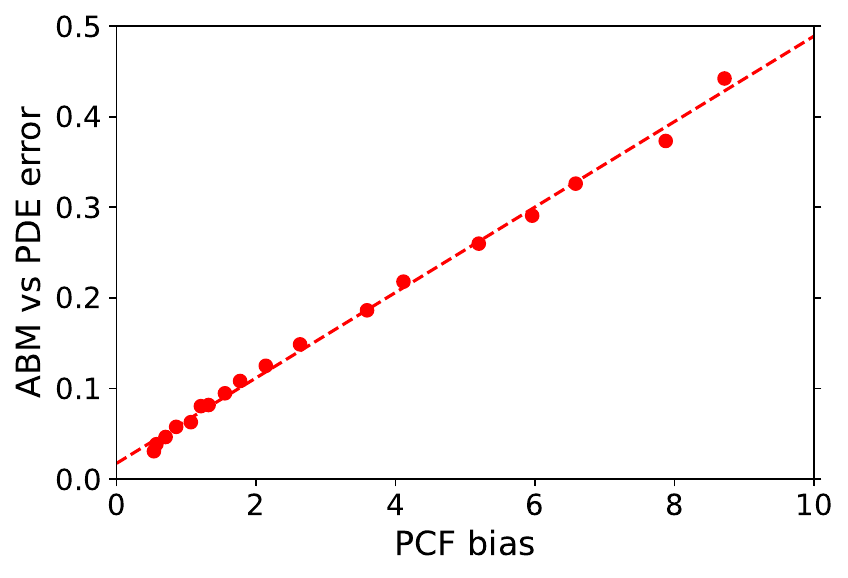}
}

\caption{We simulate the Myopic Remain ABM with $K=10$ stages with various motility rates $r_m$ between $1$ and $100$ and fixed proliferation rate $r_p = 1$ for $t=7$ time units. At each motility rate, we perform $20$ repeats, and calculate the average PCF $\hat f(d)$ at $t=1$. For each ABM realisation, we calculate the error at the final time of the simulation, $E(7)$, between the ABM solution. We considered the average error over all repeats. In Panel (a), we compare the motility rate to the error, in panel (b) we compare the motility rate to the PCF bias given by $|\hat f (d) -1|$, and in panel (c) we compare the PCF bias to the error. For each panel, we have plotted the line of best fit through our numerical simulations.}
\label{Comparing PCF-motility-error behaviour}
\end{figure}

In general, it is true that PCFs that demonstrate weaker spatial correlations are correlated with better agreement between the average discrete ABM behaviour and the continuum PDE solution. To show this, we considered the average column error between the ABM and PDE, $E(t)$,which can be calculated as below:
\begin{equation}
    E(t) = \frac{1}{Y}\sum_{i=1}^{Y} |\hat L_i(t) - C(t)|, 
\end{equation}
where $L_i(t)$ is the average density of column $i$ at time $t$ of the ABM, and $C(t)$ is the solution solution to the corresponding continuum PDE at time $t$ (which is constant in space, due to our choice of initial conditions). We observe that in Figure \ref{Comparing PCF-motility-error behaviour} that larger motility rates lead to a smaller error between the discrete and continuum model. Furthermore, we can also see that the PCF acts as a good predictor for the ABM-PDE error.

In this section, we have observed through numerical simulations that when the motility rate of cells is low, there is poor quantitative agreement between the average ABM behaviour (over many realisations) and the related continuum PDE. This is because under parameter regimes where motility is low, the assumptions we made in Section \ref{Sec III: The effects of Cell motility} to derive continuum models do not hold. In order to quantify the agreement between an ABM and the corresponding continuum PDE, we introduced the PCF, which enabled us to measure the strength of pairwise correlations in an ABM for a given motility rate. Under a uniform initial condition, we demonstrated that when the PCF differs from 1, the agreement between the continuum PDE and the average ABM behaviour is poor, whereas we observe good agreement when the PCF is approximately equal to 1 for all pairwise distances. We will now consider non-uniform initial conditions which lead to travelling waves, and study the behaviour of our ABMs and corresponding PDEs in this context.

\section{Travelling waves} \label{Sec IV: Travelling Waves}
In tissue invasion barrier assays performed by Streichan et al. \cite{streichan2014scc}, cell invasion was shown to be density dependent. At high local densities, cells were shown to arrest and may reactivate when space is available due to cells moving into the free space created when the barrier is removed. Models capturing this behaviour have been previously explored. For example, Falc\'{o} et al. \cite{falco2025qcc} proposed a coupled system of two PDEs which each incorporated density dependent terms to model the evolution of cell densities labelled by the FUCCI cell-cycle construct \cite{sakaue2008vsd}, with red cells in $G_0/G_1$ and green cells in $S/G_2/M$. Using this model, they derived wave-speeds for the travelling waves. Vittadello et al. \cite{vittadello2018mmc} studied continuum PDEs corresponding to a three-stage Remain model, which included an intermediate labelled stage representing early $S$-phase cells. Similarly, they derived wave-speeds for this specific system of PDEs. 

In this section, we will use the systems of PDEs~(\ref{Reset: PDEs (hold)}), (\ref{Remain: PDEs (hold)}), (\ref{Myopic reset: PDEs (hold)}), (\ref{Myopic remain: PDEs (hold)}) to model cell invasion for a generic $K$-stage model. Recall that these equations correspond to the different combinations of reset/remain and non-myopic/myopic behaviour. We first give theory which allows us to calculate the minimum speed of a steady-state wavefront for these continuum PDE models. We then run numerical simulations to confirm that the theoretical predictions hold for our systems of PDEs. We also determine under which parameter regimes these theoretical predictions are accurate for the corresponding ABMs. For different ABMs, we compare how travelling waves form and discuss why the behaviour in each ABM gives rise to these characteristics.

\subsection{Theoretical wave-speeds}
We study the theoretical wave-speeds of the systems of PDEs~(\ref{Reset: PDEs (hold)}), (\ref{Remain: PDEs (hold)}), (\ref{Myopic reset: PDEs (hold)}), (\ref{Myopic remain: PDEs (hold)}). We consider $1$-D travelling waves by taking the average column density of $2$-D  PDEs initialised with initial conditions that are translationally invariant in the $y$-direction and with appropriate boundary conditions. Given a continuous $2$-D concentration over time $C(x,y,t)$, we can write the average column density as 
\begin{equation}
    \hat C(x,t) = \frac{1}{Y}\int_0^{Y} C(x,y,t) \ dy.
    \label{Average column density}
\end{equation}
Using equation~(\ref{Average column density}) we can rewrite the systems of PDEs~(\ref{Reset: PDEs (hold)}), (\ref{Remain: PDEs (hold)}), (\ref{Myopic reset: PDEs (hold)}), (\ref{Myopic remain: PDEs (hold)}) in terms of $\hat C$ by integrating over $y$ and then dividing through by $Y$. This allows us to study a reduced $1$-D system of PDEs. If we further assume that the initial condition $C^{(s)}(x,y,0)$ is translationally invariant in $y$ and we have zero-flux boundary conditions on the horizontal boundaries then $C(x,y,t) = \hat C(x,t)$ for any value $y$. Then, we can explicitly write the $1$-D systems of PDEs by replacing every $C$ by $\hat C$ and $\nabla^2C$ by $\secpartder{\hat C}{x}$. For example, the system of PDEs~(\ref{Myopic remain: PDEs (hold)}) for the Myopic Remain model can be written
\begin{equation}
    \begin{cases}
        \partder{\hat{C}^{(1)}}{t} 
        = D \left[(1-\hat{C})\secpartder{\hat{C}^{(1)}}{x} + \hat{C}^{(1)}\secpartder{\hat{C}}{x}\right] 
        + 2\lambda \hat{C}^{(K)}(1-\hat{C}^4)  - \lambda \hat{C}^{(1)} , \\[6pt]
        \partder{\hat{C}^{(s)}}{t} 
        = D \left[(1-\hat{C})\secpartder{\hat{C}^{(s)}}{x} + \hat{C}^{(s)}\secpartder{\hat{C}}{x}\right] 
        +  \lambda \hat{C}^{(s-1)} - \lambda \hat{C}^{(s)}, \qquad \text{if }s=2,..,K-1, \\[6pt]
        \partder{\hat{C}^{(K)}}{t} 
        = D \left[(1-\hat{C})\secpartder{\hat{C}^{(K)}}{x} + \hat{C}^{(K)}\secpartder{\hat{C}}{x}\right] 
        +  \lambda \hat{C}^{(K-1)} - \lambda \hat{C}^{(K)}(1-\hat{C}^4). \\
    \end{cases}
    \label{1-D PDE example}
\end{equation}
For convenience, we will simply write $\hat C = C$ throughout the rest of this section.

If we linearise the reduced $1$-D system of PDEs~(\ref{1-D PDE example}) about the unstable steady state $C^{(s)} = 0$, we obtain the following linear system of PDEs
\begin{equation}
    \begin{cases}
        \partder{C^{(1)}}{t} = D \secpartder{C^{(1)}}{x} + 2\lambda C^{(K)} - \lambda C^{(1)}, \\
        \partder{C^{(s)}}{t} =  D \secpartder{C^{(s)}}{x} + \lambda C^{(s-1)} - \lambda C^{(s)} \qquad \text{if } s = 2,..,K.
    \end{cases}
    \label{Linearised PDEs}
\end{equation}
In fact, the linearisation of the other $1$-D reductions of the systems of PDEs~(\ref{Reset: PDEs (hold)}), (\ref{Remain: PDEs (hold)}), (\ref{Myopic reset: PDEs (hold)}) gives the same linear system of PDEs~(\ref{Linearised PDEs}). This suggests that, whenever the continuum limits accurately describe the discrete ABMs, the predicted wave-speeds will be identical under the different cell proliferation models with the same number of stages $K$. 

We assume that solutions take the form $C(x,t) = C(x-vt)$ and satisfy $\underset{x \rightarrow -\infty}{\lim} C(x)  = 1, \underset{x \rightarrow \infty}{\lim} C(x)  = 0$. This means we have travelling wavefront which move rightwards with wave-speed $v$. Using the front propagation method by \citet{van2003front}, \citet{gavagnin2019isc} show that the minimum wave-speed in a $K$-stage model is
\begin{equation}
    v = 2\sqrt{D\lambda (2^{1/K} - 1)}.
    \label{Theoretical wave-speed}
\end{equation}
Taking the number of stages to be $K=1$, we recover the classical invasion speed in the FKPP model, $2\sqrt{D\lambda}$. Note that increasing the number of stages $K$ decreases the minimum wave-speed $v$. Assuming the continuum limit of an ABM is a good approximation to the average of the agent-based dynamics, this suggests that cells proliferating under a multi-stage model invade more slowly than cells proliferating under an exponential model.

From section \ref{Sec III}, we know that the motility rate must be sufficiently high relative to the proliferation rate for the continuum descriptions to give good approximations to the average behaviour of the ABMs. In the next section, we will perform numerical simulations to investigate the travelling wavefronts which form when the continuum PDEs provide poor agreements to the ABMs. We also verify that the theoretical minimum wave-speed~(\ref{Theoretical wave-speed}) is valid under the assumption of sufficiently high motility rates.

\subsection{Comparing ABM wave-speeds to theoretical predictions}

\subsubsection{Low motility}
\begin{figure}[H] 
\centering
\subfigure[][]{
	\includegraphics[width=0.45\textwidth]{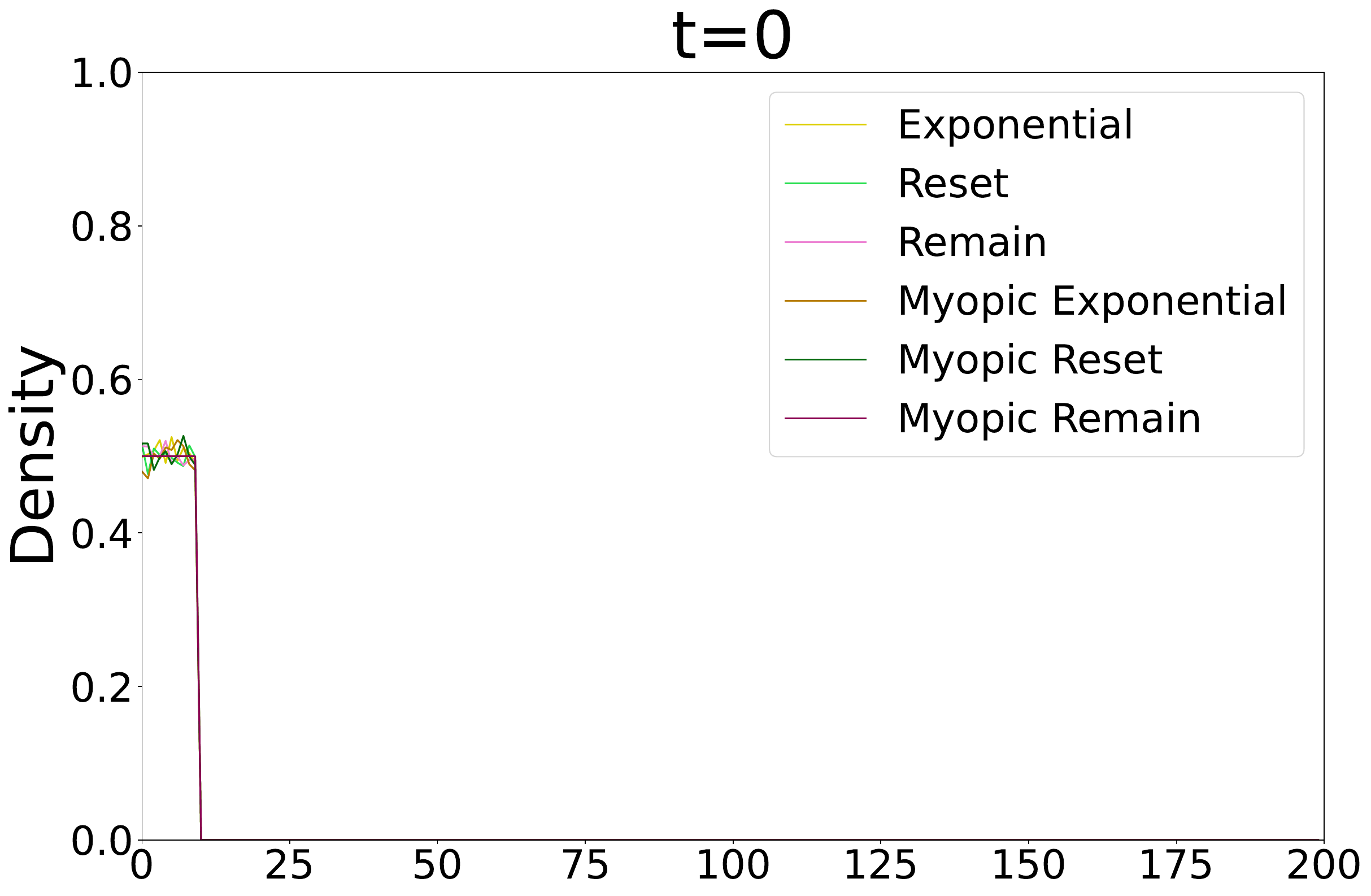}
}
\subfigure[][]{
	\includegraphics[width=0.45\textwidth]{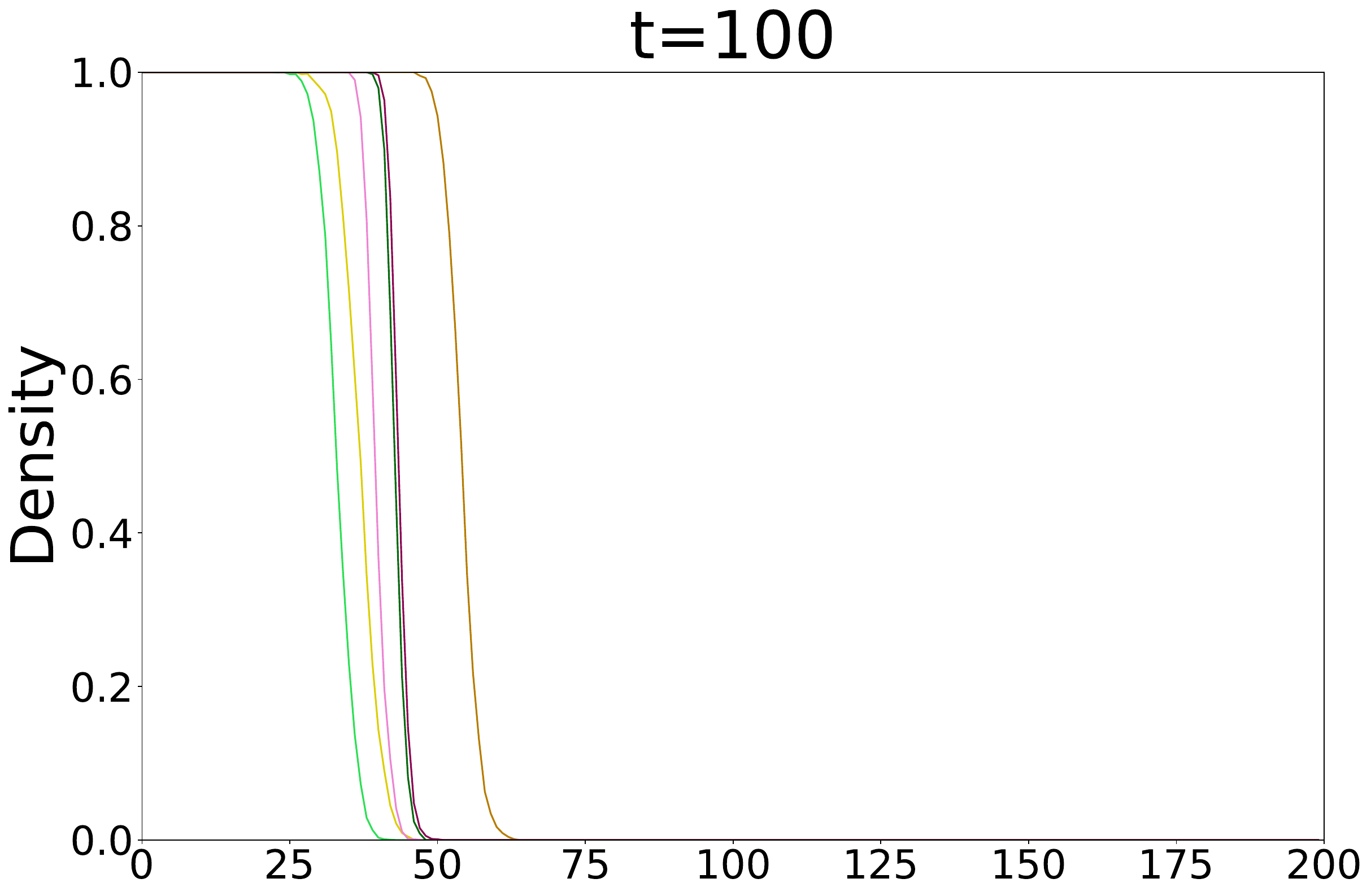}
}
\subfigure[][]{
	\includegraphics[width=0.45\textwidth]{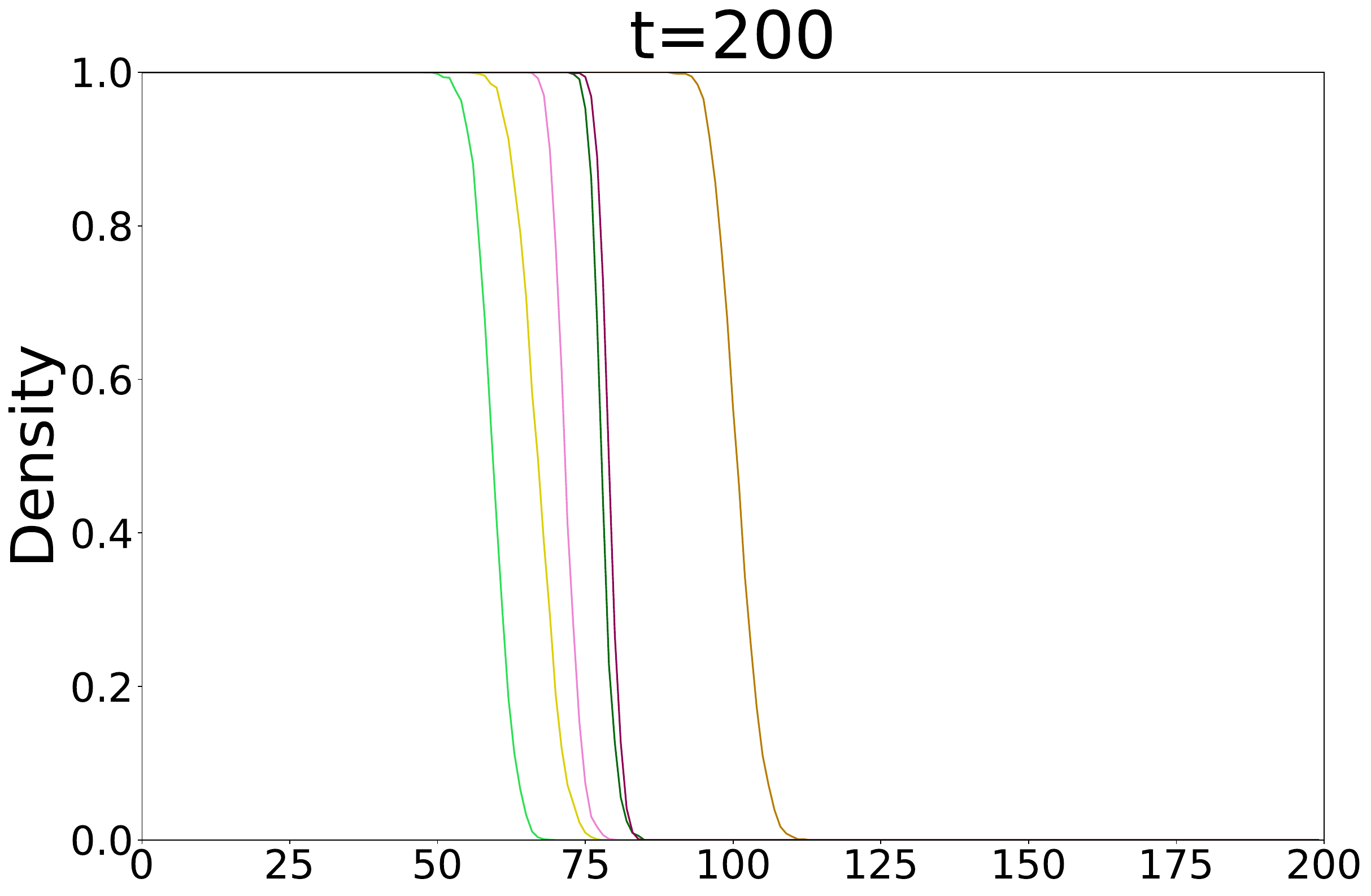}
}
\subfigure[][]{
	\includegraphics[width=0.45\textwidth]
    {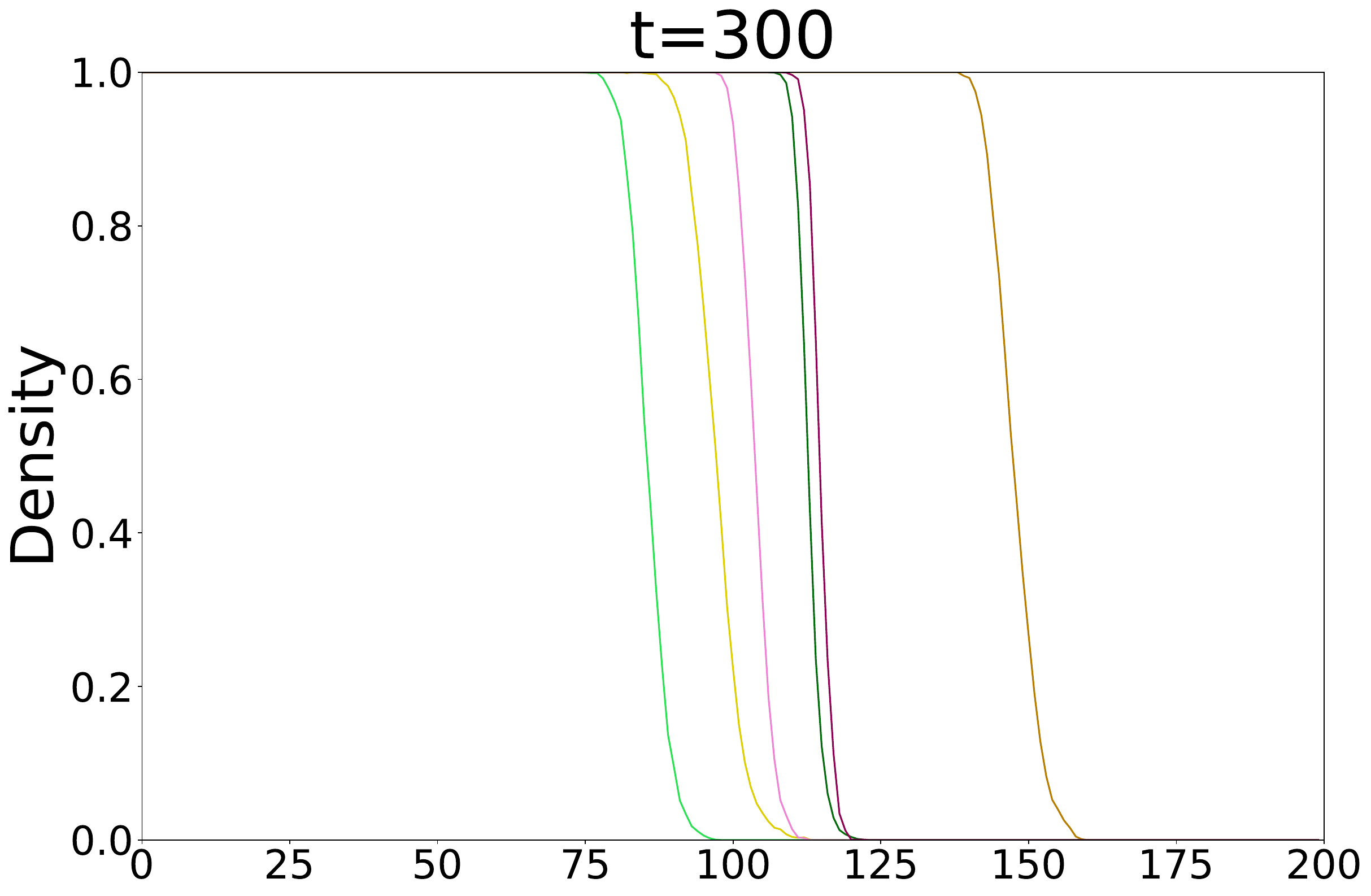}
}
\caption{ABM simulations of travelling wavefronts on a $200 \times 100$ lattice. We set the proliferation and motility rates to be equal $r_m = r_p = \sqrt{0.0233\times4}$, and the multi-stage models have $K=10$ stages. We initialise our ABMs so that the first $10$ columns are uniformly populated with stage $1$ cells at $50\%$ density (see (a)). At the horizontal boundaries, we impose periodic boundary conditions, and at the vertical boundaries we impose reflecting boundary conditions. We simulate over $t=300$ time units, and take the average column density over $20$ realisations for each model. In each sub-figure, we plot the column density for each ABM, distinguishing between different models using different colours. Panel (a) displays the initial conditions. Panels (b), (c), and (d) display the wave-fronts at $t=100$, $t=200$ and $t=300$ respectively.}
\label{Low motility wavefronts}
\end{figure}

\begin{figure}[H]
\centering
    \includegraphics[width=0.96\linewidth]{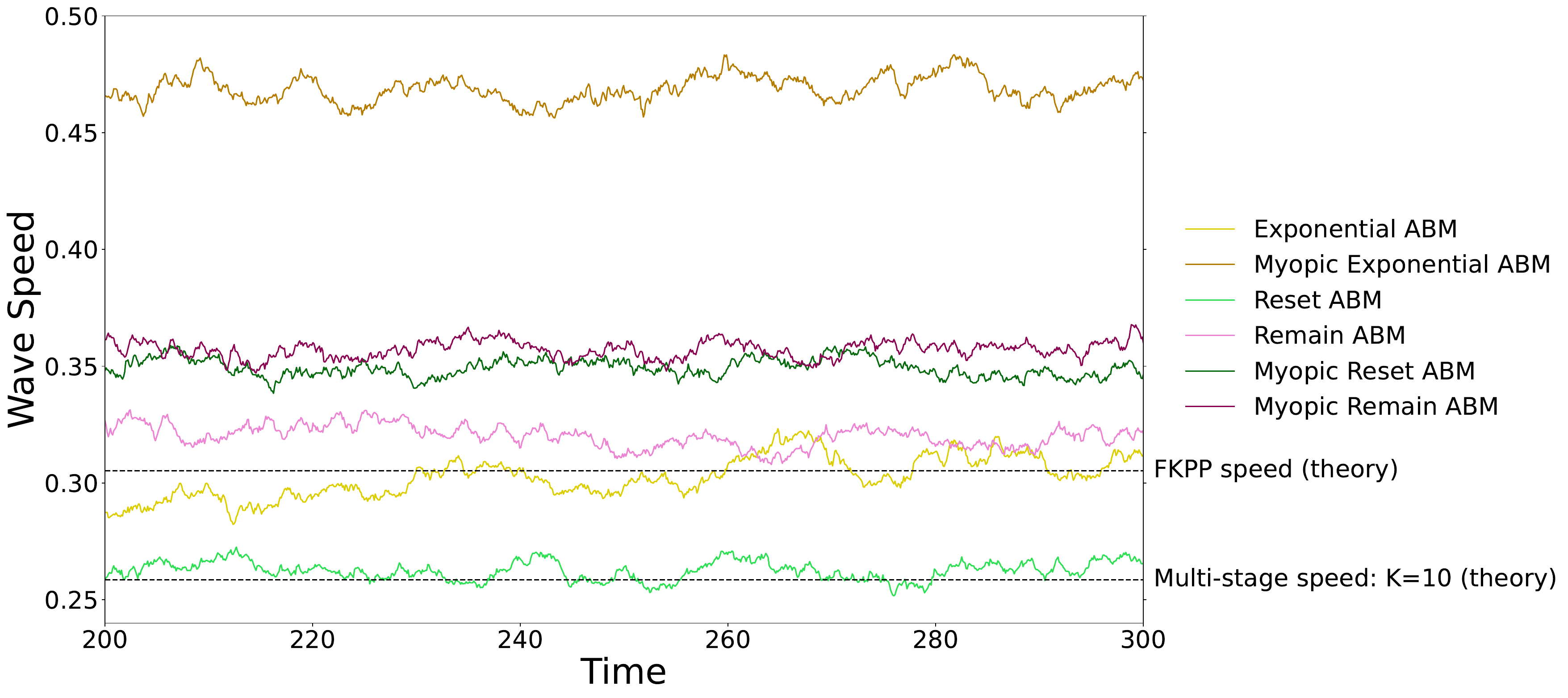}
\caption{Wave-speeds for the ABM simulations in Figure \ref{Low motility wavefronts} between $t=200$ and $t=300$. We calculate the wave-speed by finding the average mass added per unit time as in Harrison and Yates \cite{harrison2016hac}, and plot a moving average wave-speed for each ABM with time window $5$. We denote the wave-speeds of different models using the same colours as Figure \ref{myopic average density evolutions}, and plot the theoretical wave-speeds for the FKPP and 10-stage model in black dashed lines. }
\label{Low motility wave-speeds}
\end{figure}

We begin by considering a regime in which the motility rate, $r_m$, is of the same order as the proliferation rate, $r_p$ (previously the low motility regime). In Figure \ref{Low motility wavefronts}, we can see that the wave-fronts which form in each ABM have similar shapes and are very steep.  This is because cell proliferation is the major factor driving the travelling wave when the proliferation rate is comparable to the motility rate. This phenomenon has been observed in previous studies of travelling waves \cite{van2003front}. From Figure \ref{Low motility wave-speeds}, we observe that most ABMs exhibit different wave-speeds to each other. This is in contrast to the theoretical predictions of the wave-speed in equation (\ref{Theoretical wave-speed}) for high motility rates $r_m$, where we expect models with the same number of stages, $K$, to have identical invasion speeds. We can see that the wave-speed in Myopic models are all quicker than wave-speeds in Non-Myopic models with the same number of stages. We also observe that the speed of invasion in Remain models is quicker than Reset models due to cells at intermediate column densities likely having free space in the direction the travelling wave is moving. This means that cells which can attempt to proliferate more quickly after a failed division event, as is the case in the Remain model, have a proliferative advantage over cells proliferating under the Reset model. However,  the Myopic Remain and Myopic Reset models have similar wave-speeds. This is because cells exhibiting myopic behaviour only fail to proliferate when they are completely surrounded by neighbouring cells. Therefore, proliferation, and as a consequence the wave-speed, is more limited by the rate at which cells move to create free space i.e. the motility rate $r_m$. Recall further that the predicted minimum wave-speed in equation (\ref{Theoretical wave-speed}) decreases as we increase the number of stages $K$. However, in Figure \ref{Low motility wave-speeds}, the Exponential model exhibits a lower wave-speed than every model except the Reset model.

The different speeds of invasion in our different models, shown in Figure \ref{Low motility wave-speeds}, qualitatively match the rates of growth in Figure~(\ref{myopic average density evolutions}) at low to intermediate densities. As the wave-front is driven by proliferation, the wave-speed will correspond to the rate of growth of a population at a low density. Hence, the behaviour of travelling waves at relatively low motility rates can be explained by the characteristics of the proliferation models which we described in Section \ref{Sec II B}. 

As the motility rate is equal to the proliferation rate in the simulations we performed in Figure~\ref{Low motility wavefronts} and \ref{Low motility wave-speeds}, we do not expect good agreement between the average ABM behaviour and the continuum PDE limit. Indeed, in Figure \ref{Low motility PDE vs ABM wavefront} below, we can observe that the wavefront in the ABMs featuring remain or myopic behaviour agree poorly with their continuum PDE wavefronts. We can also see that the shape of the wavefronts for the Exponential and Reset ABMs do not match the shape of the wavefronts their continuum PDEs. However, there is some overlap between the profiles of the wavefronts. This is because the Reset and the Non-Myopic Exponential ABMs travel at speeds similar to their corresponding continuum limits, as we saw in Figure \ref{Low motility wave-speeds}.

In contrast, the ABMs evolve to a wave-speed greater than the minimum theoretical wave-speed of the continuum model, which results in poor agreement with the corresponding continuum limit. This is because the proliferation methods in these ABMs provide an advantage at the wave-front which is not fully captured under the mean-field continuum PDEs. In low motility regimes in our ABMs, cells at the wave-front are likely to be surrounded by $3$ cells, even at low to intermediate densities. The empty neighbouring space is likely to be located forwards of the wave-front. Cells which remain in the final stage of the cell cycle after a failed proliferation attempt can quickly re-attempt to proliferate into the empty space. Cells which exhibit myopic behaviour will always proliferate into the empty space. As the empty space is in the direction of the travelling wave-front, this has the effect of pushing the wave forwards more quickly in the ABM than the continuum PDE where the direction of proliferation does not have this impact. 
\begin{figure}[H]
\centering
\subfigure[][]{
    \includegraphics[width=0.45\textwidth]{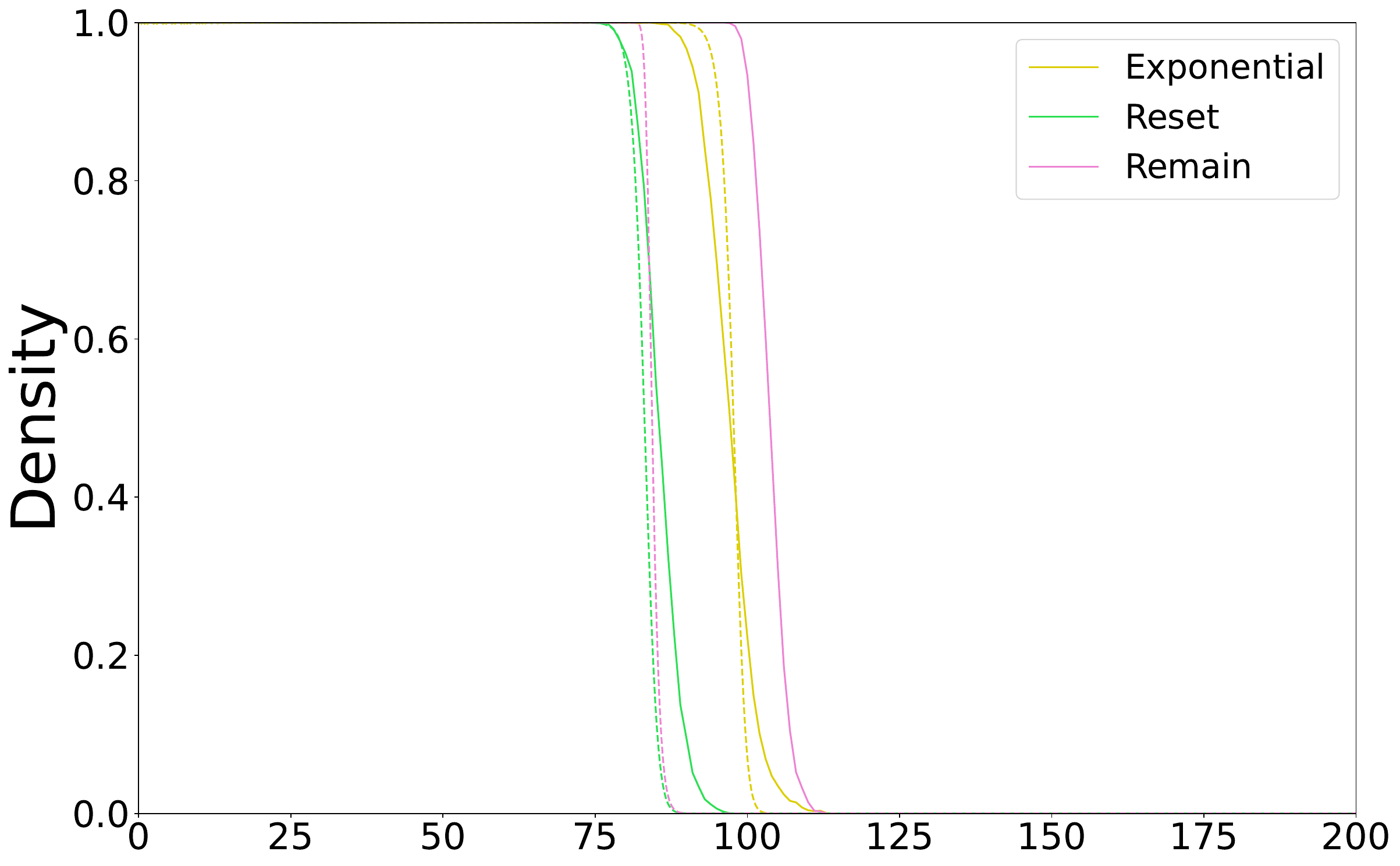}
}
\subfigure[][]{
    \includegraphics[width=0.45\textwidth]{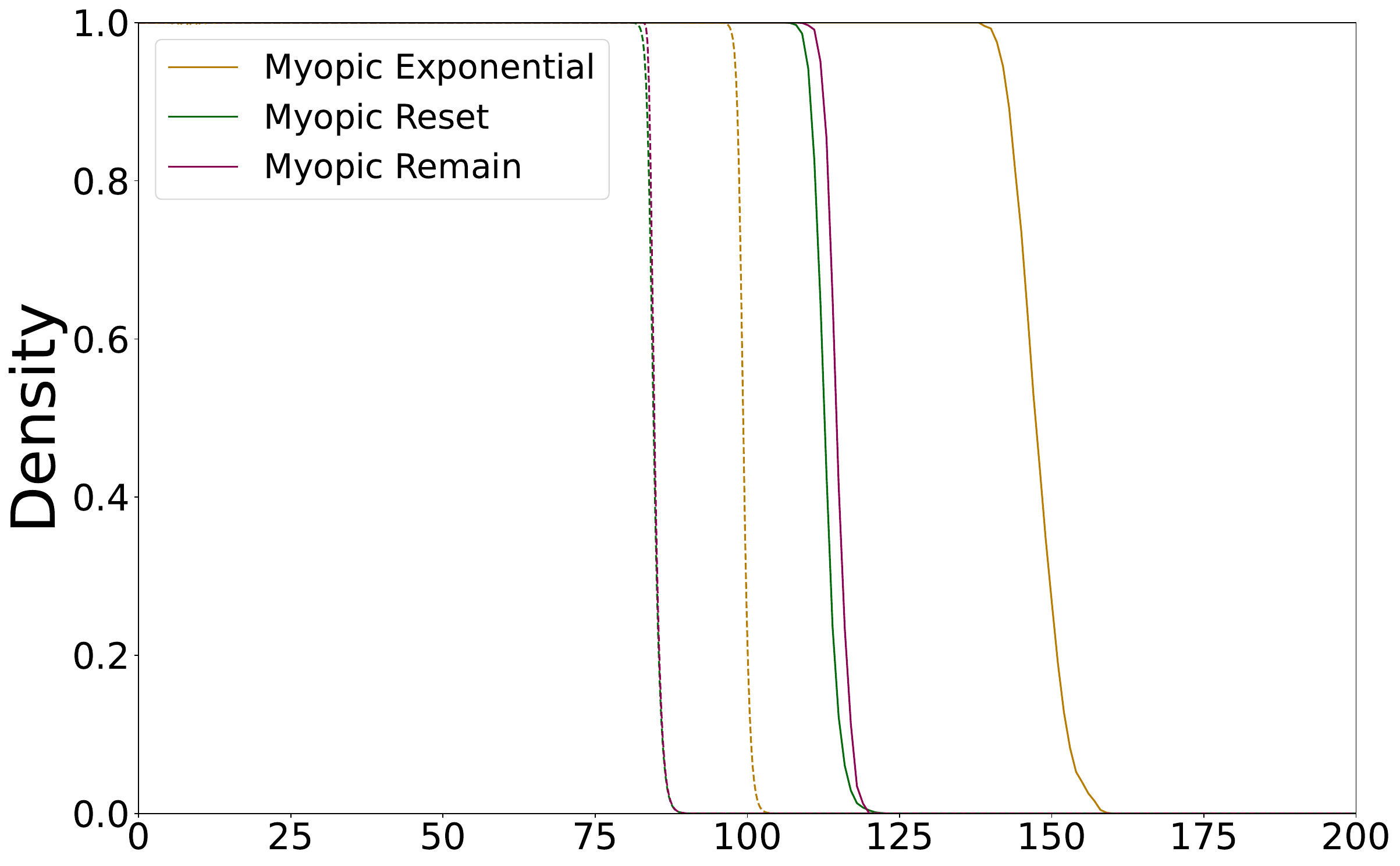}
}
\caption{Comparing the wave-fronts at time $t=300$ for each ABM to the continuum PDE solution at time $t=300$. We initialise our PDE solutions with $C^{(1)}(x,0) = 0.5$ for $x \leq 10$, $ C^{(1)}(x,0) = 0$ otherwise, and $C^{(s)}(x,0) = 0$ for $s>1$, so that the initial conditions of the PDEs match the corresponding ABM as in Figure \ref{Low motility wavefronts}(b). We impose periodic boundary conditions on the horizontal axes, and zero-flux boundary conditions on the vertical axes. To solve the system of PDEs, we use the method of lines, discretising the spatial component with a second-order finite-difference scheme using spatial grid size $\delta x = 0.25$, and integrating forward in time using the \texttt{RK45} method from the \texttt{solve\_ivp} function in the \texttt{scipy} package in Python.
Panels (a) and (b) compare the non-myopic and myopic ABM wavefronts, respectively, with their corresponding continuum PDE wavefronts. In each plot, we distinguish PDE and ABM wavefronts using dashed and solid lines respectively.}
\label{Low motility PDE vs ABM wavefront}
\end{figure}

\subsubsection{High motility}
\begin{figure}[H] 
\centering
\subfigure[][]{
	\includegraphics[width=0.45\textwidth]{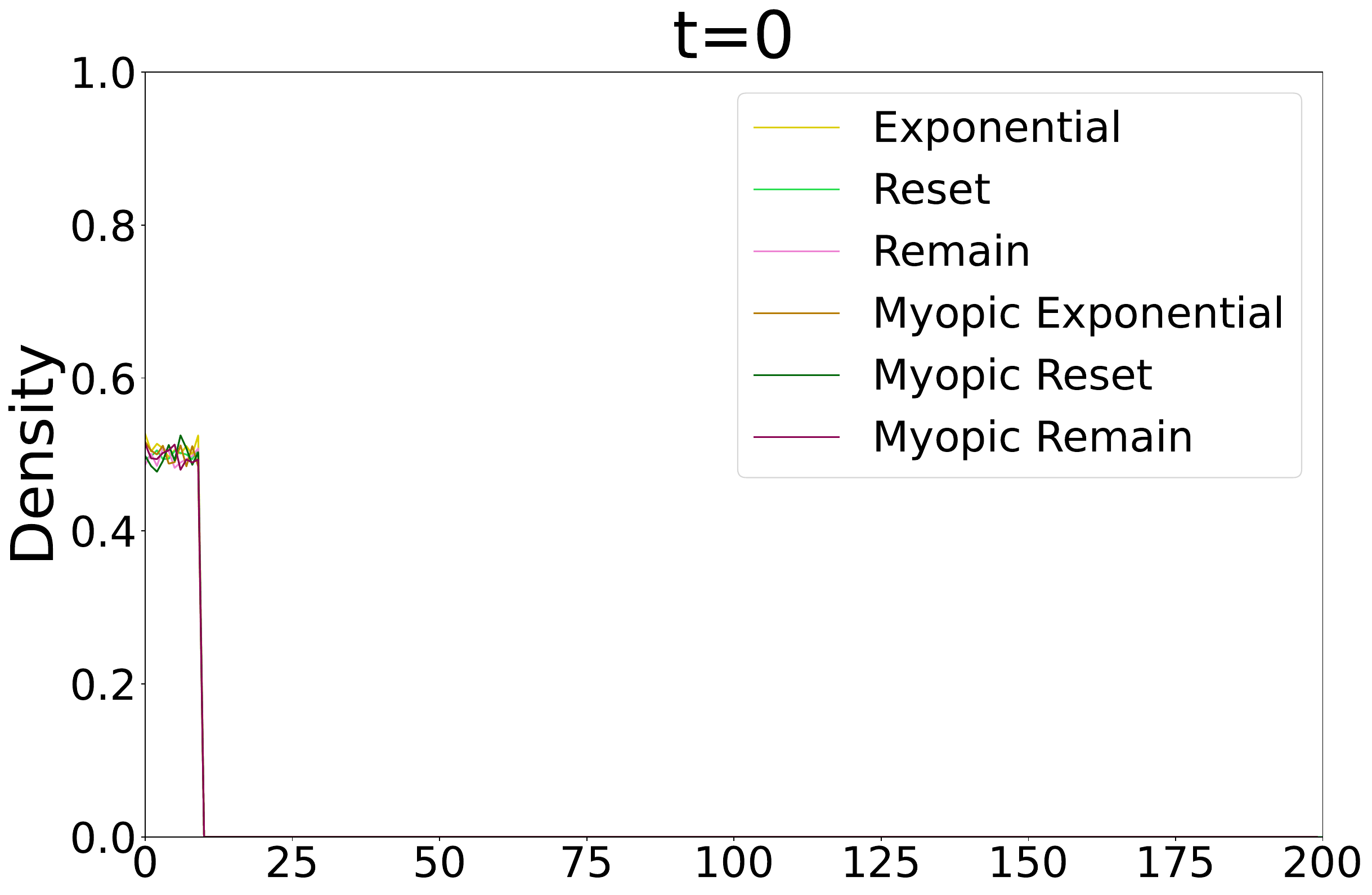}
}
\subfigure[][]{
	\includegraphics[width=0.45\textwidth]{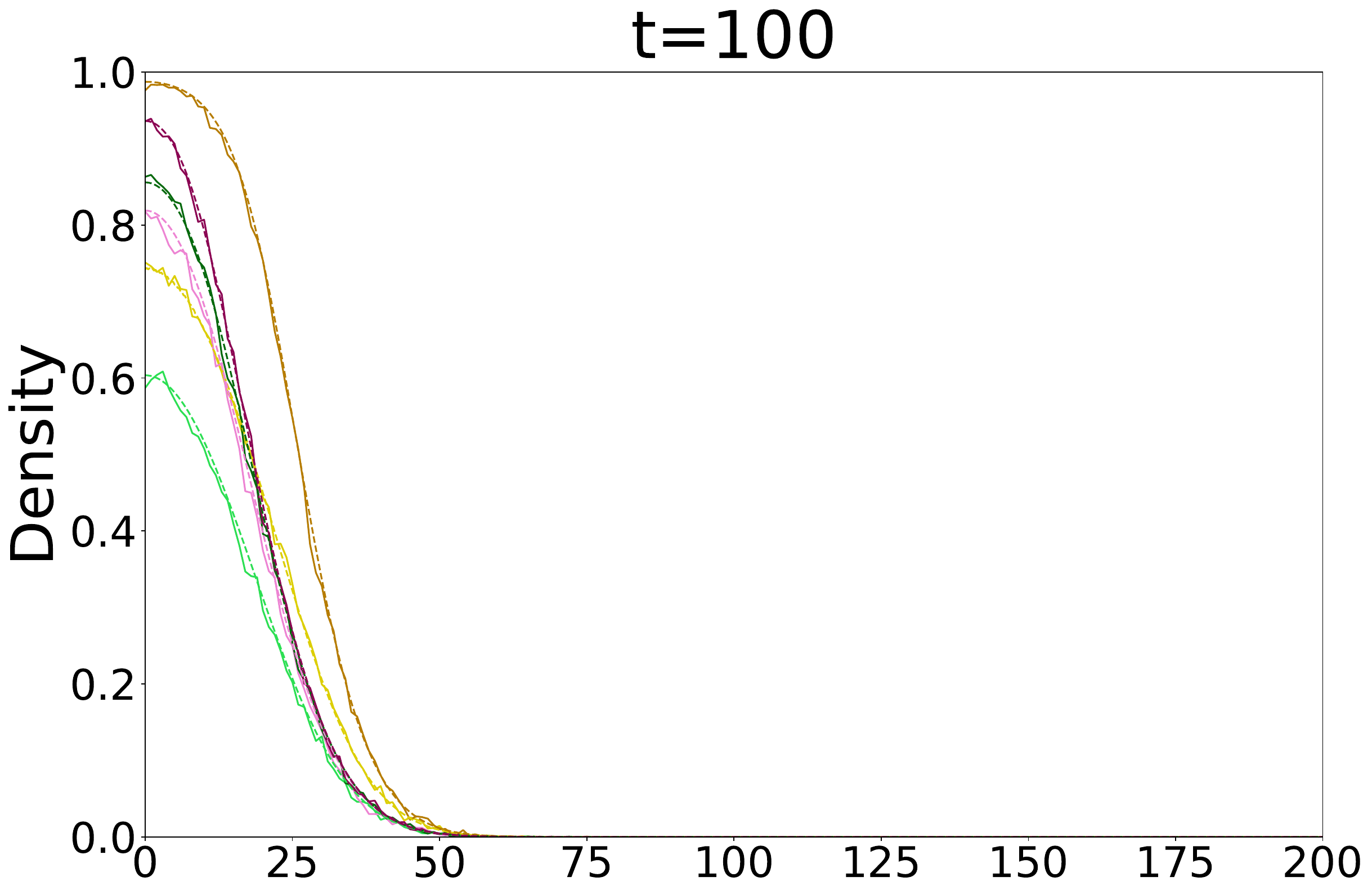}
}
\subfigure[][]{
	\includegraphics[width=0.45\textwidth]{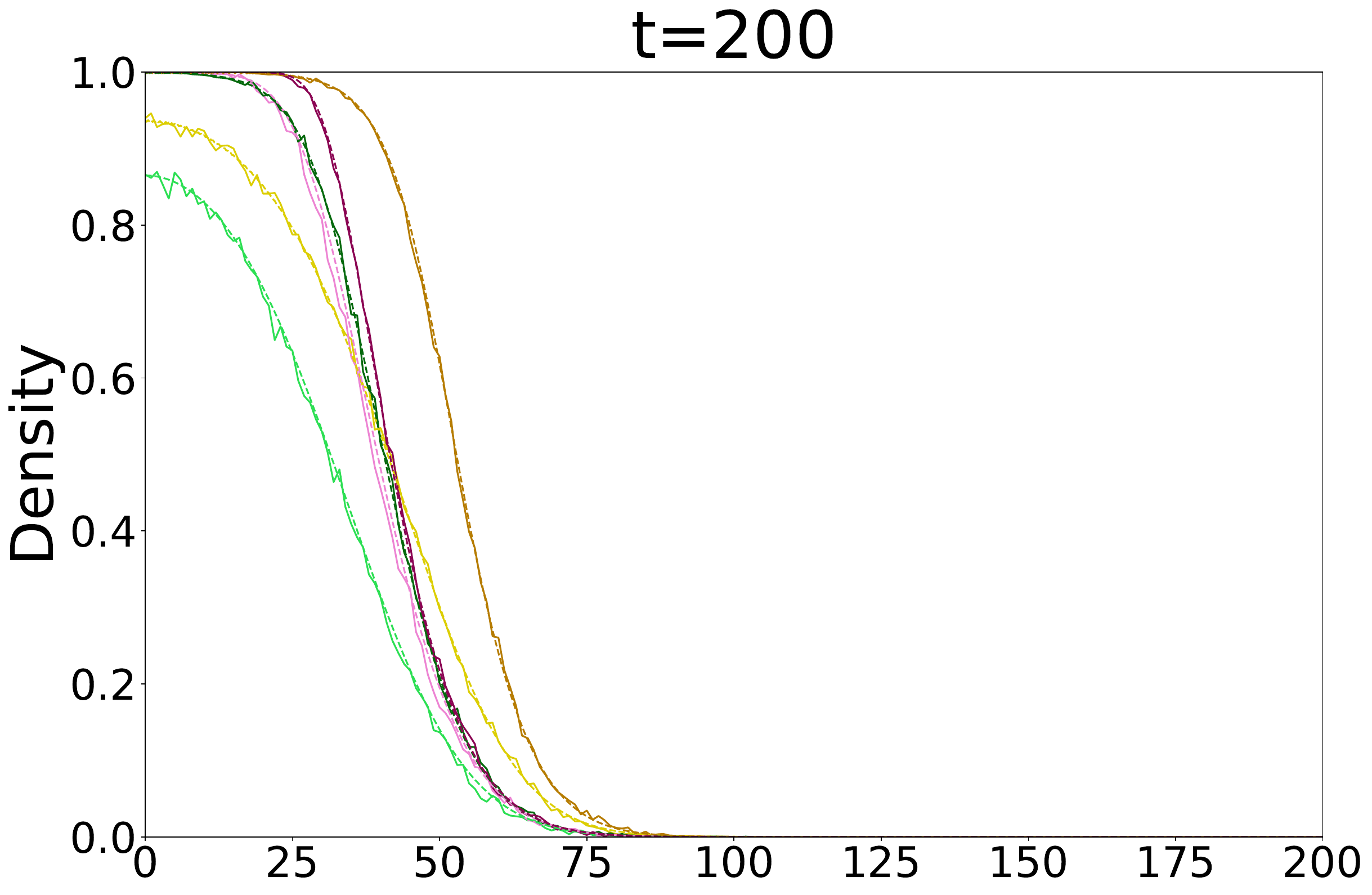}
}
\subfigure[][]{
	\includegraphics[width=0.45\textwidth]{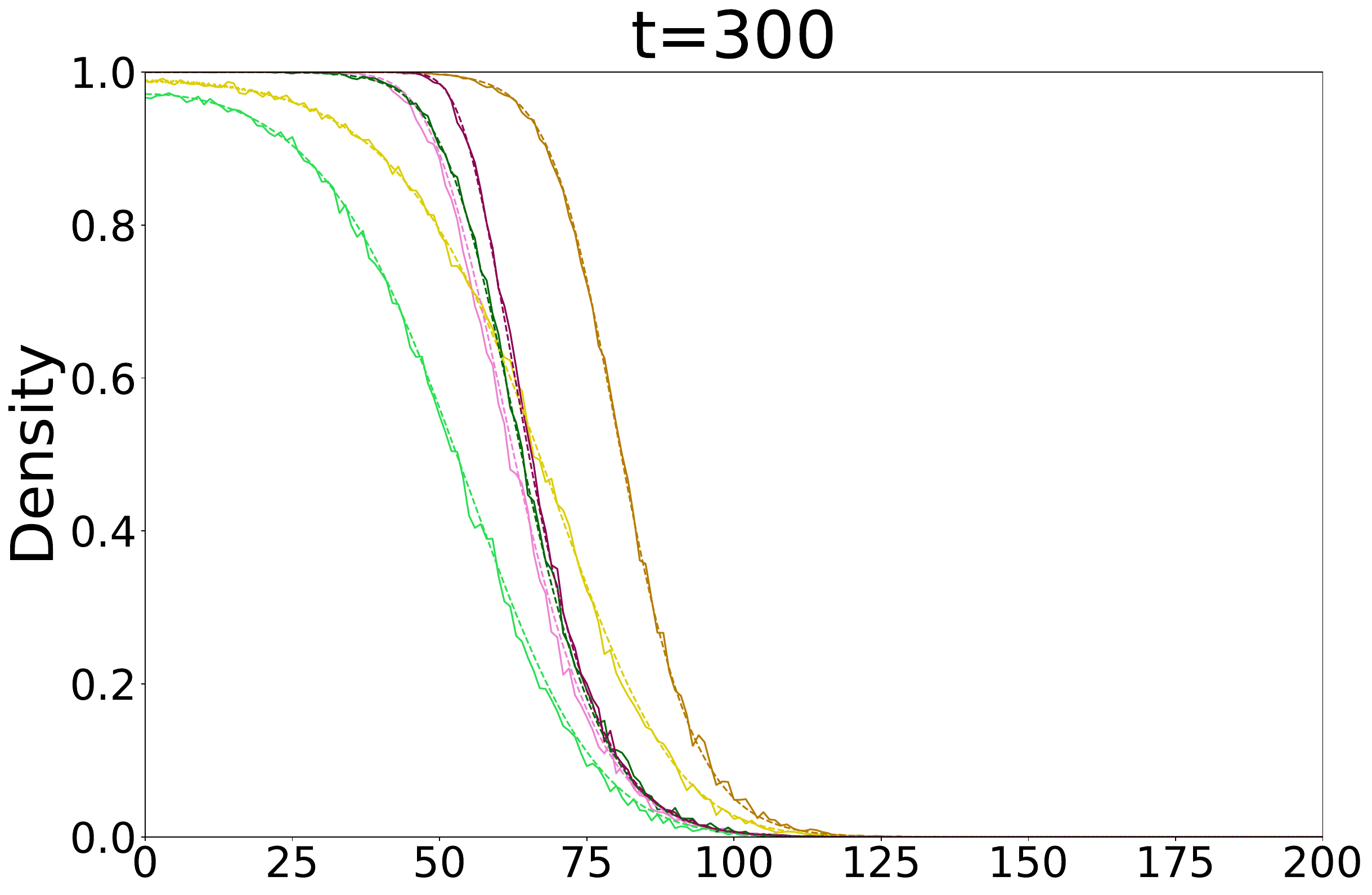}
}
\caption{Comparing ABM simulations to PDE solutions over $300$ time units. ABM parameters are identical to the simulations from Figure \ref{Low motility wavefronts} but with motility rate $r_m= 4$ and proliferation rate $r_p = 0.0233$. We initialise and solve the PDEs identically to the simulations in Figure \ref{Low motility PDE vs ABM wavefront}, but with parameters updated to be consistent with the different motility and proliferation rates. We distinguish between different cell cycle models using different colours and denote PDE and ABM wavefronts using dashed and solid lines respectively.}
\label{High motility PDE vs ABM Wavefront t300}
\end{figure}

We now consider a parameter regime in which the motility rate is high relative to the proliferation rate. In Figure \ref{High motility PDE vs ABM Wavefront t300}, we see that even after $300$ time units, the Exponential and Reset models have not yet reached a steady state wavefront. We also note that the wavefronts across all models under this high motility scheme are shallower than in the wavefronts under a low motility scheme in Figure \ref{Low motility wavefronts}. This is because the lower proliferation rate means that cells are less quick to fully occupy columns. We also observe that the agreement between the PDE and ABM wavefronts is very good across all models at every time point plotted. The improved agreement is due to the large motility rate relative to the proliferation rate which means that the continuum limits provide a better description of the discrete ABMs, as discussed in Section \ref{Sec III: The effects of Cell motility}.

\begin{figure}[H] 
\centering
\subfigure[][]{
	\includegraphics[width=0.45\textwidth]{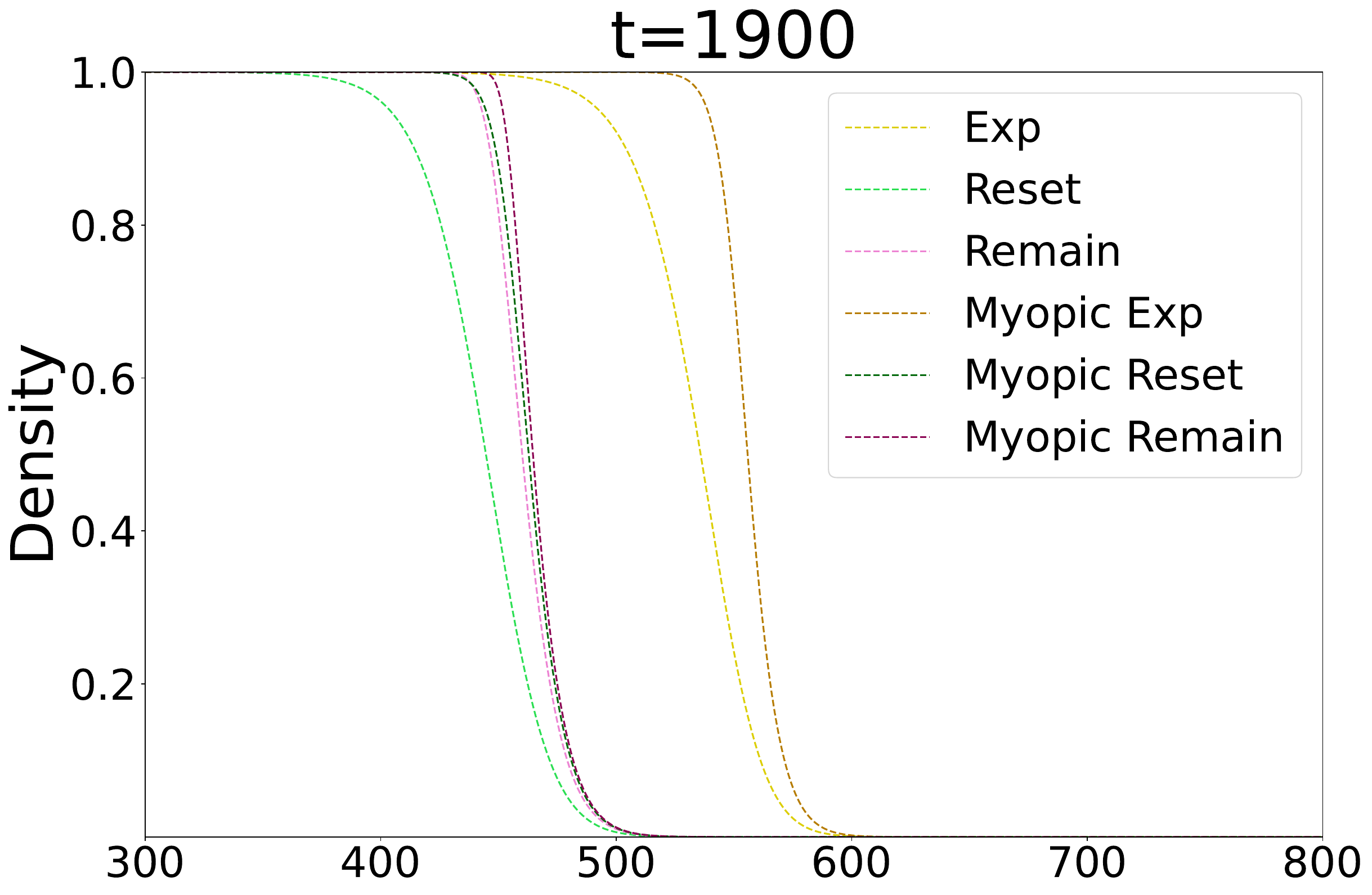}
}
\subfigure[][]{
	\includegraphics[width=0.45\textwidth]{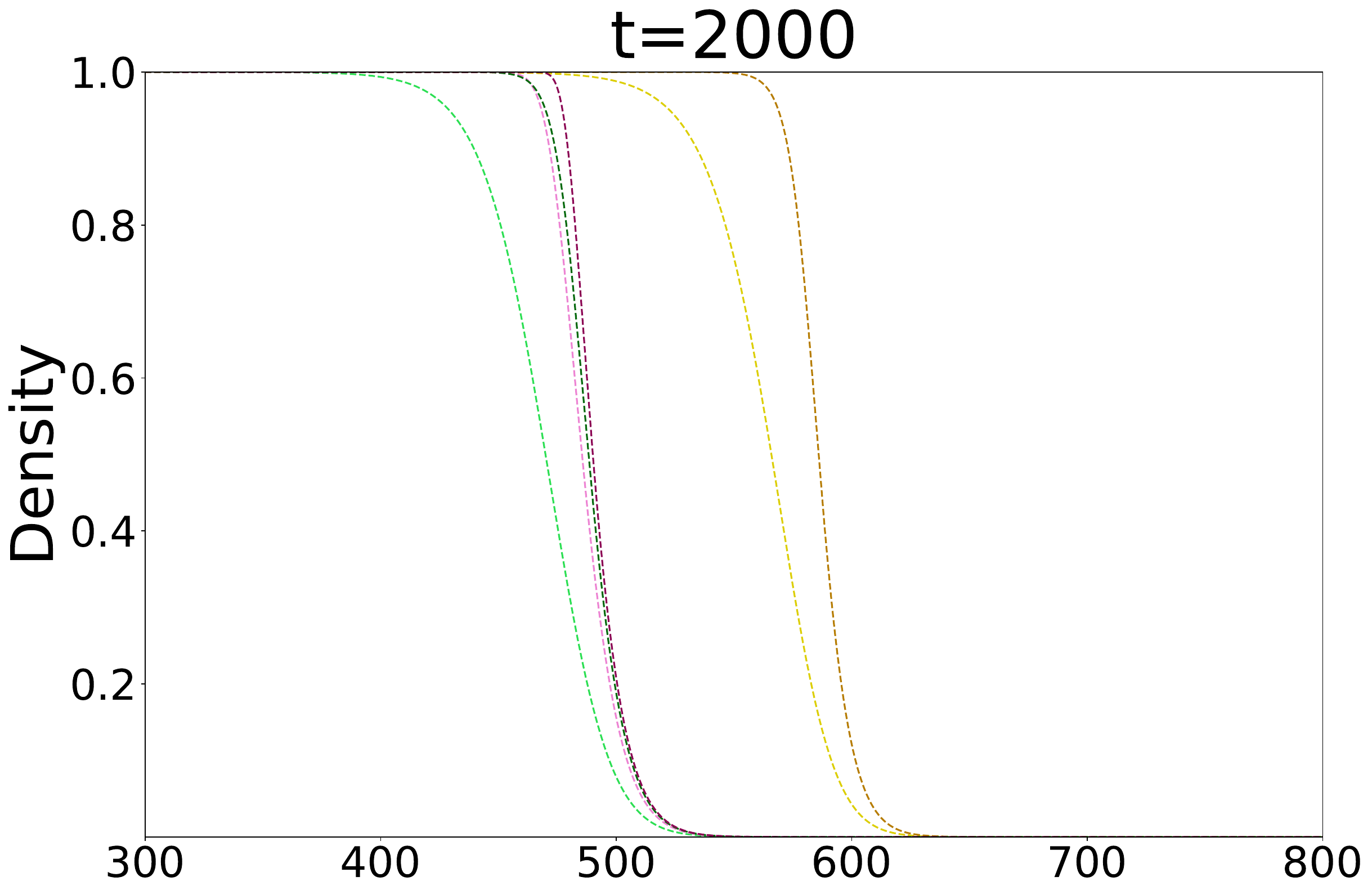}
}
\caption{PDE solutions at $t=1900$ and $t=2000$ with parameters and profiles initialised identically to Figure \ref{High motility PDE vs ABM Wavefront t300}. Panels (a) and (b) show PDE solutions at $t=1900$ and $t=2000$ respectively. As in previous figures, different models are represented using different colours.}
\label{High motility PDE Wavefront t2000}
\end{figure}

In Figure \ref{High motility PDE Wavefront t2000}, we display asymptotic wavefronts after a long period of time. We can see that models with no proliferative advantage (Non-Myopic Reset and Non-Myopic Exponential) have a shallower wave-profile than models with a proliferative advantage. This is because these models are less quick to completely fill a column than cells proliferating under different models. In Figure \ref{High motility wave-speeds} below, we can see that the wave-speeds of every multi-stage model are close. Similarly, the wave-speeds for both Exponential models are close to identical. This is because the wave-speed of the continuum PDEs are determined by the linearised system (\ref{Linearised PDEs}), which are identical for models with the same number of stages $K$. As the continuum PDEs provide almost exact agreements to their discrete ABMs, we would expect that travelling waves in ABMs under high motility regimes are characterised purely by the number of stages $K$ in the cell cycle.

We also note that the average wave-speeds for every model under the PDE are the minimum theoretical wave-speed up to numerical error. The relative error between the theoretical and numerically solved PDE wave-speeds can be well approximated using a result from Brunet and Derrida \cite{brunet1997svf}. The numerical error is caused by floating point errors in Python, which causes the wave-front to be cut-off at small values. 

\begin{figure}[H]
    \centering
    \includegraphics[width=0.96\linewidth]{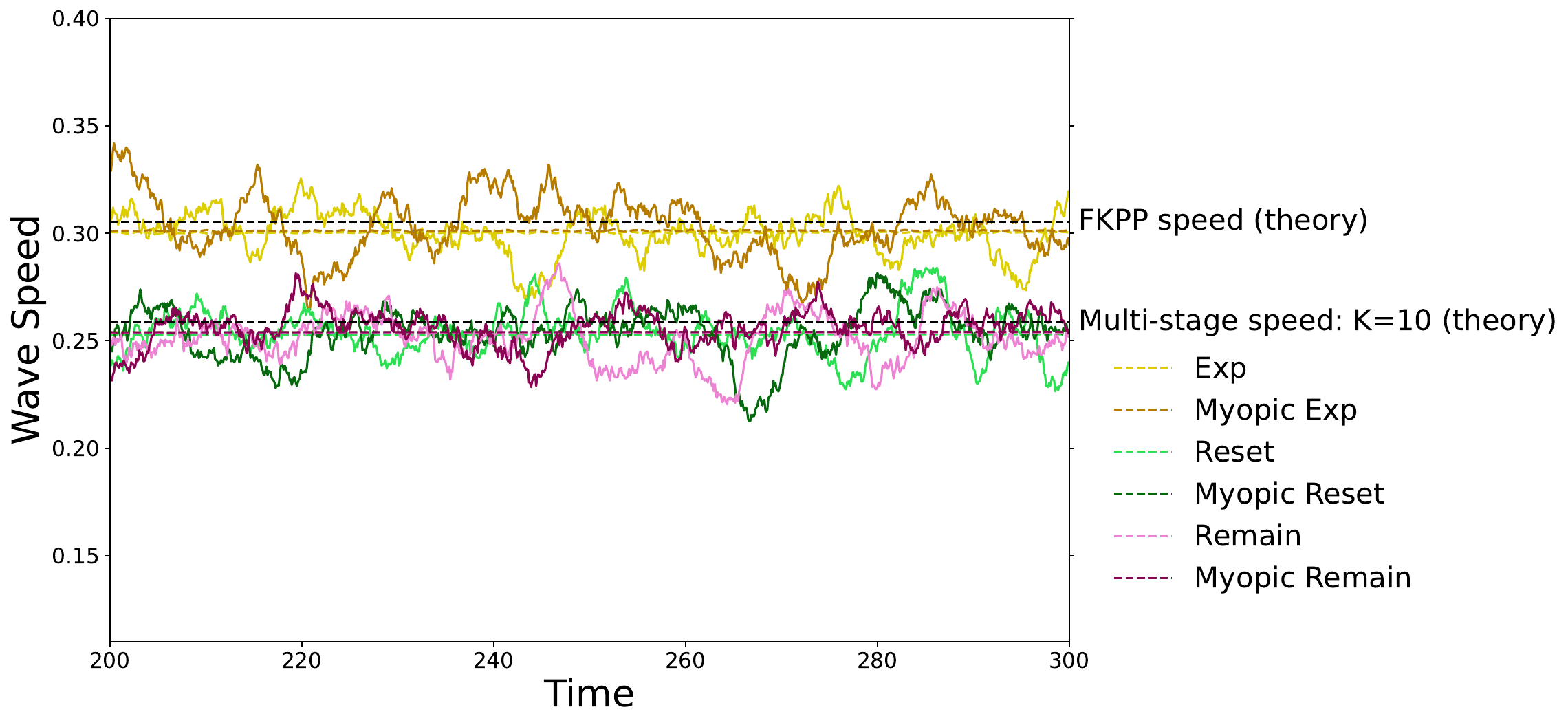}
    \caption{Comparing ABM and PDE wave-speeds. We calculated the PDE wave-speeds using the simulations from Figure \ref{High motility PDE Wavefront t2000}. We used the steady-state wave profiles from Figure \ref{High motility PDE Wavefront t2000} to initialise ABM simulations on a $800 \times 20$ lattice and simulated over $100$ time units. We plot the moving average wave-speeds over $20$ realisations for each of our models. We distinguish between PDE and ABM wave-speeds using dashed and solid lines respectively. As before, we differentiate between different models using different colours. }
    \label{High motility wave-speeds}
\end{figure}

In this section, we have presented theoretical predictions of the wave-speed of travelling wavefronts for our different models. Using numerical simulations, we confirmed that the theory holds well for parameter regimes in which ABMs and their corresponding continuum limits agree well. We have also demonstrated that even under low motility rates relative to the proliferation rate, the theoretical minimum PDE wave-speed provides a lower bound for the corresponding ABM wave-speeds.
\section{Discussion}
In this work, we developed a framework for modelling the cell cycle which is capable of capturing realistic cell-cycle time distributions and accommodating a range of density dependent cellular behaviours. Through discrete agent-based and continuum PDE modelling approaches (the PDEs being derived directly from the ABMs), we investigated how different cellular behaviours affect population-level growth. We incorporated several distinct ways in which  cells arrest their growth in high density environments consistent with experimental observations showing that cell-cycle progression is constrained by the overall cell population \cite{streichan2014scc}. 

We compared the growth dynamics of a cell population under these different modelling frameworks. Under the Reset model, a cell that attempts, but fails to proliferate is sent back to the first stage of the multistage cell cycle and must undergo the entire cell cycle again before reattempting to proliferate. In contrast, under the Remain model, cells enter a quiescent state in high local density environments and retain the ability to repeatedly attempt proliferation without the requirement to revisit earlier stages in the cell cycle. Through numerical experiments, we demonstrated that this leads to faster growth in the Remain model than in the Reset model. This result is consistently observed across both ABM and corresponding PDE simulations. Moreover, the capacity for cells to arrest in the final stage of the cell cycle allows for the Remain model to increase its density more quickly than the Exponential model at higher densities. This behaviour contrasts with non-spatial theory, in which multi-stage models grow at a slower rate than the Exponential model \cite{yates2017msr}. 

We also presented the Myopic model, a novel adaptation which allows cells to sense their local environment for the purposes of proliferation. Numerical simulations demonstrated that the Myopic property confers a greater proliferative advantage to cells than even the most rapidly growing non-myopic model (the Remain model) throughout the entire growth process. Again, this behaviour was consistent in both ABM and corresponding PDE simulations. Together, these results highlight that classifying cell-cycle times in volume-constrained proliferating cell populations is highly sensitive to the assumed underlying model of cell proliferation. 

Although the ABM and PDE frameworks share similar qualitative behaviour in growing cell populations, we demonstrated that, under certain parameter regimes, the PDE exhibits poor quantitative agreement with its corresponding ABM. In particular, the continuum PDE approximation overestimates cell density relative to the discrete ABM. We showed that this discrepancy arises because of the requirement that the cell motility rate must be sufficiently large relative to the proliferation rate for the continuum description of our ABMs to be valid. When the cell motility rate is insufficiently high in our ABMs, spatial correlations form which means that lattice sites are no longer independent of each other. This phenomenon leads to slower growth rates of cells in ABMs than the corresponding continuum limits. Numerical experiments with uniformly distributed initial conditions confirm that increasing cell motility, under a fixed proliferation rate, leads to faster population growth in ABMs. In contrast, solutions of the continuum PDEs are independent of the motility rate in initially uniformly distributed environments.

We further demonstrated that the cell motility rate strongly influences the spatial distribution of cells in the ABM. When motility rates are insufficiently high, pairwise spatial correlations emerge, which we quantify using the PCF. These observations show that on-lattice ABMs of cell proliferation can capture both spatial structure and growth dynamics that are not accessible to continuum approaches. Moreover, the PCF provides a quantitative measure of agreement between an ABM and its corresponding continuum limit. Given experimentally observed parameters, this tool can be used to assess whether a continuum PDE description is appropriate, or whether a finer agent-based framework is required to accurately capture cell dynamics.

Continuum PDEs and discrete ABMs are both common approaches for studying cell invasion. We have shown that under parameter regimes in which the continuum PDE description is valid, the speed of cell invasion under an ABM matches established theory. Under parameter regimes for which the PDE description becomes invalid, the speed of invasion of our ABM wavefronts exceeds the theoretical minimum predicted by the corresponding PDE. We have also shown that the shape of the asymptotic wavefront solution is dependent on the cell proliferation model we choose. Cell proliferation models which confer a proliferative advantage, i.e. the Remain and Myopic models, form steeper wavefronts than models without such an advantage (Non-Myopic Reset/Exponential). 

Further work could extend the results presented in this paper in several directions. One natural avenue is to fit average cell-cycle times in our cell-proliferation models to experimental data from density-dependent growing cell populations, and to quantify how different modelling assumptions influence the inferred cell-cycle times. Throughout this paper, we have shown that volume exclusion plays a crucial role in driving differences between cell-cycle models, and that neglecting density-dependent effects is therefore likely to impact estimates of cell-cycle duration. 

Direct calibration of parameters within our full modelling framework is challenging due to the complexity of the resulting PDE systems. In contrast, parameter fitting is easier when we neglect volume exclusion. Given a specified underlying cell-cycle model, a potential area of future work might be to determine how to infer the true cell-cycle time of a cell when experimental data are fitted using models that assume no density dependence. There is also capacity to improve the models we have presented to more accurately reflect the true biology of a cell. Typically, cells arrest prior to entering S-phase \cite{donker2022mgc}. However, under the Remain model, cells arrest in the final stage of the cell-cycle model. Instead, we could incorporate a checkpoint at an intermediate stage, from which cells may only progress if there is available space. We have also made the basic assumption that cell movement is diffusive. However, there are many example of cells which demonstrate persistence in their movement and so are not exhibiting classical diffusion on short time scales \citep{selmeczi2005cmp, wu2014tdc, heyn2024mesenchymal}. There are also cells which move non-randomly in response to environmental cues \citep{artemenko2014mtp, espina2022dmc}. These non-diffusive movement modes may have an impact on the PDE agreement but also allow behaviours such as collective cell migration, cell aggregation, and anisotropic spreading, which cannot be captured by the assumption of purely diffusive movement, to be modelled. It is also possible to extend our modelling framework to incorporate cell-cell adhesion or cell swapping \cite{noureen2025mas}. Including these extensions to our model could provide a highly realistic representation of both cell proliferation and movement.

In this work, we have demonstrated that different cell cycle modelling assumptions can lead to markedly different behaviour at the population level. Within the ABM framework, we showed that the Remain and Myopic models produce distinct growth dynamics, and that multi-stage models can, under certain density-dependent conditions, exhibit faster population expansion than simpler exponential models. These findings may have important experimental implications, particularly when inferring cell cycle time distributions from data collected in crowded environments. While the continuum framework performed well when the underlying mean-field assumptions are satisfied, we showed that parameter choices in the ABMs strongly influence the cellular spatial structure, which in turn affect the agreement between discrete and continuum descriptions. Our results can be used to determine whether continuum or individual based approaches are appropriate for modelling cell populations. 

\section{Acknowledgements}
JCD is supported by a scholarship from the EPSRC Centre for Doctoral Training in Statistical Applied Mathematics at Bath (SAMBa), under the project EP/Y034716/1. CAY and CAS are supported by a Leverhulme Grant Ref. RPG-2024-104. This co

\section{Code availability}
Code employed in this paper is accessible at \\ \url{https://github.com/jcd65/msm-volume-exclusion-cell-models}

\appendix
\section{Deriving the continuum models for Myopic ABMs} \label{Appendix: Derivation}
In this appendix we give more details on the derivations of the continuum equations for the Myopic Reset (\ref{Myopic reset: PDEs (hold)}) and Myopic Remain (\ref{Myopic remain: PDEs (hold)}) models. In the master equations (\ref{Myopic reset stage 1 PDE derivation}) and (\ref{Myopic remain stage 1 PDE derivation}), we can separate out the contributions due to cell movement, cell stage progression, and cell proliferation. As the contributions arising from cell movement and cell cycle progression are independent of the proliferative mechanism, they are identical to those in the non-myopic models. Therefore, we need only consider the contribution due to proliferation. Let $p(t+\tau) - p(t)$ denote the contribution to the change in cell density at lattice site $(i,j)$ over a small time step $\tau$ due solely to proliferation events under a Myopic model. Then $p(t+\tau) - p(t)$ takes the form:
\begin{align}
    p(t+\tau) - p(t) &= \lambda \tau \left( C_{i-1,j}^{(K)}P((i,j)\mid(i-1,j)) +
    C_{i+1,j}^{(K)}P((i,j)\mid(i+1,j)) 
    \right)
    \notag \\
    &+ \lambda \tau \left( C_{i,j-1}^{(K)}P((i,j)\mid(i,j-1)) +
    C_{i,j+1}^{(K)}P((i,j)\mid(i,j+1))\right),
    \label{Appendix - proliferative term}
\end{align}
where $P((i,j)\mid (i',j'))$ denotes the probability that a myopic cell at site $(i',j')$ chooses to attempt to proliferate into site $(i,j)$. We will show that the proliferative term evolves according to the following ODE:
\begin{equation}
    \ordder{p}{t} = \lambda C^{(K)}(1-C^4),
    \label{Appenix: prolif contributon}
\end{equation}
under the assumption that as $\tau, \Delta \rightarrow 0$, we hold $\lambda$ constant. 

If we expand the terms $C_{i\pm1, j}, C_{i,j\pm1}$ around site $(i,j)$ in equation (\ref{Appendix - proliferative term}), the we obtain the following equation
\begin{align}
    p(t+ \tau) - p(t)
    &= \tau\lambda C_{i,j}^{(K)} \sum_{i',j'} P((i,j) \mid (i',j')) \notag \\
    &+ \tau O(\Delta) 
    \label{Appendix: intermediate}
\end{align}
where $\sum_{i',j'} P((i,j) \mid (i',j'))$ is the sum over all possible neighbouring sites so we have that $(i',j') \in \{ (i\pm 1, j), (i, j\pm 1) \}$. We will now simplify equation (\ref{Appendix: intermediate}), by expanding the terms $P((i,j) \mid (i',j'))$, and finding a simpler expression for the sum of the probabilities. 

The probabilities $P((i,j)\mid (i\pm 1,j)), P((i,j)\mid (i,j\pm 1))$ in equation (\ref{Appendix - proliferative term}) can be written,
\begin{align}
P((i,j)\mid(i\pm 1,j)) 
&= (1 - C_{i,j}) \times\sum_{l=0}^{3} \mathcal{N}_l(i\pm1, j), \notag \\
P((i,j)\mid(i ,j \pm 1)) 
&= (1 - C_{i,j}) \times\sum_{l=0}^{3} \mathcal{N}_l(i, j\pm1),
\label{Appendix: Myopic proliferation probability}
\end{align}
where each $\mathcal{N}_l(i',j')$ represents the probability that a cell at site $(i',j')$ with $l$ occupied neighbours chooses to proliferate into site $(i,j)$. For the horizontal neighbours of site $(i,j)$, recall that we can write $\mathcal{N}_l = \mathcal{N}_l(i\pm1, j)$ as below:

 \begin{align}
     \mathcal{N}_0 &= \frac{1}{4} \left( 
    (1 - C_{i\pm2,j})(1 - C_{i\pm1,j-1})(1 - C_{i\pm1,j+1}) 
\right) \notag \\
    \mathcal{N}_1 &= \frac{1}{3} 
    C_{i\pm2,j}(1 - C_{i\pm1,j+1})(1 - C_{i\pm1,j-1}) \notag \\
    &+ \frac{1}{3}C_{i\pm1,j-1}(1 - C_{i\pm2,j})(1 - C_{i\pm1,j+1})  \notag \\
    &+ \frac{1}{3}C_{i\pm1,j+1}(1 - C_{i\pm2,j})(1 - C_{i\pm1,j-1}) \notag \\
    \mathcal{N}_2 &= \frac{1}{2} 
    C_{i\pm2,j}C_{i\pm1,j-1}(1 - C_{i\pm1,j+1}) \notag \\
    &+ \frac{1}{2}C_{i\pm2,j}C_{i\pm1,j+1}(1 - C_{i\pm1,j-1})  \notag \\
    &+ \frac{1}{2}C_{i\pm1,j+1}C_{i\pm1,j-1}(1 - C_{i\pm2,j}) \notag \\
    \mathcal{N}_3 &= C_{i\pm2,j}C_{i\pm1,j-1}C_{i\pm1,j+1},
    \label{Appendix: (i+1,j)}
 \end{align}
and we have similar forms for the terms $\mathcal{N}_l(i, j\pm1)$. We now give the Taylor expansion of the terms $\mathcal{N}_l $ for each $l=0,1,2,3$ about site $(i,j)$. This gives the following expansions,
\begin{align}
    \mathcal{N}_0 &= \frac{1}{4}(1-C_{i,j})^3 + O(\Delta)\notag \\
    \mathcal{N}_1 &= C_{i,j}(1-C_{i,j})^2 + O(\Delta) \notag \\
    \mathcal{N}_2 &= \frac{3}{2}C_{i,j}^2(1-C_{i,j}) + O(\Delta) \notag \\
    \mathcal{N}_3 &= C_{i,j}^3 + O(\Delta).
    \label{appendix: mathcal Ns}
\end{align} 
Using equation~(\ref{Appendix: Myopic proliferation probability}), this gives explicit forms for $P((i,j) \mid (i\pm1,j))$ and $(P(i,j) \mid (i,j\pm1))$. Summing these four probabilities gives the following equation:
\begin{align}
    &\sum_{i',j'} P((i,j) \mid (i',j')) = \notag \\ 
    &(1-C_{i,j})^4 + 4C_{i,j}(1-C_{i,j})^3 +6C_{i,j}^2(1-C_{i,j})^2 + 4C_{ij}^3(1-C_{i,j}), \notag \\
    &= (1-C_{i,j}^4).
    \label{Appendix: summing probabilities}
\end{align}

 Dividing equation~(\ref{Appendix - proliferative term}) by $\tau$ gives the following (when expanding to $O(\Delta)$):
\begin{align}
    \frac{p(t+\tau) - p(t)}{\tau} &= \lambda  \left( C_{i-1,j}^{(K)}P((i,j)\mid(i-1,j)) +
    C_{i+1,j}^{(K)}P((i,j)\mid(i+1,j)) 
    \right)
    \notag \\
    &+ \lambda \left( C_{i,j-1}^{(K)}P((i,j)\mid(i,j-1)) +
    C_{i,j+1}^{(K)}P((i,j)\mid(i,j+1))\right) \notag \\
    &= \lambda  C_{i,j}^{(K)} \sum_{i',j'} P((i,j) \mid (i',j')) \notag \\
    &+ O(\Delta) \notag \\
    &= \lambda C_{i,j}^{(K)}(1-C_{i,j}^4) + O( \Delta),
    \label{appendix: final master eq}
\end{align}
where we used equations (\ref{Appendix: intermediate}) and (\ref{Appendix: summing probabilities}) to simplify the resulting expressions.

If we take the limit of equation (\ref{appendix: final master eq}) as $\tau, \Delta \rightarrow 0$, then we obtain the following ODE
\begin{equation}
    \ordder{p}{t} = \lambda C^{(K)}(1-C^4).
    \label{Appenix: prolif contributon again}
\end{equation}
This is precisely the ODE~(\ref{Appenix: prolif contributon}). 

Recall that the other contributions to the PDE for stage-$1$ cells under the Myopic Reset and Myopic Remain models from the master equations (\ref{Myopic reset stage 1 PDE derivation}, \ref{Myopic remain stage 1 PDE derivation}) involve cell movement and cell progression terms which are not impeded by the proliferative mechanism and so  are the same as those of the non-myopic models. This allows us to use the diffusive and proliferative terms in PDEs~(\ref{Reset: PDEs (hold)}) and (\ref{Remain: PDEs (hold)}) to obtain the following PDE for stage $1$ cells under the Myopic Reset model:
\begin{equation}
    \label{appendix: stage 1 myopic reset}
            \partder{C^{(1)}}{t} = D [(1-C)\nabla^2C^{(1)} + C^{(1)}\nabla^2C] + \lambda C^{(K)}(2-C^4)  - \lambda C^{(1)}.
\end{equation}
Similarly, for the Myopic Remain model, adapting equation (\ref{Remain: PDEs (hold)})  we have the resultant PDE
\begin{equation}
    \label{appendix: stage 1 myopic remain}
            \partder{C^{(1)}}{t} = D [(1-C)\nabla^2C^{(1)} + C^{(1)}\nabla^2C] + 2\lambda C^{(K)}(1-C^4)  - \lambda C^{(1)}.
\end{equation}

\section{PCF plots for ABMs other than Myopic Remain} \label{Appendix: B PCF}
In Figure \ref{Appendix: PCF comparisons} below, we plot the pair correlation functions (PCFs) for ABMs other than the Myopic Remain model. Recall that given an $(X \times Y)$ lattice, we can calculate the PCF at distance $d$ in repeat $m$ using the following equation:
\begin{equation}
f^m(d) = \frac{(XY-1)|C^M(d)|}{2dN(N-1)},
    \label{Appendix: PCF}
\end{equation}
where $N$ is the number of cells in the lattice, and $| C^M(d)|$ is the number of pairs of cells separated by distance $d$. This definition is still valid for our different ABMs, as each simulation uses the same boundary conditions (periodic). We considered the average PCF $\hat f(d)$ over multiple realisations of each ABM.

As in the Myopic Remain model, we see that for motility rate $r_m=1$, each ABM demonstrates strong pairwise correlations for small distances $d$. At motility rate $r_m=100$, we see the strength of correlation decreases, but is still present in every model as in the Myopic Remain model. When we increase the motility rate to $r_m=1000$, this effectively removes the pairwise correlations in every model, as $\hat f(d) \approx 1$ for all distances $d$. There are slight differences in the value $|\hat f(d) - 1|$, in each model, but it is clear that the strength of correlations between lattice sites is approximately the same under every ABM.
\begin{figure}[H] 
\centering
\subfigure[][]{
	\includegraphics[width=0.45\textwidth]{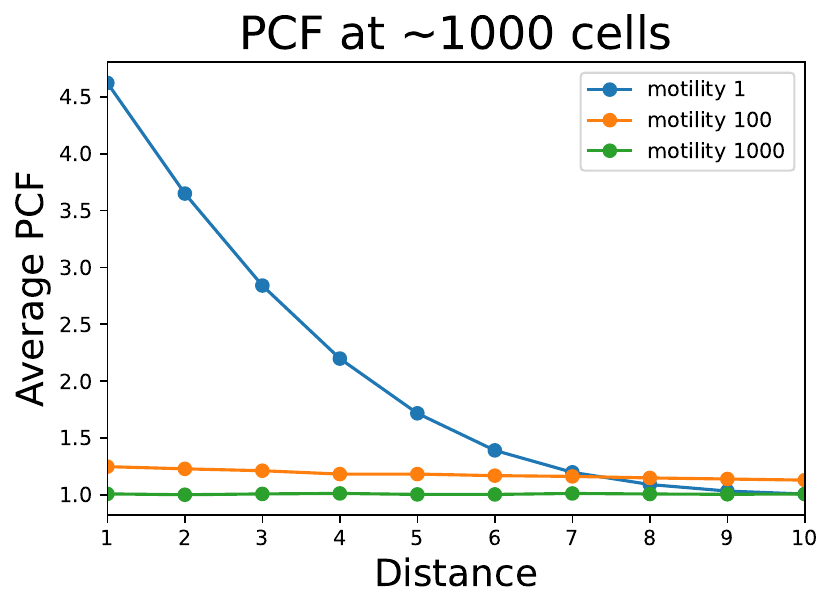}
}
\subfigure[][]{
	\includegraphics[width=0.45\textwidth]{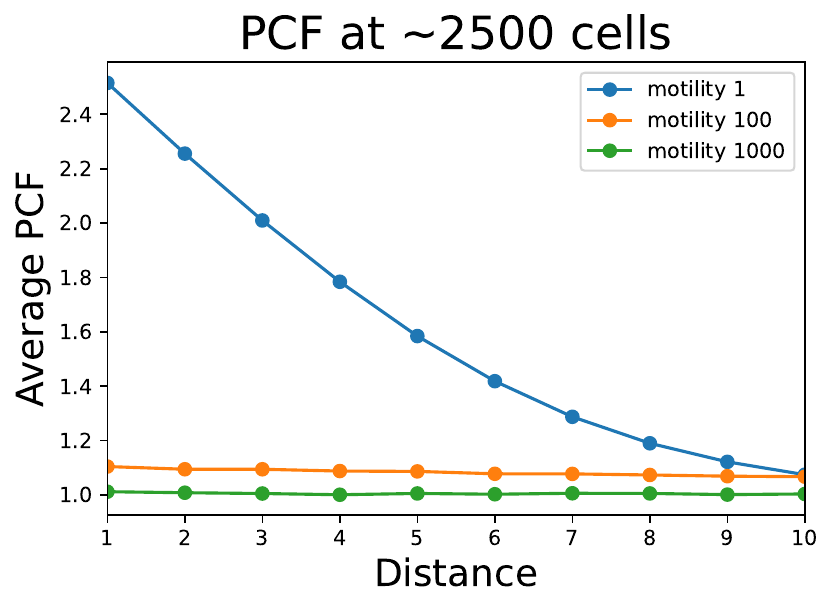}
}
\subfigure[][]{
	\includegraphics[width=0.45\textwidth]{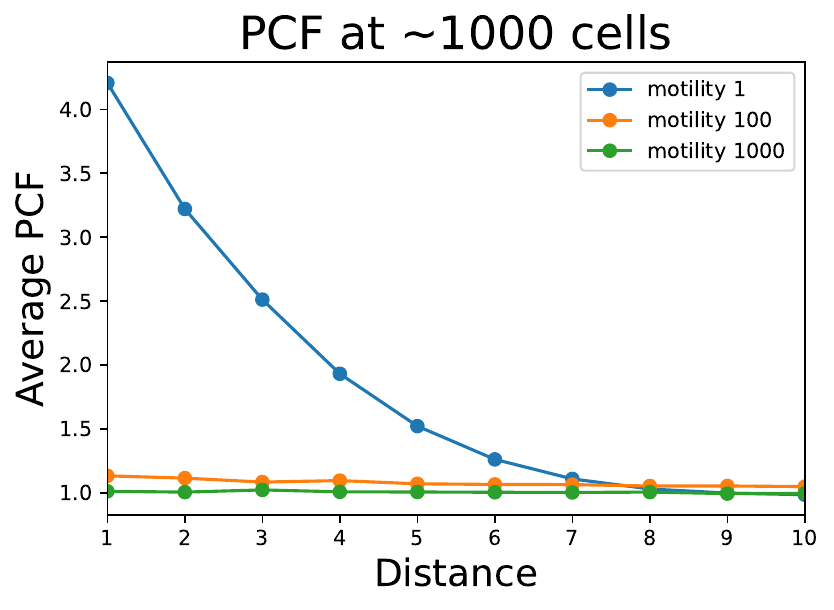}
}
\subfigure[][]{
	\includegraphics[width=0.45\textwidth]{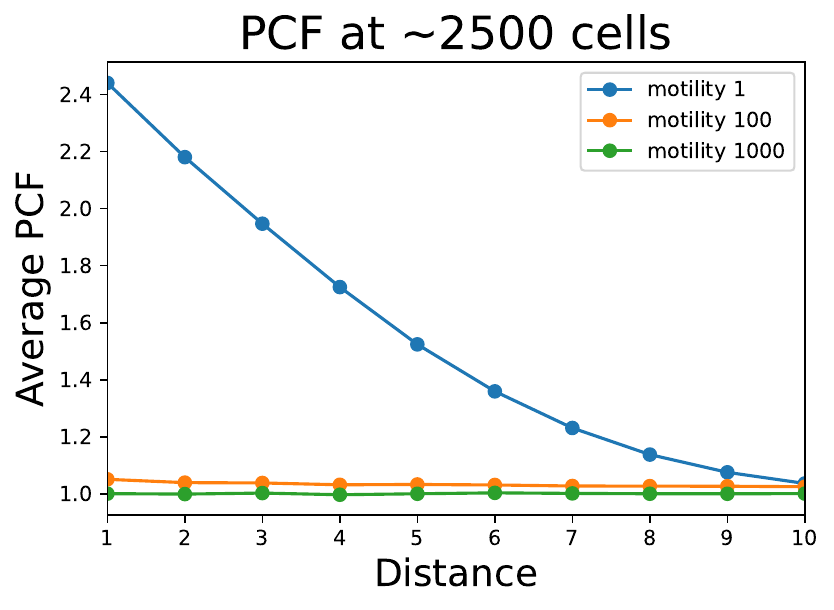}
}
\subfigure[][]{
	\includegraphics[width=0.45\textwidth]{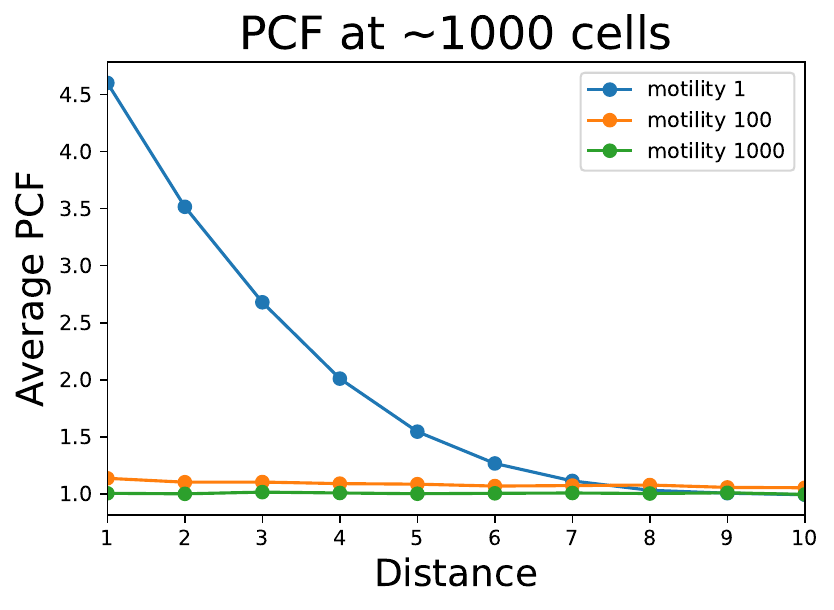}
}
\subfigure[][]{
	\includegraphics[width=0.45\textwidth]{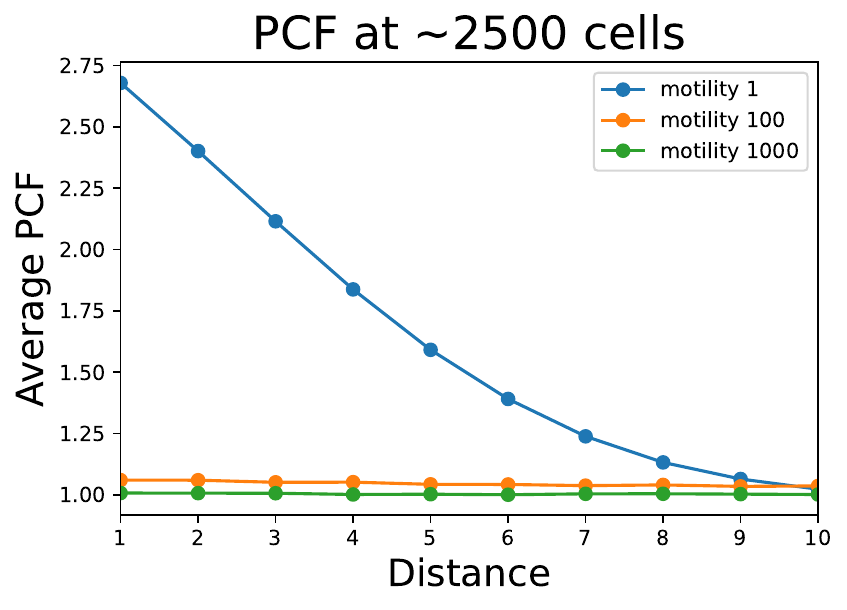}
}
\caption{Plotting the PCFs $\hat f(d)$ over distances $d=1,2,...,10$ for non-Myopic models averaged over $20$ realisations of different ABMs for motility rates $r_m=1,100,1000$ with a fixed proliferation rate $r_p = 1$ and number of stages $K=10$. For each model, in each realisation, we record the time point where the ABM first exceeds $1000$ and $2500$ cells, and use this to calculate the average PCF. Panels (a) and (b) show the results of the Exponential model,  panels (c) and (d) show the results of the Reset model, and panels (e) and (f) show the results of the Remain model.}
\label{Appendix: PCF comparisons}
\end{figure}

\begin{figure}[H] 
\centering
\subfigure[][]{
	\includegraphics[width=0.45\textwidth]{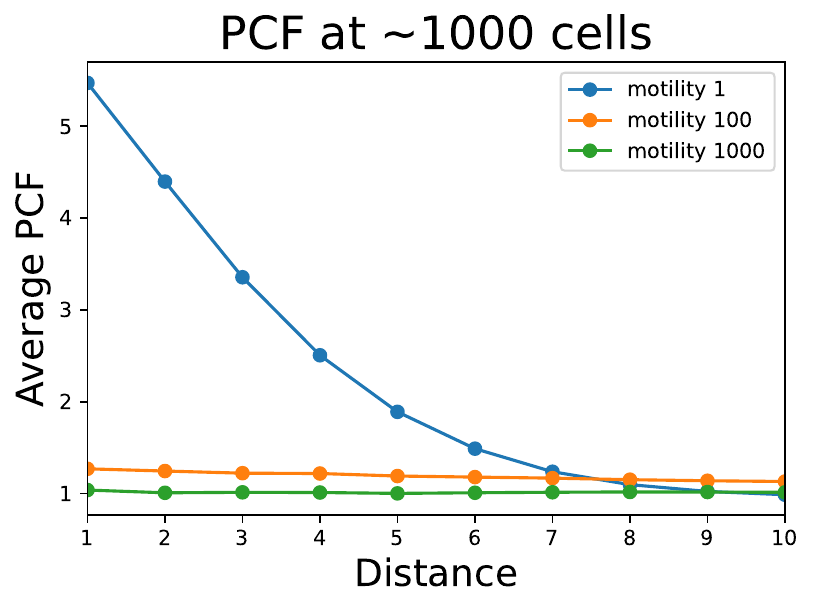}
}
\subfigure[][]{
	\includegraphics[width=0.45\textwidth]{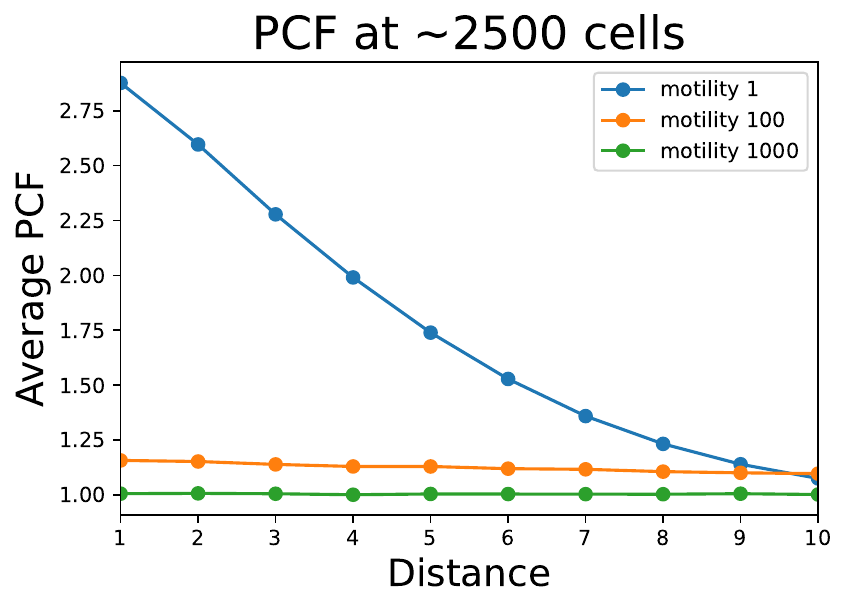}
}
\subfigure[][]{
	\includegraphics[width=0.45\textwidth]{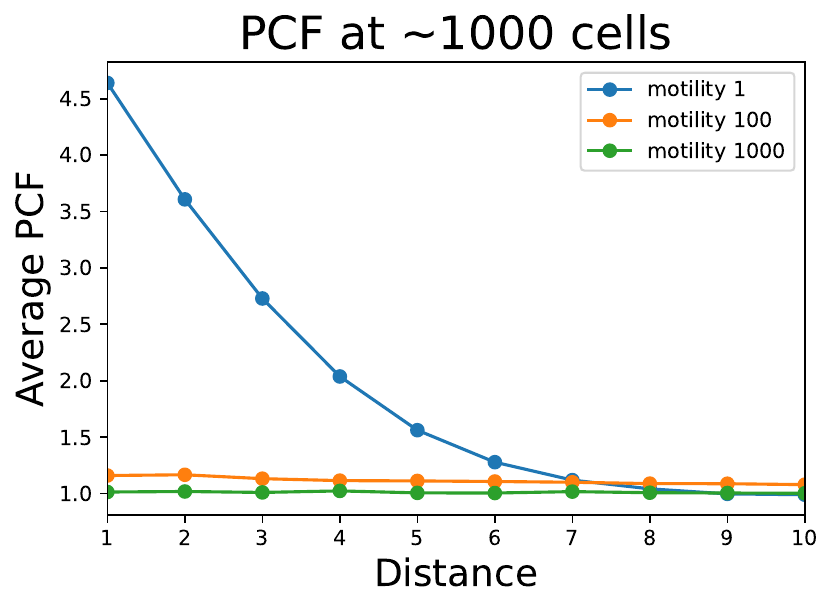}
}
\subfigure[][]{
	\includegraphics[width=0.45\textwidth]{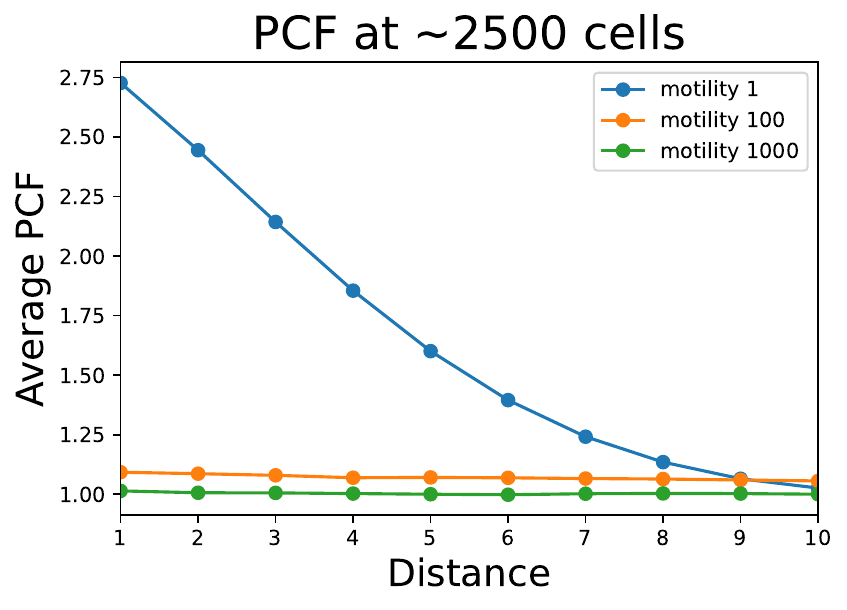}
}
\end{figure}

\begin{figure}[H] 
\centering
\caption{Plotting average PCFs for Myopic models, as in Figure \ref{Appendix: PCF comparisons}. Panels (a) and (b) show the results of the Myopic Exponential model, panels (c) and (d) show the results of the Myopic Reset model.}
\label{Appendix: PCF comparisons 2}
\end{figure}
\bibliographystyle{unsrtnat}
\bibliography{references}
\end{document}